\documentclass{emulateapj}

\newcommand{\citepeg}[1]{\citep[{e.g.,}][]{#1}}

\shorttitle{The Swift GRB Host Galaxy Legacy Survey}
\shortauthors{Perley et al.}

\begin{document}

\title{The Swift Gamma-Ray Burst Host Galaxy Legacy Survey--- \\I. Sample Selection and Redshift Distribution}

\def\cit{1}
\def\mail{*}
\def\dark{2}
\def\eso{3}
\def\puc{4}
\def\mia{5}
\def\iaa{6}
\def\harvard{7}
\def\gsfc{8}
\def\jssi{9}
\def\ipac{10}
\def\einstein{11}
\def\arizona{12}
\def\mpe{13}
\def\iceland{14}
\def\warwick{15}
\def\edinburgh{16}
\def\leicester{17}

\author{D.~A.~Perley\altaffilmark{\cit,\dark,\mail},
T.~Kr\"uhler\altaffilmark{\eso,\dark},
S.~Schulze\altaffilmark{\puc,\mia},
A.~de~Ugarte~Postigo\altaffilmark{\iaa},
J.~Hjorth\altaffilmark{\dark}, \\
E.~Berger\altaffilmark{\harvard},
S.~B.~Cenko\altaffilmark{\gsfc,\jssi},
R.~Chary\altaffilmark{\ipac},  
A.~Cucchiara\altaffilmark{\gsfc},
R.~Ellis\altaffilmark{\cit},
W.~Fong\altaffilmark{\einstein,\arizona},
J.~P.~U.~Fynbo\altaffilmark{\dark}, \\
J.~Gorosabel\altaffilmark{\iaa}, 
J.~Greiner\altaffilmark{\mpe},
P.~Jakobsson\altaffilmark{\iceland},
S.~Kim\altaffilmark{\puc,\mia},
T.~Laskar\altaffilmark{\harvard},
A.~J.~Levan\altaffilmark{\warwick},
M.~J.~Micha{\l}owski\altaffilmark{\edinburgh}, \\
B.~Milvang-Jensen\altaffilmark{\dark}, 
N.~R.~Tanvir\altaffilmark{\leicester},
C.~C.~Th\"one\altaffilmark{\iaa}, and
K.~Wiersema\altaffilmark{\leicester}
}


\altaffiltext{\cit}{Department of Astronomy, California Institute of Technology, MC 249-17, 1200 East California Blvd, Pasadena CA 91125, USA}
\altaffiltext{\dark}{Dark Cosmology Centre, Niels Bohr Institute, University of Copenhagen, Juliane Maries Vej 30, 2100 K{\o}benhavn {\O}, Denmark}
\altaffiltext{\eso}{European Southern Observatory, Alonso de C\'ordova 3107, Vitacura, Casilla 19001, Santiago 19, Chile}
\altaffiltext{\puc}{Instituto de Astrof\'isica, Facultad de F\'isica, Pontificia Universidad Cat\'olica de Chile, Vicu\~{n}a Mackenna 4860, 7820436 Macul, Santiago 22, Chile}
\altaffiltext{\mia}{Millennium Institute of Astrophysics, Vicu\~na Mackenna 4860, 7820436 Macul, Santiago, Chile}
\altaffiltext{\iaa}{Instituto de Astrofísica de Andaluc\'ia (IAA-CSIC), Glorieta de la Astronom\'ia s/n, E-18008, Granada, Spain}
\altaffiltext{\harvard}{Harvard-Smithsonian Center for Astrophysics, 60 Garden Street, Cambridge, MA 02138, USA}
\altaffiltext{\gsfc}{Astrophysics Science Division, NASA/Goddard Space Flight Center, Greenbelt, MD 20771}
\altaffiltext{\jssi}{Joint Space Science Institute, University of Maryland, College Park, MD 20742}
\altaffiltext{\ipac}{US Planck Data Center, MS220-6, Pasadena, CA 91125, USA}
\altaffiltext{\einstein}{Einstein Fellow}
\altaffiltext{\arizona}{Steward Observatory, University of Arizona, 933 North Cherry Avenue, Tucson, AZ 85721}
\altaffiltext{\mpe}{Max-Planck-Institut f\"ur extraterrestrische Physik, Giessenbachstrasse, 85748 Garching, Germany}
\altaffiltext{\iceland}{Centre for Astrophysics and Cosmology, Science Institute, University of Iceland, Dunhagi 5, 107 Reykjav\'ik, Iceland}
\altaffiltext{\warwick}{Department of Physics, University of Warwick, Coventry CV4 7AL, UK}
\altaffiltext{\edinburgh}{Scottish Universities Physics Alliance, Institute for Astronomy, University of Edinburgh, Royal Observatory, Edinburgh, EH9 3HJ, UK}
\altaffiltext{\leicester}{Department of Physics and Astronomy, University of Leicester, University Road, Leicester, LE1 7RH, UK}
\altaffiltext{\mail}{e-mail: dperley@dark-cosmology.dk}

\slugcomment{Accepted to ApJ 2015-10-31}

\begin{abstract}
We introduce the \textit{Swift} Gamma-Ray Burst Host Galaxy Legacy Survey (``SHOALS''), a multi-observatory high-redshift galaxy survey targeting the largest unbiased sample of long-duration gamma-ray burst hosts yet assembled (119 in total).   We describe the motivations of the survey and the development of our selection criteria, including an assessment of the impact of various observability metrics on the success rate of afterglow-based redshift measurement.   We briefly outline our host-galaxy observational program, consisting of deep Spitzer/IRAC imaging of every field supplemented by similarly-deep, multi-color optical/NIR photometry, plus spectroscopy of events without pre-existing redshifts.  Our optimized selection cuts combined with host-galaxy follow-up have so far enabled redshift measurements for 110 targets (92\%) and placed upper limits on all but one of the remainder.  About 20\% of GRBs in the sample are heavily dust-obscured, and at most 2\% originate from $z>5.5$.  Using this sample we estimate the redshift-dependent GRB rate density, showing it to peak at $z\sim2.5$ and fall by at least an order of magnitude towards low ($z=0$) redshift, while declining more gradually towards high ($z\sim7$) redshift.  This behavior is consistent with a progenitor whose formation efficiency varies modestly over cosmic history.   Our survey will permit the most detailed examination to date of the connection between the GRB host population and general star-forming galaxies, directly measure evolution in the host population over cosmic time and discern its causes, and provide new constraints on the fraction of cosmic star-formation occurring in undetectable galaxies at all redshifts.
\end{abstract}


\section{Introduction}
\label{sec:intro}

Long-duration gamma-ray bursts (GRBs) are relativistic, jetted explosions of very massive stars at the end of their lives.\footnote{They are physically distinct from short ($T_{90} < 2$ sec) gamma-ray bursts, which are believed to be associated with merging compact objects \citep{Berger+2014} and are not discussed in this work.}  The peak luminosities of GRBs and their afterglows exceed those of the most luminous supernovae, galaxies and quasars by orders of magnitude \citepeg{Bloom+2009} throughout the electromagnetic spectrum.  As a result GRBs are routinely detected from great distances: the median redshift of \textit{Swift} GRBs is $z\sim2$ \citep{Jakobsson+2006,Fynbo+2009} and GRBs have been detected up to a redshift of $z=8-9$ \citep{Tanvir+2009,Salvaterra+2009,Cucchiara+2011}.

GRBs are therefore fundamentally cosmological objects, and their study is intimately coupled with that of high-redshift galaxies and cosmic evolution.
GRB afterglows make excellent probes of individual galaxy sightlines, a technique that has been extensively exploited to characterize the interstellar medium in distant star-forming galaxies---chemical abundances, kinematics, and dust properties---to a level of detail unmatched by any other technique \citepeg{Prochaska+2007,Prochaska+2009,Eliasdottir+2009,Delia2011,Zafar+2012,DeCia+2013,Thoene+2013,Sparre+2014,Friis+2015,Cucchiara+2015}.  
They may also be useful for studies of galaxy and cosmic evolution in a broader sense.  Because GRBs originate from short-lived massive stars, they stochastically sample the sites of cosmic star formation in proportion to their relative contributions to the cosmic total, providing (in principle) a means of estimating the importance of different epochs or different galaxy populations to the Universe's stellar mass assembly \citepeg{Totani1997,Wijers+1998,Blain+2000,Porciani+2001}. This includes populations difficult to study by other means: very low-luminosity galaxies \citep{Berger+2002,Schulze+2015,Greiner+2015}, very dusty galaxies \citep{Djorgovski+2001,RamirezRuiz+2002,Berger+2003,Tanvir+2004,LeFloch+2006,Michalowski+2008,Perley+2009a,Perley+2013a,Perley+2015a,Rossi+2012,Greiner+2011,Hunt+2014,Schady+2014,Kohn+2015}, very high-redshift galaxies \citep{Lamb+2000,Tanvir+2012,Trenti+2012}, and possibly even Population III stars \citep{Bromm+2006,Campisi+2011b}.

These broader questions of using GRBs as probes of galaxy evolution and cosmology have provided major drivers for GRB research over the past two decades and in particular since 2004, when the launch of the \textit{Swift} satellite \citep{Gehrels+2004} revolutionized the field with its ability to detect and instantly localize GRBs in large numbers ($\sim$100 per year).  The vast (and still-expanding) \textit{Swift} legacy data-set, including follow-up provided by the world-wide ground-based community, has made possible the study of GRBs, their environments, and their host galaxies in sufficiently large numbers to investigate these topics statistically.

The cosmic star-formation rate is not necessarily the only factor expected to affect the GRB rate, however.  The initial chemical composition of a star influences its fate, and many models specifically predict that the GRB rate in metal-poor environments should be different---typically, higher---than in metal-rich ones \citepeg{MacFadyen+1999,Fryer+2005,Hirschi+2005,Yoon+2005,Woosley+2006}.  Alternatively, variations in the IMF or close-binary fraction could affect the GRB rate \citep{Dave2008,Wang+2011}, as might dynamical interactions in dense clusters \citep{vandenHeuvel+2013}.  If present, these effects would have to be taken into account when attempting to employ GRBs to investigate broader cosmological issues.

A non-constant GRB production efficiency (defined as the GRB rate relative to SFR) is supported by several studies of the GRB host population at $z \lesssim 1$ showing known GRB hosts to be smaller, bluer, less massive, and less chemically enriched on average than typical star-forming galaxies or core-collapse supernova hosts \citep{Fruchter+2006,LeFloch+2006,Stanek+2006,Modjaz+2008,Svensson+2010,Levesque+2010mz,Graham+2013,Boissier+2013,Kelly+2014,Vergani+2015}.  Furthermore, a number of studies have concluded that the comoving GRB rate density (as inferred from the redshift distribution) evolves differently from the star-formation rate density, showing an excess of GRBs originating from $z\gtrsim3$ compared to what would be expected if the GRB rate was a fixed fraction of the star-formation rate in all types of galaxy and at all redshifts \citepeg{Daigne+2006,Le+2007,Guetta+2007b,Salvaterra+2007,Yueksel+2008,Kistler+2008}.

Even so, the actual nature of the connection between the GRB rate and star-formation rate is not yet well-determined.  It is not certain whether the nature of the low-$z$ population is best explained by a dependence on metallicity or on another factor which correlates with it, such as star-formation rate intensity \citep{Kocevski+2011,Kelly+2014}.  Other authors dispute whether the GRB rate varies significantly with host galaxy or with redshift at all (e.g., \citealt{Wanderman+2010,Campisi+2011a,Elliott+2012,Michalowski+2012,Hunt+2014,Kohn+2015}).  

There are several complicating factors that affect these studies.   Many of the best-studied hosts are those of the so-called low-luminosity GRBs that dominate the rate locally ($z<0.25$)---but which have properties very different (orders of magnitude in energetics, intrinsic rate, and degree of beaming) from the cosmological GRBs that are detected at high redshifts \citep{Pian+2006,Guetta+2007a,Wiersema+2007,Virgili+2009,Bromberg+2011}.  Despite their much greater observed numbers, the hosts of the distant, cosmological GRBs are more difficult to study and have been explored less consistently and in much less detail, and major selection biases underly our ability to produce representative samples.  The studies of the cosmological GRB population that have been conducted to date have in all cases been small ($<20$ objects), heavily selection-biased, or characterized the hosts only minimally (observations in 1--2 photometric bands, often to limited depth.)  

In this paper we introduce and summarize the largest and most ambitious survey of the GRB host galaxy population conducted to date, designed to move past these earlier limitations by constructing a sample which is \emph{large}, \emph{unbiased}, and \emph{thoroughly observed} at a variety of wavelengths across the electromagnetic spectrum.  Our effort, which we designate the \textit{Swift} gamma-ray burst HOst gAlaxy Legacy Survey (or ``SHOALS''), provides the most complete view (in wavelength, redshift, and sample construction) yet of the environments in which GRBs explode in order to rigorously examine all aspects regarding the connection between gamma-ray bursts and galaxies across cosmic history.  What is the GRB rate history?   What is the distribution of properties among galaxies that host GRBs, how does it compare to the properties of galaxies thought to dominate the Universe's star-formation activity, and how does this change with redshift?  What fraction of the population originates from obscured regions or from dusty galaxies?  What fraction originates from galaxies too faint to be detected at all?  What can we learn by combining properties measured by observing GRB afterglows in absorption (dust columns, kinematics, and gas/metal columns) with those learned by studying their hosts in emission (mass, star-formation rate, morphology, and nebular abundance measurements)?   Our survey seeks to answer these and similar questions.

This paper provides an introduction to the survey, including an explanation of our selection methodology (\S \ref{sec:sample}--\ref{sec:selection}), a summary of our observational campaign (\S \ref{sec:observations}),  presentation of our new redshift determinations (\S \ref{sec:redshift}), as well as tables detailing the properties of the sample.  Using these observations, which establish the largest highly-complete GRB redshift sample to date, we provide a measurement of the inferred cosmic GRB rate density (\S \ref{sec:results}) unbiased by selection effects, and discuss its relation to the cosmic star-formation rate density (\S \ref{sec:discussion}).  We summarize these results and the project in general in \S \ref{sec:summary}.

Subsequent papers will summarize science results associated with the host-galaxy campaign, an ongoing effort to provide complete SED measurements for all our targets.  In particular, our Spitzer observations and the NIR luminosity and stellar mass distribution inferred from will be presented in Paper II \citep{Paper2}, which is submitted concurrently with this work.  Additional papers will focus on other science questions, including the UV luminosity distribution and its evolution as inferred from the optical photometry, careful determination of the host-galaxy physical properties (SFR, mass, etc.) from SED fitting and an assessment of their impact on the GRB rate, detailed analysis of correlations between properties of a host and properties of its afterglow, the spectroscopic properties of the host galaxies, an investigation into candidate hosts of foreground absorbing systems seen along GRB sightlines, and further topics.

\section{Sample Considerations}
\label{sec:sample}

\subsection{Biases Affecting Redshift Measurement}

\textit{Swift} detected 803 long-duration ($T_{90} > 2$s) GRBs from the start of the mission through to the end of 2014, and approximately 320 of these events have measured redshifts: a potentially enormous sample to draw from.  However, from the point of view of characterizing the host population or redshift distribution, attention must be paid to systematic biases as well as raw number statistics.  In particular, the subset of \textit{Swift} GRBs with measured redshift is not only incomplete (40\% of the population) but very likely to be biased:  the ease of providing a redshift measurement for a GRB is expected to be dependent on properties of its host galaxy and on the redshift itself \citep{Coward+2013}.

The vast majority of GRB redshifts in the \textit{Swift} era are provided by absorption spectroscopy of the optical afterglow, and GRB afterglows show great variety in their optical brightnesses, especially at early times \citep{Akerlof+2007,Kann+2010}.   In particular, the afterglows of some events are sufficiently faint ($R \gtrsim 22$ mag, even during the first hour) that they cannot be detected by the small-to-moderate aperture telescopes typically used to identify the optical afterglow before triggering large spectrographs (and in some cases, even deep optical imaging at large telescopes fails to detect an afterglow).  These are typically referred to as ``dark'' GRBs \citepeg{Groot+1998,Fynbo+2001}\footnote{Alternatively, a GRB can be defined as ``dark'' based on the degree of optical faintness relative to X-ray wavelengths (e.g. via $\beta_{\rm OX}$; \citealt{Jakobsson+2004,vanderHorst+2009})---which isolates optically-absorbed bursts specifically, including very luminous afterglows that can shine through thick dust screens but excluding GRBs whose faintness is intrinsic.  Since this work is focused on afterglow sample completeness (and not necessarily its underlying causes) the former, ``absolute'' definition is generally more relevant here.  We employ the term loosely to refer to GRBs with optical afterglows that are sufficiently faint that afterglow-based redshift recovery is unusually challenging or impossible.}; redshift catalogs based primarily on publicly-reported afterglow redshifts necessarily exclude them and are therefore intrinsically optically-biased.  This bias removes GRBs in dusty environments or at very high redshifts, two of the most interesting regimes we might wish to use GRBs to explore, and without which our view of the redshift distribution and host population are severely incomplete.

Fortunately there are other means available to us to recover the GRB redshifts even without an optical detection: GRB afterglows are luminous across the entire electromagnetic spectrum, and the afterglow \emph{position} can often be recovered via observations at near-infrared, X-ray, or radio wavelengths, permitting the host galaxy to be identified and its redshift measured spectroscopically in emission (or, less optimally, photometrically).  Historically, subarcsecond follow-up at wavelengths outside the optical band was much less routine than optical follow-up and the problem was nearly intractable from a statistical point of view: dark GRB hosts and redshifts could be recovered in a few special cases \citepeg{Djorgovski+2001} but most of them were lost forever without an accurate localization.  

This situation was transformed by \textit{Swift}.  The satellite is equipped with an on-board focusing X-Ray Telescope (XRT; \citealt{Burrows+2005}) and the capability of automatically slewing to a field immediately after detecting an event with its Burst Alert Telescope (BAT; \citealt{Barthelmy+2005}), meaning that early time, high spatial-resolution X-ray coverage is available for every event the BAT detects.   XRT observations of Swift GRBs almost always detect the afterglow: 98\% of long GRBs are detected by XRT \citep{Burrows+2007}, and if events for which the XRT follow-up was delayed are excluded this figure rises to 100\%.  
So even without an optical counterpart it is almost always possible---with sufficient effort---to recover (or place a deep upper limit on) the host galaxy using the $\sim1.5\arcsec-2.0\arcsec$ X-ray positions routinely delivered by the XRT \citep{Butler2007,Goad+2007,Evans+2009}.  These rapidly-available positions also make it much easier to identify faint optical counterparts (or to recognize their absence and trigger deeper optical follow-up or multi-wavelength observations) for bursts whose ``darkness'' is borderline.

Host-galaxy spectroscopy (and host-galaxy observations in general) is observationally expensive and can only be carried out for a limited number of objects: recovery of \emph{all} missing \textit{Swift} redshifts is unfeasible.  Moreover, absorption redshifts can be missed for a variety of reasons other than intrinsic darkness:  GRBs occur at random times and directions and some cannot be followed if, for example, the location is too close in projection to the Sun or the Galactic plane.  Indeed, about half of \textit{Swift} GRBs are probably missed for mundane reasons.  Several studies have estimated an intrinsic dark fraction of about 20\% \citep{Cenko+2009,Greiner+2011,Melandri+2012}, so 80\% of \textit{Swift} GRBs should have afterglows bright enough in principle for optical spectroscopy, yet in only 30\% of cases are redshifts actually reported.  Expending effort and observational resources to recover hosts and redshifts for events missed for these reasons would not be informative.  

At the same time, for the reasons discussed above, we cannot simply neglect GRBs without afterglow redshifts without also discarding obscured or high-$z$ events from the sample.  Two general approaches are available to characterize the complete population.  The first is to simply take the known-$z$ population and attempt to ``correct'' it by also observing a limited number of GRBs that are definitely dark (have deep, early optical limits) and combining the samples.  This has been adopted in the past \citepeg{Kruehler+2011,Rossi+2012,Perley+2013a,Hunt+2014} and among other things has clearly shown that GRBs with very faint optical afterglows probe a different (dustier and more massive) host population from GRBs with bright optical afterglows.  However the actual contribution of dark GRBs to the whole is not precisely determined, in part because the matter is much more complicated than an either/or distinction between ``bright'' and ``dark'' bursts:  the distribution of afterglow brightnesses is a continuum, as is (we expect) the likelihood of redshift measurement as a function of brightness.  ``Very'' versus ``somewhat'' dark bursts may likewise have different host and redshift distributions.

An alternative approach, and the one we adopt in this work, is to carefully down-select the \textit{Swift} sample to remove GRBs that occurred under circumstances that were not optimal for ground-based follow-up and isolate a sub-set for which the afterglow redshift completeness is close to the expected maximum achievable value of about 80\% (the remaining 20\% being dark bursts).   This basic technique was first exploited to study unbiased afterglow demographics and redshift distributions \citep{Jakobsson+2006,Fynbo+2009,Cenko+2009,Perley+2009a,Greiner+2011}, and more recently has been successful in addressing the properties of the GRB host galaxy population as well.  Most notably, the multi-year VLT-based ``TOUGH'' project (The Optically Unbiased GRB Host Survey; \citealt{Hjorth+2012}) used a series of observability cuts to isolate a sample of 69 objects out of the broader \textit{Swift} sample and has achieved a redshift completeness of close to 90\% after intensive optical spectroscopy of the host galaxies of those events lacking bright afterglows \citep{Jakobsson+2012,Kruehler+2012,Schulze+2015}.  ``BAT-6'' \citep{Salvaterra+2012} is an effort with a similar size, design, and redshift completeness (58 objects at $\gtrsim$90\% completeness.)

Though these projects have been informative in charactering the host population in various ways \citepeg{MilvangJensen+2012,Michalowski+2012,Schulze+2015,Vergani+2015} and have provided the first nearly-unbiased estimates of the redshift distribution \citep{Jakobsson+2012}, from the perspective of measuring actual redshift \emph{evolution} of the GRB host population, TOUGH and BAT-6 are limited by their modest sample sizes.  For example in the case of TOUGH, the 69 events span a redshift range of $0 < z < 6.3$ over which an enormous degree of cosmic evolution has occurred: to provide a snapshot of the host population at a particular epoch, and compare to other epochs, would require splitting the sample into at least 5--6 redshift bins, each of which would have only a few host galaxies.  This is enough to make broad statements, but less than necessary to characterize the parameter distribution at any point in history. 

Larger samples are therefore needed.  In principle this could simply be done by extending the year cut-off affecting earlier samples closer to the present (e.g. TOUGH was limited to events from 2005--2007 based on the time the survey was conducted; this cutoff could be extended to include more recent bursts).   However, we can now do better:  while the detailed selection criteria for earlier samples were developed based on well-informed guessing about the factors that affect the success rate of afterglow follow-up, we now have the benefit of hindsight in the form of a large \textit{Swift} GRB population from which to establish \emph{in practice} what parameters and values maximize the benefit to our science goals.   

As a result, before proposing for and executing our survey we devoted significant attention to examining the impact of a variety of criteria on the size and completeness level of the resulting sample, deciding on a final set of optimized parameters based on these investigations.  As the associated findings are relevant to understanding the design of our survey and may also be useful to investigators considering even more ambitious projects in the future, we describe these in detail in the next section.

\subsection{Additional Considerations}

Since it is essential to produce an unbiased population, we consider \emph{only} observational criteria that are not expected to be physically connected to the burst's environment in any way.  Given the cosmological nature of GRBs, factors related to time or sky location are all clearly unconnected to the GRB host environment and therefore fair to consider.  Properties intrinsic to the GRB itself (measurements of the prompt emission or afterglow) are less obvious.
Factors specific to the prompt emission should not be biased with respect to environment, since in the favored internal/external shock paradigm \citepeg{Sari+1997} the physical mechanism producing the GRB prompt emission is unrelated to the nature of the environment it formed in.\footnote{This does assume that long GRBs all represent a single class of object, rather than a superposition of multiple types with different progenitors and prompt-emission properties.}  
On the other hand, the GRB afterglow is produced by interaction of the GRB ejecta with the circumprogenitor environment \citep{Meszaros+1997} and its light may be attenuated by gas and dust within its host galaxy, so we do \emph{not} consider any afterglow-related property.  

Because our selection cuts are developed a-posteriori, it is important to note that even the application of nominally unbiased cuts can produce some degree of bias in a discrete sample, depending on how many features are considered and how precisely fine-tuned the cut threshold is optimized.  If the cuts are drawn from a large potential parameter space and optimized directly against the data with no separate training set, the choice of cut parameters and their values may become driven by stochastic effects---artificially driving up the completeness and re-introducing biases associated with redshift measurement.   To minimize this risk, we only examine features for which a link between the variable and redshift completeness is at least potentially expected from fundamental considerations, and only cut on features where significant dependence is observed.  We also only consider round-number, discrete thresholds in determining the cut value.  (An alternative would have been to divide the sample into a training set and the applied sample, but as the number of known-redshift GRBs remains modest---a few hundred---this would overly restrict the sample size of both sets.)

Choosing how many cuts to apply, and how stringently to cut, also requires striking a balance between redshift completeness and overall statistical size.  Even when starting with an initial sample of many hundreds of \textit{Swift} events, discrete cuts are necessarily blunt and probabilistic tools and many known-redshift events will be lost in trying to increase the completeness significantly from its very low initial value of 30\%.  We set as a reasonable goal obtaining at least 100 events, and at least 65\% \emph{initial} (pre-host-follow-up) redshift completeness.  The target sample size is needed to provide a significant expansion over earlier host surveys and to ensure that the sample is large enough such that, even if subdivided into 5 or more redshift bins, there are enough targets per bin to reasonably statistically constrain the parameter distribution.   The completeness goal is informed largely by the success of TOUGH in increasing its redshift completeness from a similar starting level to its current $\sim$90\%, which is about the level that is necessary for systematic considerations (preferentially missing redshift measurements of faint host galaxies) not to dominate the statistical ones.   (We had no trouble meeting these goals, with our final cuts producing a sample of 119 events at pre-host redshift completeness of 68\%.)

Redshift completeness is not the only motive for applying cuts to the \textit{Swift} sample---it is also desirable to maximize overlap with existing observations of GRBs and their hosts.  High-quality early afterglow observations, even if not leading to a successful redshift measurement, can produce a more secure host identification, a redshift upper-limit, and useful complementary information about the GRB sightline; previous host-galaxy observations reduce the observational demands needed to complete the survey.

Because our Spitzer/IRAC campaign (\S \ref{sec:iracobs}) was conducted during Cycle 9 (November 2012 through September 2013), our targets were necessarily restricted to bursts occurring before that period, so we only considered events up to the end of October 2012 for inclusion.

\section{Sample Selection and Analysis of Factors Influencing Redshift Completeness}
\label{sec:selection}

\begin{figure*}
\centerline{
\includegraphics[scale=0.8,angle=0]{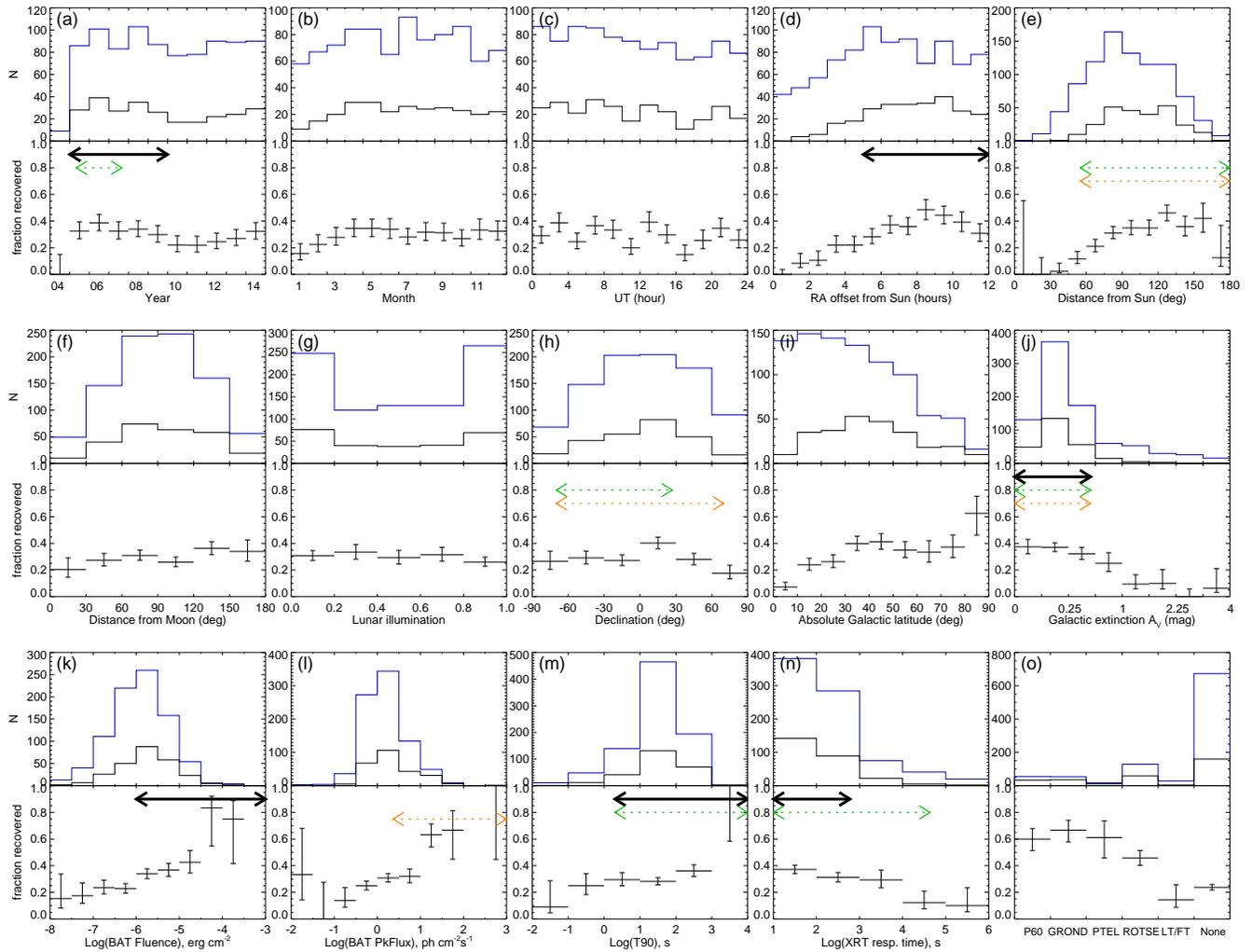}} 
\caption{Redshift recovery numbers and fractions for GRBs as a function of various observational parameters.  We only consider rapidly-reported redshifts (in nearly all cases from absorption spectroscopy of the GRB afterglow) and do not include redshifts from host observations at late times.  The blue histogram indicates all triggered \textit{Swift} GRBs while the black histogram indicates \textit{Swift} GRBs with known redshift.  The lower panel shows the ratio of the two; error bars are calculated using the binomial Bayesian method of \cite{Cameron2011}.  Cuts employed by the SHOALS survey are indicated by thick black arrows.  We also show the cuts employed by two other host-oriented uniform samples: TOUGH (green arrows) and BAT6 (orange arrows).}
\label{fig:zrecovery}
\end{figure*}

With the above considerations in mind, we investigated the impact of many different observables to identify what combination of unbiased cuts would produce the highest final redshift completeness.  These are explained in detail in the following sub-sections, with the results summarized in Figure \ref{fig:zrecovery}.  Our data were drawn from the \textit{Swift} GRB table\footnote{http://swift.gsfc.nasa.gov/archive/grb\_table/}, supplemented by our own corrections to redshifts or classifications where appropriate, using published sources and our own observations.

For the purposes of this analysis we did \emph{not} treat redshifts that were determined at late times via host-galaxy observations as ``known'', since the latter subset may be affected by conscious (and potentially biased) follow-up considerations.   We did, however, consider emission-line redshifts that were promptly reported in the GCN circulars.  (For low-redshift, low-luminosity GRBs the host galaxy may be of comparable or greater brightness than the optical afterglow by the time the latter is observed, so an emission-based redshift may be reported even if an absorption redshift may have been possible in absence of host emission lines.)  

We consider only GRBs detected onboard the satellite, excluding ground or slew-survey triggers (which are distributed only after many hours' delay and are rarely followed).

\subsection{Time of Explosion}

The amount and quality of observational follow-up directed towards a given GRB afterglow is affected by terrestrial timescales in many ways.  Observatories are geographically clustered (most large optical facilities are located in Hawaii, the western United States, Chile, or the Canary Islands) and the time of day a burst explodes (as well as its location) affects which observatories can follow it and how quickly.   Seasonal weather patterns also influence the likelihood of whether a given facility is able to follow-up a GRB (and, if so, how effectively).   On longer timescales, other factors may come into play:  interest in GRB follow-up among different observational groups has both waxed and waned over \textit{Swift}'s lifetime, while ground-based facilities important to GRB follow-up have been commissioned (e.g. GROND, RATIR, GTC, X-shooter) and decommissioned (e.g., PAIRITEL, LOTIS) over the same period.

In Figure \ref{fig:zrecovery}a-c\footnote{Figure \ref{fig:zrecovery} shows all GRBs through the end of 2014 in order to provide the most up-to-date view of the population, although our selection criteria were developed in late 2012.  The trends are the same in either case.} we plot the redshift completeness as a function of calendar year, calendar month, and UT hour.  While the trends are weak, at least some clear signatures are evident.   First, the redshift completeness peaked in the early years of \textit{Swift} (40\% between 2005--2009) but has fallen since then (27\% since 2010), probably because follow-up of ``routine'' bursts has become less common over the years and also because of smaller allocations of time to rapid-response spectroscopy.  This trend is large enough that a year cut significantly improves the redshift completeness of the sample; restricting ourselves to older bursts also has the benefit that the large amount of time elapsed has provided more opportunity for previous host-galaxy campaigns to acquire ground-based data.  We therefore restrict the base sample to events in the five-year 2005--2009 period.  (Some post-2009 events are later included under a separate criterion; \ref{sec:groundobs})

A seasonal effect may also exist;  the redshift completeness drops significantly between January and March (26\%) compared to the rest of the year (35\%).  We investigated whether this dependence may be different for Northern and Southern hemisphere bursts and found the trend to be present in both data sets.  Despite the strength of this effect we elected not to cut on it, since the complementary benefits are less clear than for the other cuts employed.   Nevertheless, considering the impact observed despite the crudeness of a raw seasonality cut, the trend we observed here provides good reason to consider more specific cuts, based on actual weather conditions at various sites, in the future.

We see no clear secular dependence on the time of day and did not cut on this parameter.

\subsection{Sun Angle}

Ideally, \textit{Swift}'s pointing would always be centered close to 180 degrees away from the Sun, such that every burst detected by the BAT could be followed immediately by any telescope in the night hemisphere of Earth and remain above the horizon until sunrise.  In reality, operational considerations (in particular the cooling requirements of the XRT) require that the pointing direction of \textit{Swift} is frequently far from this ideal and many GRBs are detected close in projection to the Sun, making ground-based follow-up challenging or impossible.    

In Figure \ref{fig:zrecovery}d-e we plot the redshift completeness against angular separation from the Sun as well as against separation in right-ascension only.  It is clear that Sun angle has a large impact on redshift completeness, with about $\sim$40\% of bursts in the anti-Solar hemisphere successfully recovered but extremely poor completeness for bursts nearest the Sun ($<$10\% within 60$^{\circ}$).

Previous surveys (TOUGH, BAT6) used a Sun distance cut of $>55^{\circ}$ and no right-ascension cut.  However, based on the plots above it is clear that somewhat more stringent cuts can significantly further improve the redshift completeness with only modest additional down-selection:  the redshift efficiency remains very poor for events closer than $\sim75^{\circ}$ or for bursts with a right ascension difference of less than 5 hours.  We employ a minimum right-ascension separation of 5 hours, and in principle an even more stringent cut could have been used at the expense of reducing the sample size.  We make no formal cut on angular separation, but note that the right-ascension cut alone excludes all GRBs closer than $54^{\circ}$ from the Sun.

\subsection{Lunation and Moon Angle}

While the location of the burst relative to the Sun has a large impact on the ability to follow up a burst, lunar considerations appear to be negligible.  In Figure \ref{fig:zrecovery}f-g we plot the recovery rate against lunar illumination and moon distance, respectively.  We see only weak, low-significance trends in either case, and so we do not use lunar information to restrict our sample.  
(While this result may seem surprising, we note that bursts very close to the full Moon are also very nearly anti-Sun, and the improved observability may counteract the higher sky background.  We also note that bursts \emph{very} close to the moon are automatically excluded from our sample, though not from Figure \ref{fig:zrecovery}, on the basis of the lack of XRT follow-up.)

\subsection{Declination}

Bursts that occur close to the celestial equator can be observed by telescopes in both hemispheres, whereas those occurring closer to the poles are difficult or impossible to follow from the opposing hemisphere and (in extreme cases) cannot be observed by equatorial-mount telescopes at all.  On the other hand, high-declination bursts remain above the horizon for longer than equatorial ones.  In practice, the declination dependence of redshift completeness appears to be weak (Figure \ref{fig:zrecovery}h): no significant declination trend is seen in the Southern hemisphere, and while a fairly strong trend is evident in the Northern hemisphere it becomes much less significant after our preferred $A_V$ and $T_{90}$ cuts (\S\ref{sec:cutav}--\S\ref{sec:prompt}) are applied, indicating that it may be coincidental.  We therefore did not apply a declination cut.

\subsection{Galactic Latitude and Extinction}
\label{sec:cutav}

The position of a burst as projected against the plane of our own Galaxy strongly impacts our ability to recover its optical afterglow, mostly due to extinction by interstellar dust within the Milky Way (but crowding and source confusion with foreground stars may also play a role).  This effect can be clearly seen in Figures \ref{fig:zrecovery}i and \ref{fig:zrecovery}j where the recovery fraction is plotted against both the absolute Galactic latitude $|b|$ and the foreground extinction $A_V$ (as calculated from the \citealt{Schlegel+1998} dust maps\footnote{We use the original Schlegel et al.\ maps in preference to more recent work \citepeg{Peek+2010,Schlafly+2011} for consistency with earlier GRB host selection efforts.}).  For extinction columns in excess of $\gtrsim0.5$ mag, and for latitudes less than $|b| \lesssim 10$ degrees, the recovery fraction plummets.

We employ a maximum foreground extinction of $A_V < 0.5$ mag (Schlegel) in SHOALS, which of course in practice also cuts most events near the Galactic plane.  This is the same criterion used by both TOUGH and BAT6.

\subsection{Prompt Emission Properties: Fluence, Flux, and Duration}
\label{sec:prompt}

The GRB prompt emission is not thought to depend directly on the properties of the circumprogenitor environment, so incorporating prompt-emission properties in our selection should not bias the sample---although such selections \emph{will} necessarily impact properties such as the intrinsic redshift distribution of the sample versus that of the \textit{Swift} parent population (we expect brighter bursts to be closer, on average).  More importantly, GRB prompt-emission brightness is also observed to be correlated with GRB afterglow brightness \citep{Gehrels+2008,Nysewander+2009,Kann+2010}: partly because luminous bursts tend to release more kinetic energy into their surroundings and produce brighter afterglows, and also simply because closer bursts (at fixed luminosity) will be brighter at all wavelengths.  We therefore expect that the brighter afterglows associated with a population of bright GRBs should make their redshifts easier to measure.

The impact of adding a prompt-emission cut on redshift completeness was first clearly demonstrated by \cite{Salvaterra+2012}: the BAT6 sample reached $\sim$80\% redshift completeness even without any host-galaxy follow-up after application of a fairly stringent cut on the peak photon flux in addition to their observability criteria (after host follow-up this rose to $\sim$90\%).  A flux/fluence cut is also desirable from the point of view of better understanding events with optical nondetections, since it is easier to establish limits on the spectral index $\beta_{\rm OX}$ \citep{Jakobsson+2004,vanderHorst+2009} and determine if the afterglow is obscured in the presence of a brighter X-ray afterglow.  Finally, imposition of an explicit cut on prompt emission properties serves to replace the complicated \textit{Swift} triggering criteria \citep{Band2006} with a simple well-defined threshold that can be more easily modeled for the purpose of measuring intrinsic rates and redshift/luminosity distributions (\S \ref{sec:zdist}).\footnote{We wish to emphasize that cutting on fluence may in principle affect the redshift distribution---but does so in a controlled way that improves, rather than harms, our ability to accurately measure the intrinsic burst rate.  The BAT sample is always flux/fluence-limited; our cut replaces the complex on-board trigger criteria with a well-defined selection criterion that we can correct for.  This issue is discussed further in \S \ref{sec:completeness}.}

We plot the dependence of the redshift completeness on the BAT 15-150 keV energy fluence and 1\,s peak photon flux in Figures \ref{fig:zrecovery}k-l.  The redshift completeness rises sharply for brighter bursts.  Both fluence and peak flux show trends of similar magnitude, although in practice we found that using a fluence cut produced a slightly higher redshift completeness: after all other cuts were applied, the redshift completeness was 4\% higher for a fluence cut than for a peak flux cut chosen to produce the same final sample size.  A fluence cut is also expected to affect the redshift distribution less than a peak-flux cut would, since the cosmological dimunition in luminosity is offset by time dilation (see also \S \ref{sec:zdist}).  As a result we choose to cut on fluence ($S_{\rm 15-150 keV} > 10^{-6}$ erg cm$^{-2}$).

We exclude short-duration bursts from the sample, requiring $T_{90} > 2$ seconds.  This was done not for its effects on redshift completeness, but rather to remove short-duration gamma-ray bursts from the sample, as these appear to have a physically distinct origin associated with compact object mergers and a very different host population \citepeg{Fong+2013a}.  Strictly, this cut actually has no impact: our fluence cut strongly disfavors short events and removes all events with $T_{90} <$ 2 sec on its own.   However, we also remove two events with $T_{90} > 2$ sec (GRBs 060614 and 080503) whose prompt-emission light curves resemble those of short-duration events with extended emission (while their exact origins are debated, these events have properties very different from ordinary long-duration GRBs: see e.g., \citealt{Fynbo+2006b,GalYam+2006,Gehrels+2006,Perley+2009b} for further discussion).

Unlike the other properties discussed in this section, the BAT measurements are subject to measurement uncertainties and in principle values can change upon re-analysis.   Our selection was conducted using values in the \textit{Swift} GRB table, which is collated from values reported in the BAT refined analysis GCN Circulars  (which in turn are based on analysis conducted by the BAT team 1--2 days after the event).  Further analysis and improved measurements has been carried out only for a subset of these (e.g. \citealt{Sakamoto+2011}).  Since we wished to examine the entire \textit{Swift} sample using the same data we used the GCN values to establish our selection.  Fortunately, the differences are usually very small (a few percent) and the choice has negligible impact on the sample properties.   Only four new GRBs would have entered the sample had we used the updated values on bursts where they are available:  060605 at $z$=3.78 \citep{Ferrero+2009}, 060124 at $z$=2.300 \citep{Fynbo+2009}, 090102 at $z$=1.547 \citep{GCN8766}\footnote{The change in status for GRB 090102 is not due to reanalysis but is the result of a typo in the \textit{Swift} GRB table.}, and 090728 (unknown-$z$; possibly an $R$-dropout and at $z\sim6$; \citealt{GCN9726}).  No GRBs currently in the sample would be dropped from it.

\subsection{Rapid Observability to Swift}
\label{sec:groundobs}

Some events which the \textit{Swift} BAT detects cannot be rapidly observed by the XRT due to pointing considerations or technical difficulties.  In such cases, the lack of a prompt X-ray position reduces some of \textit{Swift}'s advantages relative to previous satellites.  Optical afterglow follow-up is more difficult because observers must identify an (often-faint) new source within the full 3$^{\prime}$ radius initial BAT error circle, potentially causing afterglow identifications to be delayed or missed entirely and resulting in lower redshift completeness (a trend which can be seen in Figure \ref{fig:zrecovery}n).  Furthermore, because later-time X-ray observations do not guarantee a detection, host-galaxy follow-up and redshift identification may be strictly impossible in some cases if early observations are not carried out.  For both these reasons, we require a prompt XRT response ($<600$ sec).

\subsection{Rapid Observability to Specific Ground Facilities}

Many of the factors described in earlier sections (Sun angle, etc.) are chosen based on their impact on ground-based observability.  Of course, other factors we have not explicitly considered also affect whether or not a ground-based telescope can actually respond to a burst, including weather conditions, the instrument set(s) available that night, and whether or not the telescope was under maintenance.

In principle it would be desirable to explicitly incorporate site-dependent weather and telescope downtime at the most important facilities into our calculations as a formal sample cut.  This may be a promising avenue for establishing future samples (in particular for more recent bursts, given the lower degree of redshift completeness overall since 2010), but as this information is not readily available we did not consider it in designing our survey.  However, we can be even more precise by considering whether or not several leading ground-based imaging response facilities actually \emph{were} successfully triggered by \textit{Swift} for early-time observations.   Such an approach has already been applied extensively by our team and others to produce uniform optical afterglow samples \citep{Cenko+2009,Perley+2009a,Greiner+2011} and the same procedure can be applied to GRB hosts.

Figure \ref{fig:zrecovery}o shows the recovery rate for events observed ``rapidly'' by various GRB-oriented follow-up facilities:  the Palomar 60-inch telescope (P60; bursts from 2005--2012; \citealt{Cenko+2009} and work in prep.), the Gamma-Ray Burst Optical/Near-infrared Detector (GROND; 2007--2012; \citealt{Greiner+2011} and Kr\"uhler et al., priv. comm.), the Peters Automated Infrared Telescope (PAIRITEL 2005-2010; \citealt{Morgan+2014b}), the Robotic Optical Transient Search Experiment (ROTSE-III; 2005-2012; taken from a search of the GCN circulars), and the combination of Faulkes and Liverpool telescopes (2005-2007; \citealt{Melandri+2008}), compared to events observed by none of these facilities on a rapid timescale.  Except in the case of the (small) Liverpool/Faulkes sample the triggering of early-time observations is, unsurprisingly, correlated with modest increases in recovery rate.

Except possibly in the case of the ROTSE events (whose redshift completeness is lower than desired), these samples are not large enough for our purposes, even if applied with no other cuts.  Therefore instead of \emph{requiring} ground-based follow-up by one of these facilities, we treat rapid observations to either P60 or GROND as an \emph{alternative} criterion allowing these bursts to bypass the three afterglow-observability requirements (year, hour-angle, and XRT-response time), such that a GRB nominally failing one or more of these criteria can nevertheless enter the sample as long as either of these two telescopes observed it at early times.  GROND or P60 observations, however, do \emph{not} provide immunity to criteria not connected with ground observability, in particular the fluence and extinction cuts which remain in effect across the entire sample.

To maximize overlap with the TOUGH sample \citep{Hjorth+2012}, for which extensive host-galaxy data is already available, we offer a similar bypass to events within that sample (but, again, with the requirement that they must still satisfy the fluence cut.)

\subsection{Additional Host Observability Criteria}
\label{sec:addl}

Two additional criteria are added to ensure that host-galaxy follow-up is possible.  First, we require that a $<2\arcsec$ (at 90\% confidence) position be available, since uniquely identifying a host galaxy in an area larger than this carries significant risk of misidentification or ambiguity.  Because all of our events were observed by the XRT at early times when the burst was bright (or by a sensitive ground-based optical facility) a position of this accuracy is almost always available, and this cut removes only a single event (GRB 080613B) from consideration.  In principle this exclusion creates a very weak bias (if this event had an optical afterglow we would have included it: two other events, GRBs 071021 and 061110B, have $>2\arcsec$ XRT position uncertainties but were included thanks to available optical/IR follow-up)---but given its small impact in practice we ignore this effect.

We also require that the host position not be contaminated by a bright foreground object in the form of a Galactic star or intervening galaxy.  This decision is somewhat subjective (stars and galaxies contaminate many fields to various degrees, especially in the Spitzer imaging due to its large PSF), and in the case of a contaminating galaxy requires some degree of deep follow-up to have been obtained in the first place to recognize that the foreground galaxy is not the host.  Events which we exclude on the basis of a contaminating bright star are GRBs 050716, 060923C, 071003, 080129, 080212, 080229A, 080905B, and 101023A.  Events excluded due to a contaminating galaxy are GRBs 080319C (for which ground-based spectroscopy and HST imaging shows the source closest to the afterglow location to be a superposition of the host and a foreground system at $z=0.81$) and 081028 (for which no ground-based imaging or spectroscopy is available, but for which the observed IRAC magnitude of the source underlying the afterglow position is inconsistent with any galaxy at that redshift, so is probably one of the foreground absorbers mentioned in \citealt{GCN8434}).

\subsection{Summary and Sample Properties}

We summarize our final selection criteria\footnote{While final for the purposes of this study, our sample is readily extendable by loosening or expanding the afterglow observability criteria.  In hypothetical future cases where it becomes necessary to be specific, we will refer to the uniform sample established by the specific criteria outlined in Table \ref{tab:criteria} as the SHOALS09+ sample. In this paper we will simply refer to it as the SHOALS sample.} in Table \ref{tab:criteria}.  Characteristics of the sample are presented in Table \ref{tab:characteristics}, and a table of key properties of the sample relevant to our study (positions and redshifts; see subsequent sections) is presented in Table \ref{tab:grbinfo}.

As these numbers indicate, our selection criteria were highly successful in isolating a large, well-observed sample of \textit{Swift} GRBs: 119 targets,  68\% of which had redshifts measured before any late-time host follow-up had been conducted.   Including redshifts from late-time follow-up (both preceding our efforts and including our host-galaxy campaign; see \S \ref{sec:redshifts}) we have achieved a  completeness close to 90\% (89\% considering only secure spectroscopic redshifts, 92\% including photometric redshifts, 94\% including $N_H$-bracketed and a tentative single-line spectroscopic redshift, and 99\% including upper limits) so far on a sample twice the size of previous efforts.  Forthcoming observations will likely increase the completeness even further.

Approximately 20--30 events ($\sim$20\% of the sample) can be classified as ``dark'', with the exact number depending on the definition employed.  25 GRBs have no unambigous optical ($0.3-1.0 \mu$m) afterglow detection reported in the GCN circulars or elsewhere, and in only four of these cases could this be readily attributed to lack of deep or early follow-up (see Table \ref{tab:grbinfo} for details.)   On the other hand, 12 events \emph{with} optical detections have red colors indicative of significant dust attenation, in some cases a great deal of dust attenuation  (e.g., $A_V \sim 3-4$ mag for GRBs 080607, 090709A, and 100621A: \citealt{Perley+2011,Cenko+2010,Greiner+2013})---these events probably lie in the same physical class as optically-undetected dark bursts, but optical detections were secured thanks to particularly efficient follow-up and/or a very intrinsically luminous afterglow.

Since the optical afterglow observations for many of the GRBs in our sample have not been thoroughly analyzed beyond the quick reports given in the GCN circulars, it is of course possible that some additional events are (modestly) obscured without us being aware of it.   Nevertheless, as our obscured fraction is consistent with other recent estimates \citep{Perley+2009a,Greiner+2011,Covino+2013} it is likely that the events we have identified constitute the large majority of all events that were dust-obscured in our sample.   The afterglow properties of the sample (including quantitative metrics of darkness such as $\beta_{\rm OX}$; \citealt{Jakobsson+2004}) will be revisited in more detail in forthcoming papers.

\begin{deluxetable*}{llllll}  
\tabletypesize{\small}
\tablecaption{Summary of Selection Criteria}
\tablecolumns{5}
\tablehead{
\colhead{No.} &
\colhead{Type} & 
\colhead{Short description} & 
\colhead{$\Delta N$\tablenotemark{a}} &
\colhead{$\Delta C$\tablenotemark{b}}
}
\startdata
1         &   GRB properties      & Onboard \textit{Swift}/BAT trigger before October 2012 w/XRT observations           &     &         \\
2         &   GRB properties      & $T_{90} > 2$s and not an SGRB+EE\tablenotemark{c}                          &$-$2  & +0.4\%  \\
3         &   GRB properties      & $S_{15-150 {\rm keV}} > 10^{-6}$ erg cm$^{-2}$                             &$-$69 & +9.6\%  \\
4         & Host/afterglow visibility  & Galactic $A_V < 0.5$ mag                                              &$-$35 & +9.7\% \\
5         & Afterglow visibility  & Amenable to follow-up: (5a-i and 5a-ii and 5a-iii), OR 5b, OR 5c           &$-$143& +17.1\% \\
\ \ 5a-i  & Afterglow visibility  & XRT observations within 10 minutes                                         &$-$10 & +2.2\%  \\
\ \ 5a-ii & Afterglow visibility  & Between 2005--2009 (inclusive)                                             &$-$36 & +8.1\%  \\ 
\ \ 5a-iii& Afterglow visibility  & Sun hour angle separation $>$5 hours                                       &$-$13 & +5.7\%  \\
\ \ 5b    & Afterglow visibility  & Automatically triggered P60 within 1000~s or GROND within 1 hour           &  +18 & +0.8\%  \\
\ \ 5c    & Afterglow visibility  & Satisfies TOUGH criteria \citep{Hjorth+2012}                               &  +11 &$-$1.4\%  \\
6         & Host/afterglow visibility & No known foreground star or galaxy contaminating the position          &$-$7  & +2.2\%  \\
7         & Host visibility       & $<2 \arcsec$ position available                                            &$-$1  & +0.6\%  \\
\enddata
\label{tab:criteria}
\tablenotetext{a}{Number of GRBs affected by this criterion, if applied \emph{after} all other criteria; i.e., these numbers indicate the (negative of) the change in sample size if the criterion in question were removed and the sample reconstructed using all other criteria still in place.  In the case of criteria 5b and 5c, numbers are positive since these criteria enable GRBs to be included despite failing one of the 5a criteria.}
\tablenotetext{b}{Increase in redshift completeness after cutting on this criterion, if applied after all other criteria.  This indicates the (negative of) our change in pre-host-followup redshift completeness if the given criterion was dropped.}
\tablenotetext{c}{A short-duration GRB with extended emission.  Only GRB\,080503 and GRB\,060614 fall under this category among targets not cut by other criteria.}
\end{deluxetable*}

\begin{deluxetable}{ll}  
\tabletypesize{\small}
\tablecaption{Sample Characteristics}
\tablecolumns{2}
\tablehead{
\colhead{} &
\colhead{}
}
\startdata
 Total sample size                & 119        \\
 Number of ``early'' redshifts    &  81 (68\%) \\
 Number of redshifts to date      & 110 (92\%) \\
 Number with redshift limits      & 118 (99\%) \\
 Mean redshift                    & 2.18       \\
 Redshift quartiles               & 1.26, 2.06, 2.77   \\
 Redshift range                   & 0.03--6.29 \\
\enddata
\label{tab:characteristics}
\end{deluxetable}

\section{Observations and Data Analysis}
\label{sec:observations}

A primary goal of our survey is to produce high-quality, multi-filter SEDs for all host galaxies within the sample, enabling the construction of rest-frame luminosity functions at any wavelength and the measurement of important physical parameters (mass, SFR, etc.)\ via SED fitting.  These efforts are still ongoing, and will be described in full in subsequent papers.  Instead, we briefly outline our general observational strategy and its motivations.

\subsection{IRAC Observations}
\label{sec:iracobs}

Spitzer observations are key to our effort, extending the wavelength coverage by a factor of two relative to previous host galaxy surveys and providing access to physically distinct information to what is possible from ground-based observations alone.   IRAC's capabilities in its shortest-wavelength filters (3.6$\mu$m and 4.5$\mu$m) are undiminished even in its warm mission, and the instrument remains sufficiently sensitive to detect typical galaxies out to $z\sim5$.  Furthermore, the luminosity of a galaxy at these wavelengths (which always probe wavelengths redward of the Balmer break across this redshift range) is determined primarily by a single parameter (its stellar mass) with only modest dependence on age and extinction---providing a means of directly interpreting IRAC observations even without the supporting data we are amassing.  (This contrasts with the situation at the rest-frame UV wavelengths probed by ground-based optical imaging, since a potentially very luminous galaxy can appear quite faint in these bands if it is heavily dust-obscured.)

Observations of many targets in our sample were already present in the Spitzer Legacy Archive; the remaining targets were observed as part of our Cycle 9 Large Program at 3.6 $\mu$m with an exposure time chosen depending on the redshift (typically between 0.5--5 hours; see Paper II for details).  Observations at 4.5 $\mu$m were also acquired for some targets, in particular for those at unknown redshift and for GRBs designated as ``dark'', in order to provide better photometric redshift estimates and to model dust extinction in high-redshift galaxies.

\subsection{Optical and Near-Infrared Observations}
\label{sec:otherobs}

Spitzer observations alone provide an estimate of a galaxy's total stellar mass, but do not constrain the other properties of a galaxy, including the nature of the young stellar population that (presumably) produced the GRB.   In addition, because of Spitzer's large PSF size ($\sim$2$\arcsec$) it is not straightforward to uniquely identify the host galaxy based on Spitzer observations alone, even in possession of a precise afterglow localization:  deep observations approach the confusion limit, and it is not always clear whether an extended source at the afterglow location represents an extended host-galaxy or a blend of the host galaxy and a foreground object.  

For both these reasons, we have obtained a large volume of imaging at optical and near-IR wavelengths of all of our targets.  Given the wide range in redshifts and luminosities of the galaxies targeted by our survey and its all-sky nature (requiring different observational facilities to cover the northern and southern regions, and observing runs scattered throughout the year) this follow-up is necessarily heterogeneous and usually tailored to each individual target.   Typically, we try to obtain at least one deep ($R_{\rm lim} \sim 26$) ground-based optical (rest-frame UV at $z>1$) measurement and then obtain additional filters to the extent possible given the brightness of the target and the resources available; we also employ archival observations from a variety of previous surveys and from the Gemini and VLT archives.  

Most observations were conducted using the Low Resolution Imaging Spectrometer (LRIS; \citealt{Oke+1995}) on Keck I, with substantial imaging also coming from the Gemini Multi-Object Spectographs (GMOS; \citealt{Hook+2004}) at Gemini-North and Gemini-South, the Optical System for Imaging and low-Intermediate-Resolution Integrated Spectroscopy (OSIRIS) at the Gran Telescopio Canarias,  the Inamori Magellan Areal Camera and Spectrograph (IMACS; \citealt{Dressler+2011}) at Magellan, the Focal Reducer and Low Dispersion Spectrograph 2 (FORS-2) at the Very Large Telescope, the Wide Field Camera 3 (WFC3) aboard HST, and from the Gamma-Ray Burst Optical/Near-Infrared Detector (GROND; \citealt{Greiner+2008}) at the MPG 2.2m in La Silla.

In general, we make use of standard reduction techniques and pipelines where available to produce a stacked image, then astrometrically align the stacked images against the Spitzer imaging.  As the Spitzer PBCD imaging is by default astrometrically aligned against the 2MASS catalog \citep{2MASS}
this effectively establishes 2MASS as the astrometric reference system for the survey, and all positions reported in this work are therefore based on the 2MASS astrometric reference system.

Observational efforts are still ongoing (although nearly complete for targets above $\delta > -20^{\circ}$), so it is not yet possible to present a complete catalog of imaging acquired by the survey: this will be presented in forthcoming work following the completion of this effort.  At the present time, we have collected over 690 individual photometric data points on host galaxies within the sample, of which more than 510 represent detections (the remainder being upper limits; nonconstraining upper limits due to poor weather are excluded from these numbers).  These are supplemented by additional photometry from the literature.   Every host galaxy except for one has at least one deep optical observation to supplement the IRAC data, and all except for 21 have at least one optical or infrared detection.  Approximately half of the sample has numerous multicolor detections suitable for detailed characterization and modeling of the SED.  We expect these statistics to improve modestly as our observational efforts wrap up during the coming year.

\subsection{Spectroscopic Observations}
\label{sec:spectra}

A primary goal of the survey is to increase our spectroscopic completeness as high as is possible to remove any bias associated with redshift measurement---in particular that associated with dark bursts, but also potentially for bursts which do have detectable afterglows that are fainter than average and more difficult if not necessarily impossible to obtain absorption spectra of in time (perhaps moderately extinguished bursts or those in low density media.)

Many of the events in our sample without an absorption redshift are prominent \textit{Swift} dark bursts (or overlapped other uniformly-constructed surveys, such as TOUGH or BAT6) and as such a significant fraction of targets already had host-galaxy redshifts in the literature, primarily from \cite{Jakobsson+2012,Salvaterra+2012,Kruehler+2012,Perley+2013a}, or \cite{Kruehler+2015}.  All remaining sources showing bright ($\lesssim 24$ mag in any band) host-galaxy detections, and some fainter ones, were targeted for optical and/or NIR spectroscopy.  For most of these observations we employ the X-shooter spectrograph at the Very Large Telescope \citep{Vernet+2011}, a medium resolution cross-dispersed echelle spectrograph simultaneously covering the wavelength range between $0.3\,\mu\rm{m}$ and $2.5\,\mu\rm{m}$.  For observations associated with our program (094.A-0593, PI S. Schulze), 4 exposures of 900~s each were obtained in an ABBA nodding sequence, and reduced in a standard manner using the X-shooter pipeline provided by ESO \citep{Goldoni+2006} and using our own routines.  For Northern-hemisphere targets, we used LRIS or MOSFIRE \citep{McLean+2012} on Keck I or NIRSPEC on Keck II and reduced the data using custom routines.

\subsection{Host Identification}
\label{sec:subtraction}

The number density of field galaxies on the sky is significant at the depths involved in our survey, so unambiguously identifying the host galaxy requires as accurate and reliable a position of the originating GRB as possible.  Wherever possible, we acquired the original target-of-opportunity imaging files showing the optical or near-infrared transient, aligned these against the late-time host imaging, and re-calculated the afterglow position, which provides astrometric accuracy of typically 0.4\arcsec\ or better (significantly smaller than the PSF of either the ground-based or Spitzer imaging).  These images come from a number of sources, but common instruments are the imaging camera on the Palomar 60-inch telescope \citep{Cenko+2006}, the Nordic Optical Telescope, or publicly-available guider camera imaging from VLT or acquisition exposures from Gemini.  In some cases we downloaded imaging cutout figures posted in the GCN circulars or in published papers, and re-calculate the positions by aligning the cutouts to our host-galaxy imaging in a similar fashion.

In cases where the original images are not available, we use astrometric coordinates published in the GCN circulars or in the literature, or supplied to us by others (in particular by D. Malesani and Y. Urata).  We apply an astrometric offset (measured directly from the USNO and 2MASS catalogs) where necessary to translate from a USNO-aligned system to 2MASS: we assume published optical coordinates to be in a USNO-system if the astrometric system is not stated explicitly (except for NIR imaging which we assume to be in the 2MASS system natively), although typically these offsets are quite small ($\sim$0.1--0.3\arcsec) and do not dominate the uncertainty.  Radio or millimeter coordinates are not offset.

In some cases the only afterglow position available comes from \textit{Swift}---typically from the XRT (as is often the case for dark GRBs), although in a few cases a \textit{Swift} UVOT (UV-Optical Telescope; \citealt{Roming+2005}) position is available even though a ground-based position was not.  Positions are taken from the automated XRT analyses of \cite{Butler+2007}\footnote{http://butler.lab.asu.edu/Swift/xrt\_pos.html} and \cite{Evans+2009}\footnote{http://www.swift.ac.uk/xrt\_positions/index.php}, which provide positions with a typical accuracy of 1.5$\arcsec$.  UVOT positions are taken from the \textit{Swift} GRB table.  The coordinates are then shifted from their default USNO to the 2MASS frame, with the exception of positions noted as SDSS-aligned in the Butler tables which are not shifted.

Since in the vast majority of cases we do have a subarcsecond position available, the probability of mistaken identification of the host galaxy of any individual well-localized GRB due to a chance foreground/background alignment is low---only approximately 1\% of the sky is within 0.4$\arcsec$ of an unresolved $R<26$ mag galaxy \citepeg{Hogg+1997}.   Of course, in a large survey up to a few random alignments of this type would not be surprising, but would not significantly affect the results of the survey.  (Moreover, we can often identify and exclude them via a mismatch between host and afterglow redshifts: see \S \ref{sec:addl}).    The XRT-only, 1--2\arcsec\ positions are a source of somewhat greater concern, as the probability that a faint galaxy is present within a region of this size by chance is quite significant ($\sim$20\%) and no prior redshift is available.  This is partially alleviated by the fact that XRT-only events are typically dark, and that dark GRBs tend to originate from hosts which are much brighter than average (see e.g., \citealt{Perley+2013a} or \citealt{Kruehler+2012}) and have a low probability of chance association even considering the larger size of the XRT error circle.  Only four sources have positional uncertainties greater than 1\arcsec\ and host galaxies fainter than $R =$ 25th magnitude ($P_{\rm chance} > 0.05$): 050803, 050922B, 070621, and 070808.  Furthermore, the first two of these sources are associated with blue-dropout (likely, $z=4-5$) galaxies and the last is associated with an extremely red object---properties typical of optically-faint (dusty or high-$z$) GRB hosts but not common among galaxies selected randomly from the field.

\begin{figure*}
\centerline{
\includegraphics[scale=0.75,angle=0]{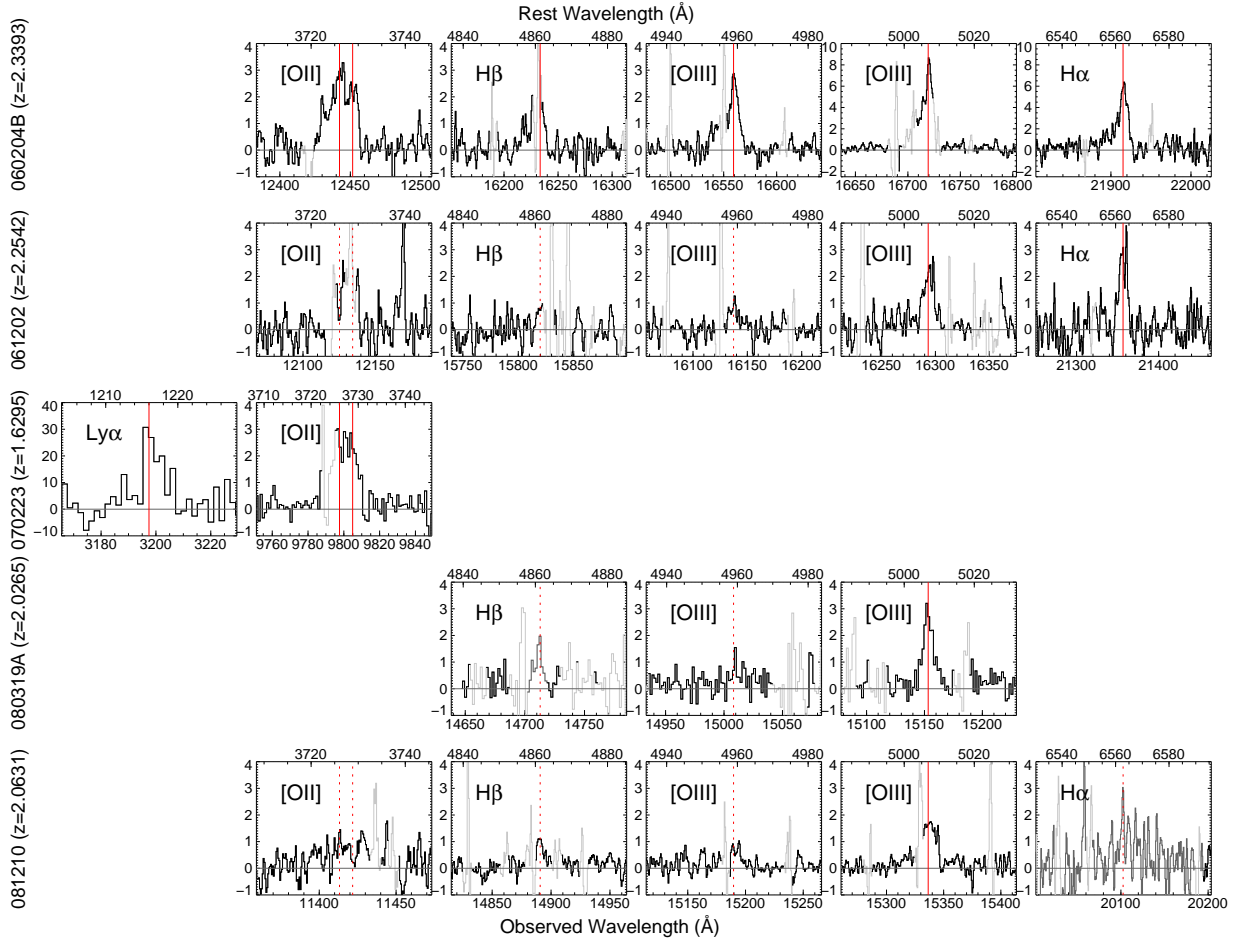}} 
\caption{New secure spectroscopic host-galaxy redshifts from our observations to date with VLT and Keck.  Portions of the spectrum plotted in grey indicate wavelength regions affected by strong night-sky OH emission or by telluric absorption.  (Many additional host spectroscopic redshifts are provided by our previously-published work and from the literature in general; see Table \ref{tab:grbinfo}.)}
\label{fig:specz}
\end{figure*}

\subsection{Host-Galaxy Photometry}
\label{sec:photometry}

Magnitudes (or upper limits) for host galaxies observed as part of our survey are measured using aperture photometry using a custom IDL wrapper around the \texttt{aper} photometric package included in the GSFC software library, or are taken from the literature.  Details of these procedures and the resulting photometry will be presented in future papers (Paper II on the IRAC observations is submitted concurrently with this paper), but are briefly summarized below.  Photometry relevant to our photometric redshifts and upper limits is presented in Table \ref{tab:grbphot}.

IRAC images at the depths relevant to our survey are at or near the confusion limit, and most of our host-galaxy targets show some degree of contamination from nearby foreground or background sources.  To mitigate this contamination we employ an iterative, partially-automated PSF-fitting routine to model and subtract nearby objects identified from the Spitzer and ground-based imaging near the object and background apertures, leaving an uncontaminated image of the host.  The host magnitudes are then measured via aperture photometry, calibrated using the zeropoint data in the IRAC handbook interpolated to the desired aperture using measurements of the instrumental PSF.

Photometry of optical/NIR images is provided in a similar manner, minus the need (in nearly all cases) to subtract contaminating sources since the host galaxy is well-isolated.  We calibrate relative to the latest release of the Sloan Digital Sky Survey \citep{DR10} wherever possible; otherwise, we provided our own secondary standards by observing the target on a photometric night alongside Landolt standards with a small telescope (P60 for northern targets or GROND for southern targets).  Magnitudes in non-SDSS filters are interpolated from the available $griz$ magnitudes using the photometric transformation equations of Lupton (2005)\footnote{http://www.sdss.org/dr12/algorithms/sdssUBVRITransform/\#Lupton2005}.  Calibration of HST photometry is performed using instrumental zeropoints.

\section{New Redshift Constraints}
\label{sec:redshift}

\subsection{Host-Galaxy Redshift Measurements}
\label{sec:redshifts}

\begin{figure}
\centerline{
\includegraphics[scale=0.55,angle=0]{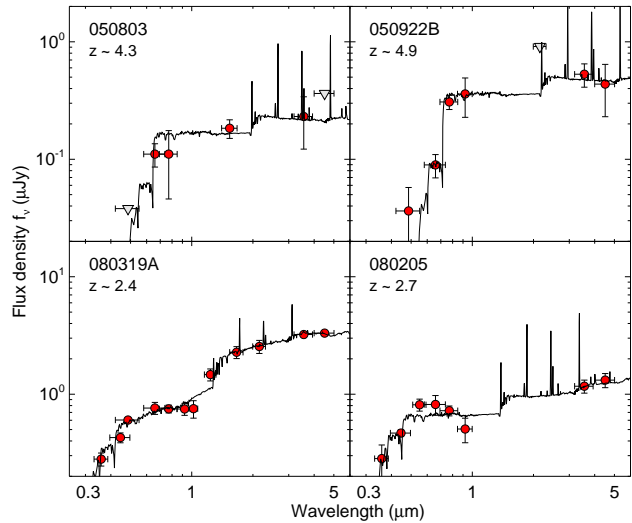}} 
\caption{New photometric redshifts inferred by the survey.   GRBs 050803 and 050922B show dropouts in the optical band indicative of a high redshift $z\sim4-5$, although in the case of the former event this will need to be confirmed by deeper $g$-band imaging.  The remaining events show a weaker flux decrement in the $U$/$B$-bands indicative of the onset of the Ly-$\alpha$ forest and (for GRB 080319A) a Balmer jump in $J$-band at a consistent redshift.  (The precise redshift of GRB 080319A was later fixed to $z=2.0265$ via emission-line spectroscopy.)}  The model SEDs shown are from EaZy at the best-fit photometric redshift.
\label{fig:photoztile}
\end{figure}

Eight new (not previously published\footnote{Some of these redshifts are also reported in \citealt{Schulze+2015} and \citealt{Kruehler+2015}, which were submitted close in time with this work by the same investigators.}) redshifts have so far been established by our program, either via the spectroscopic observations discussed in \S \ref{sec:spectra} or from photometric redshift fitting to the photometry we have gathered so far.  A summary of new redshifts provided by these methods is given below.  Zoom-ins of the spectral lines are presented in Figure \ref{fig:specz} and our photometric SEDs (showing the \texttt{EaZy} model for the best-fit redshift) are shown in Figure \ref{fig:photoztile}; the relevant data are presented in Table \ref{tab:grbphot}.

GRB050803 --- The host galaxy is marginally (3--5$\sigma$) detected in deep $R$-band imaging from the VLT, in $i$-band imaging from the GTC, and with the F160W ($H$-band) filter on HST, but no other wavelength (although a hint of a marginal detection is evident in the Spitzer imaging).  The lack of $g$-band detection is indicative of a dropout at a redshift of $z\sim4$, although not an unambiguous one.  Fitting the photometry using the EaZy photometric redshift software (\citealt{Brammer+2008}) and disallowing passive low-redshift solutions indicates a redshift of $z \sim 4.3$ (with $\pm0.4$ redshift uncertainty at 1$\sigma$ confidence, although $1.9<z<4.9$ is permitted at 95\% confidence.)  This redshift is consistent with (but slightly higher than) the result of \citealt{Schulze+2015} using the same data (but different photo-$z$ software).

GRB050922B --- We observed this galaxy with LRIS ($g$ and $R$ filters) and with OSIRIS ($i$ and $z$), complementing the existing $R$ and $K_s$ imaging from the TOUGH survey.  The host galaxy is well-detected only in $i$, $z$, and in the Spitzer filters.  A fit to the multi-filter SED (again ruling out passive low-$z$ solutions) suggests that the redshift is high, $4.2 < z < 5.2$ with a best-fit value of $z=4.9$, also in agreement with the parallel analysis of \cite{Schulze+2015}. 

GRB060204B --- X-shooter observations of the host galaxy of this GRB reveal many strong ($>8\sigma$) emission lines corresponding to H$\alpha$, [\ion{O}{3}{$_{\lambda4959}$}], [\ion{O}{3}{$_{\lambda5007}$}], and [\ion{O}{2}{$_{\lambda3727}$}], as well as probable H$\beta$,  at a common redshift of $z=2.3393$.  The velocity profile shows some structure with a prominent blue wing extending to about 300 km s$^{-1}$.

GRB061202 --- X-shooter observations of the host galaxy show strong emission lines consistent with H$\alpha$ and [\ion{O}{3}{$_{\lambda5007}$}] at a common redshift of $z=2.2543$.   Weaker emission is also seen at the locations of [\ion{O}{3}{$_{\lambda4959}$}] and H$\beta$, and possibly [\ion{O}{2}{$_{\lambda3727}$}] (although the latter is in between two bright sky lines).

GRB070223 --- Spectroscopy of the host galaxy with LRIS shows two strong emission lines at wavelengths of $\lambda=3198$ \AA\ and $\lambda=9801$ \AA, corresponding to Lyman-$\alpha$ and the [\ion{O}{2}{$_{\lambda3727}$}] doublet at a common redshift of $z=1.6295$.



GRB080319A --- We imaged this field extensively with Keck/LRIS (\textit{UBgRiZ}) and Keck/MOSFIRE (\textit{YJHK}).  The host is well-detected in every band (and with IRAC).  A fit to the SED indicates a photometric redshift of $z=2.43^{+0.20}_{-0.36}$.  We also obtained $H$-band spectroscopy of this target using MOSFIRE on the Keck I telescope on 2015-06-07 UT.  A total of 20 exposures of 120~s each were obtained.   A single, unresolved emission line is visible in the subtracted and stacked 2D spectrum at a wavelength of 15153 \AA.  This could be either [\ion{O}{3}{$_{\lambda5007}$}] at $z=2.0265$ or H$\alpha$ at $z=1.309$, but the latter case is strongly ruled out by our photometric redshift.  (An association with weaker lines such as H$\beta$ or [\ion{O}{3}{$_{\lambda4959}$}] is ruled out by the lack of additional line detections in the $H$-band spectroscopy.)  We therefore infer a redshift of $z=2.0265$.

GRB080205 --- A preliminary afterglow redshift of $z\sim4.0$ was estimated by the UVOT team on the basis of an apparent dropout in the $B$ filter in the early-time UVOT photometry \citep{GCN7253}.  However, we strongly detect the host in our ground-based $B-$band and $u$-band imaging, ruling out a redshift this high.  A photometric fit to our $uBVRiz$ and Spitzer observations imposes a maximum redshift of $z<3.08$ with a best-fit redshift of $z \sim 2.7$ (with large uncertainties; $z=2.71^{+0.25}_{-0.69}$).

GRB 081210 --- X-shooter observations of this target show a strong emission line at 15336\AA\ and some weaker features.  We identify the line as [\ion{O}{3}{$_{\lambda5007}$}] at $z=2.0631$ on the basis of probable (4$\sigma$) detections of H$\beta$ and [\ion{O}{3}{$_{\lambda4959}$}] at a consistent redshift and the fact that if this were another strong line (e.g., H$\alpha$) other lines should be detected in clean regions of the spectrum but are not observed.  At $z=2.0631$ the wavelength corresponding to H$\alpha$ is in a region of moderately strong telluric absorption.

\subsection{Redshift Upper Limits}
\label{sec:redshiftlimits}

Nine sources have thus far eluded redshift measurement.   However, even in most of these cases we can place upper limits due to the detection of optical afterglow, the detection of significant soft X-ray absorption excess above the foreground Galactic value \citep{Grupe+2007}, or the detection of an optical host galaxy.

GRB050128--We retrieved deep archival FORS-1 imaging of this source from the VLT archive; the data (associated with program 075.A-0718(A)) were taken on 2005-05-11 UT and total 3080 seconds of exposure time in the $R$-band filter.   Within the XRT error circle we securely detect a host galaxy candidate in the combined stack. The detection in this band indicates $z\lesssim5.5$.

GRB050726--The host galaxy is not detected in a very deep VLT $R$-band observation taken as part of the TOUGH project, so the host cannot be used to constrain the redshift.  However, a detection of an early afterglow was reported in the initial UVOT $V$-band exposure, suggesting $z < 3.5$.  Nondetections in subsequent UVOT $B$ and $U$-band observations suggest that the GRB may be a dropout and therefore close to this maximum redshift, but it is also quite possible that the GRB faded rapidly \citep{GCN3698}---so only an upper limit can be placed.

GRB070621--The (probable) host of this object, first identified in TOUGH, is very faint and detected only in $R$-band; while observations have been acquired in other filters with Keck and Gemini they do not reach the depths needed to detect the host.  The afterglow was not detected at any optical band despite deep early-time imaging.   The absorbing column inferred from the XRT spectrum is large, although not definitively so, and it places only a weak redshift limit of $z < 5.5$ (similar to what is implied by the putative $R$-band host detection).

GRB070808-- The afterglow of this event was detected only by XRT and the identity of the host galaxy itself is subject to some ambiguity given the coarse position; the nearest source to the center of either the DSS-refined or UVOT-refined XRT error circles was previously suggested by \cite{Hjorth+2012} as the most likely host candidate.  This putative host galaxy is very red, well-detected in IRAC and in VLT $K_s$-band imaging but only marginally or not at all in the deep optical imaging acquired so far.  Our optical photometry is not sufficient to establish a reliable photometric redshift on its own, but the red 3.6--4.5$\mu$m color effectively rules out a low-redshift solution (requiring $z>0.5$).  This GRB also has one of the largest XRT $N_H$ equivalent column excesses measured in the entire sample, which in this case actually places a more constraining limit on the redshift than any properties of the host galaxy.  Using the lower-limit $N_H$ value measured from late-time XRT data by \cite{Butler+2007} of 4.06 $\times$ $10^{21}$ cm$^{-2}$ places a maximum redshift of $z < 2.2$ according to the empirical formula of \cite{Grupe+2007}.\footnote{To critically assess the \cite{Grupe+2007} relation we repeated this procedure on the entire catalog of \textit{Swift} events with spectroscopic redshifts, including dark bursts.  With the exception of a handful of events with unrealistically large minimum $N_H$ values ($\sim10^{25}$ cm$^{-2}$, probably due to intrinsic curvature or difficulty with the automated model-fitting; at face value these would actually imply negative redshifts from the Grupe relation) we found only a single event with a redshift in excess of the predicted value: GRB 080207 (which exeeded it by only $\Delta z = 0.01$).}  Treating the upper and lower limits together provides a crude redshift estimate of $z=1.35\pm0.85$.

GRB081128-- A bright galaxy is located at the edge of the XRT error circle, although its centroid is significantly offset ($\sim1$\arcsec) from the optical afterglow position.  A combination of multi-filter imaging and spectroscopy from both LRIS and X-shooter shows it to be an early-type galaxy at $z=0.27$ with very little star-formation.  In addition, a much fainter source is evident directly underlying the afterglow in LRIS $B$, $g$, and $V$ band imaging (and marginally in $i$ and $z$-band imaging with the same instrument, and with IRAC).  While the nature of this system is not completely clear, the significant separation and color differential suggests that the bright galaxy is a foreground system unrelated to the GRB and the faint, blue object represents the true host galaxy.  We fit the photometry of the fainter source using EaZy; while a consistent redshift of $z=0.32$ is marginally favored higher redshift solutions (in particular, $z\sim3$) are also credible.  We can place only an upper limit of $z<3.4$ on the redshift.

GRB100305A--No host galaxy is detected at the XRT position in our imaging (we note that the proposed optical counterpart of \citealt{GCN10473} is well outside the final XRT error circle and the source is still present in our own Keck imaging, so is not likely to be associated with the GRB).   Two objects are just outside the edge of the error circle: one, to the southeast, is seen only in Spitzer and appears extended; the other, to the southwest, is also detected in our optical imaging.  The uncertainty about the host identification and the lack of optical afterglow precludes a definitive upper-limit based on these data.  The excess X-ray column is also low, suggesting that this may indeed be a high-redshift event that was ``missed'' on the basis of its faint afterglow.  This is the only event in the sample without an upper limit on its redshift.

GRB100802A--The only reported detection of the optical afterglow is by P60 \citep{GCN11040}.  Only an $r$-band detection was reported originally.  We stacked all of the imaging of the GRB acquired by the telescope that night and recovered detections in all four $griz$ bands, and aligned these images against our late-time Keck imaging.  A source is clearly detected in the Keck images at this location in $B$-band and (marginally) in $R$- and $i$-band, and with IRAC.  The data are not sufficient to estimate a definite photometric redshift but place an upper limit of $z<3.1$.

GRB110709B--The host galaxy (first reported by \citealt{Zauderer+2013}) was observed with X-shooter (under VLT Program 090.A-0088), integrating for 8$\times$900s using the $K$-blocking filter.  The reduced 2D frame does not show any strong lines over the spectral range, although a weak (5$\sigma$) line candidate is seen at an observed wavelength of 15479 \AA.  The most likely identification of this feature is [\ion{O}{3}{$_{\lambda5007}$}] at a redshift of $z=2.091$, as other strong-line matches are ruled out ([\ion{O}{2}{$_{\lambda3727}$}] would be resolved, while H$\alpha$ would imply emission lines of [\ion{O}{3}{$_{\lambda5007}$}] and [\ion{O}{2}{$_{\lambda3727}$}] within regions of the spectrum where we have good sensitivity and do not observe).  We consider this assignment tentative pending future follow-up without the $K$-blocking filter.  Photometrically, the redshift can be limited only to $z\lesssim5.5$ (based on the HST F606W host detection).  

GRB120308A--The host is securely detected in our LRIS $g$-band and $R$-band imaging from 2014-06-23 as well as in $I$-band LRIS imaging from 2014-05-27 and with IRAC.  We attempted a photometic redshift fit to these data; this is unable to produce a lower limit on the redshift but does establish an upper limit of $z<3.7$.

\section{Results}
\label{sec:results}

\begin{figure}
\centerline{
\includegraphics[scale=0.6,angle=0]{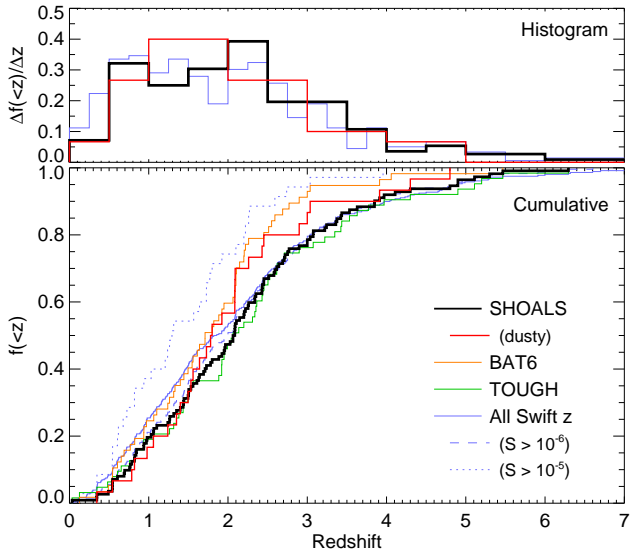}} 
\caption{Redshift distribution of SHOALS GRBs (solid black line) compared to other GRB samples.  Targets are mostly between $0.6 < z < 4.0$, with a few events at lower and higher redshifts; the overall distribution closely resembles that of known-redshift \textit{Swift} GRBs as well as other complete GRB samples (e.g. TOUGH).  Our modest fluence cut ($S_{15-150 {\rm keV}} > 10^{-6}$ erg cm$^{-2}$) does not greatly affect the redshift distribution compared to that of \textit{Swift} GRBs.}
\label{fig:zdist}
\end{figure}

\subsection{Redshift Distribution}
\label{sec:zdist}

The cumulative redshift distribution of our sample, as well as a binned fractional histogram, is plotted in Figure \ref{fig:zdist}.\footnote{Here and in subsequent analysis we place photometric redshifts (there are only six) at their best-fit values, and also use our best-guess redshift values for GRBs 070808 and 110709B.  The remaining unknown-$z$ bursts are omitted.  Removing photometric/insecure host redshifts would slightly decrease (by $\sim15$\%) the fraction of $z\sim3.5-5.5$ events in the sample and have neglibile impact on the rest of the distribution; this would not influence our results/conclusions.  If all unknown-$z$ events were very close to thir maximum redshift, their inclusion would increase the fraction of $z\sim3.5-5.5$ events slightly (by $\sim20$\%) and likewise have neglible impact on the rest of the distribution ; again this would not affect our results or conclusions.}  Remarkably, the redshift distribution of the sample is very similar to that of the overall \textit{Swift} distribution.   This suggests that, in spite of the many biases potentially affecting redshift measurement for a typical \textit{Swift} GRB, the net impact on the redshift distribution is small.  

We investigated this more closely by calculating the redshift distributions of the dust-obscured GRBs and other GRBs separately (the most prominent instrinsic factor affecting redshift incompleteness is dust obscuration, which makes afterglow-based redshift determination for $\sim$20\% of GRBs impossible in practice).   The redshift distribution for obscured bursts is shown as a red line in Figure \ref{fig:zdist}.  Its distribution closely mirrors the general SHOALS population: a mild excess of GRBs is seen at $1.5 < z < 2.5$ but a K-S test suggests this is not particularly significant ($p=0.15$ comparing the obscured GRBs in the sample versus the remaining GRBs with redshifts.)  This suggests that the fraction of cosmic star-formation that is obscured does not vary by a large amount with redshift, at least over the range in which we have reasonable number statistics ($1 < z < 4$).

The redshift distribution is also very similar to that of the TOUGH sample (a similar unbiased sample---but with no fluence cut) indicating that, even allowing for the significant overlap between the samples, the fluence cut we employed has only a relatively minor impact.  In contrast, the BAT6 sample (which is cut fairly stringently on peak flux) shows a notable skew in its redshift distribution towards lower redshifts.   While our study is not contingent on the redshift distribution of our sample matching that of \textit{Swift} GRBs overall (indeed, we would expect some differences based on the arguments in \S \ref{sec:prompt}), this indicates that our final sample is nevertheless reasonably representative of the broader \textit{Swift} population in redshift distribution.

We observe a very small fraction of the sample at high redshifts.  Only a single event is confirmed to be at $z>5.5$ (GRB 050904 at $z=6.295$), and among the 9 GRBs in the sample with no measured redshift, all but one is limited to $z<5.5$---so at most two events out of the 119 in our sample can be at $z>5.5$.  This infrequency is qualitatively consistent with other complete studies \citepeg{Perley+2009a,Fynbo+2009,Greiner+2011,Jakobsson+2012,Salvaterra+2012} but even more constraining.  Repeating the Monte-Carlo analysis technique of \cite{Perley+2009a} on our sample, we estimate that intrinsically between 0.3\% and 5\% of $S > 10^{-6}$ erg cm$^{-2}$ \textit{Swift} GRBs can originate at $z>5.5$ (95\% confidence).

While it is possible that we are preferentially missing GRBs at $z\gtrsim6$ in SHOALS relative to the all-Swift sample due to the fluence cut, as discussed earlier in this section, the impact of a fluence cut at 10$^{-6}$ on the redshift distribution appears to be small in practice.  Our results therefore suggest that the high-$z$ rate for \textit{Swift} bursts may be even lower than previously suspected (by e.g., \citealt{Perley+2009a}); a few events per year.   This underscores the challenges faced by recent efforts to observe high-redshift GRBs, and helps to explain the recent paucity of confirmed high-$z$ events.  Likewise, it contributes to our finding that the intrinsic high-$z$ GRB rate density is somewhat more modest than earlier estimates; \S \ref{sec:zrate}. 

\subsection{Sample Completeness With Respect to Prompt Emission Properties}
\label{sec:completeness}

\begin{figure}
\centerline{
\includegraphics[scale=0.55,angle=0]{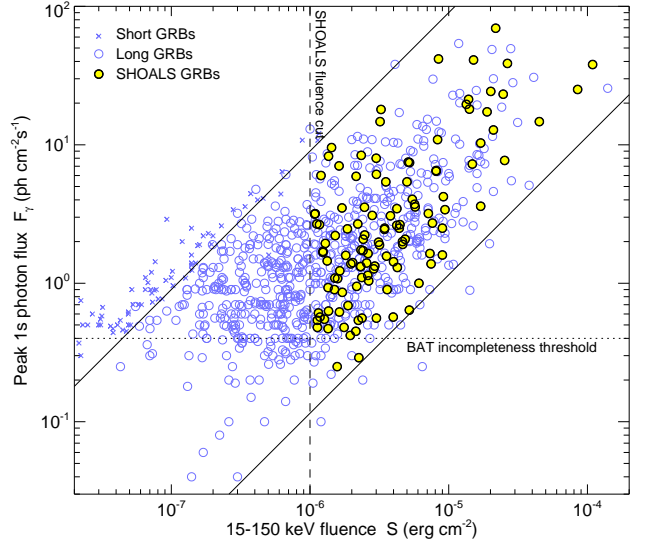}} 
\caption{Peak photon flux versus total fluence for \textit{Swift} GRBs.  Events in our sample are shown in yellow.  Although the BAT triggers (approximately) according to photon count rate and our sample is cut in fluence, due to the correlation between the two parmeters our sample nevertheless is nearly complete in fluence: only the small triangular region to the right of the fluence cut and below the incompleteness threshold is expected to contain real bursts that would have passed our fluence cut if they had successfully triggered the BAT.  Diagonal lines are of constant $F_\gamma$/$S$ and show the region containing nearly all long GRBs.}
\label{fig:fluxfluence}
\end{figure}

While we elected to cut on fluence\footnote{Here and throughout the paper, we emphasize that the flux and fluence we refer to as measured refer only to the BAT 15-150 keV band, not broad-band or bolometric values. Because GRBs occupy a wide range of intrinsic $E_{\rm peak}$ there are many GRBs with bolometric fluences well above our cut that BAT is insensitive to, but because we consider BAT-band properties alone this does not impact our analysis or conclusions.} for the reasons detailed in \S \ref{sec:prompt}, \textit{Swift}'s primary trigger mechanism is more strongly tied to a burst's peak flux, and a difficult-to-quantify incompleteness affects the satellite's ability to detect and trigger on bursts with peak flux close to its threshold---complicating, in principle, any attempts to measure the intrinsic GRB rate.  While a cut on peak flux at a value above the incompleteness level would largely eliminate this concern, the impact of our fluence cut is less straightforward: it is possible that \textit{Swift} itself may have missed some GRBs whose fluence was above our cut level but whose peak flux was too low to trigger the instrument (due to a particularly long and smooth light curve).

Even so, we have reason to expect that the sample established by our fluence cut is nearly complete.  In Figure \ref{fig:fluxfluence} we plot the peak photon flux and total measured energy fluence for all \textit{Swift} GRBs (open circles) and for our SHOALS sample (filled circles).\footnote{Figure \ref{fig:fluxfluence}, and subsequent figures/tables, show the updated \citep{Sakamoto+2011} values of flux and fluence for bursts where revised measurements of both values are available.  Our results do not differ if the original GCN measurements are used.}  Unsurprisingly, these two parameters strongly correlate in a linear fashion but show scatter at the level of $\sim$1 dex; the region of flux-fluence space inhabited by most \textit{Swift} long-duration GRBs ($98$\%) is demarcated by the solid diagonals.   The effective BAT triggering threshold is readily apparent in the data at approximately 0.4 ph cm$^{-2}$ s$^{-1}$.  For bursts with very high fluences, the BAT trigger sensitivity is clearly not an issue (e.g., no bursts with a fluence of $S > 10^{-5}$ erg cm$^{-2}$ have a peak flux anywhere near the BAT incompleteness threshold: to a good approximation, bursts this bright in the field of view should always trigger the telescope regardless of light curve shape, off-axis angle, etc.).  For our chosen cut level of $S > 10^{-6}$ erg cm$^{-2}$, instrumental incompleteness is not completely negligible, and a few events which in principle may have been bright enough to meet our selection criteria could have been missed by the BAT.   Based on the flux/fluence ratio distribution for brighter bursts where we are confident the BAT sample is complete, we expect the number of such events is relatively small ($\sim$10 or less) and not likely to have a significant impact on the conclusions presented in this paper.

\subsection{The Redshift-Dependent GRB Rate Density}
\label{sec:zrate}

\begin{figure}
\centerline{
\includegraphics[scale=0.5,angle=0]{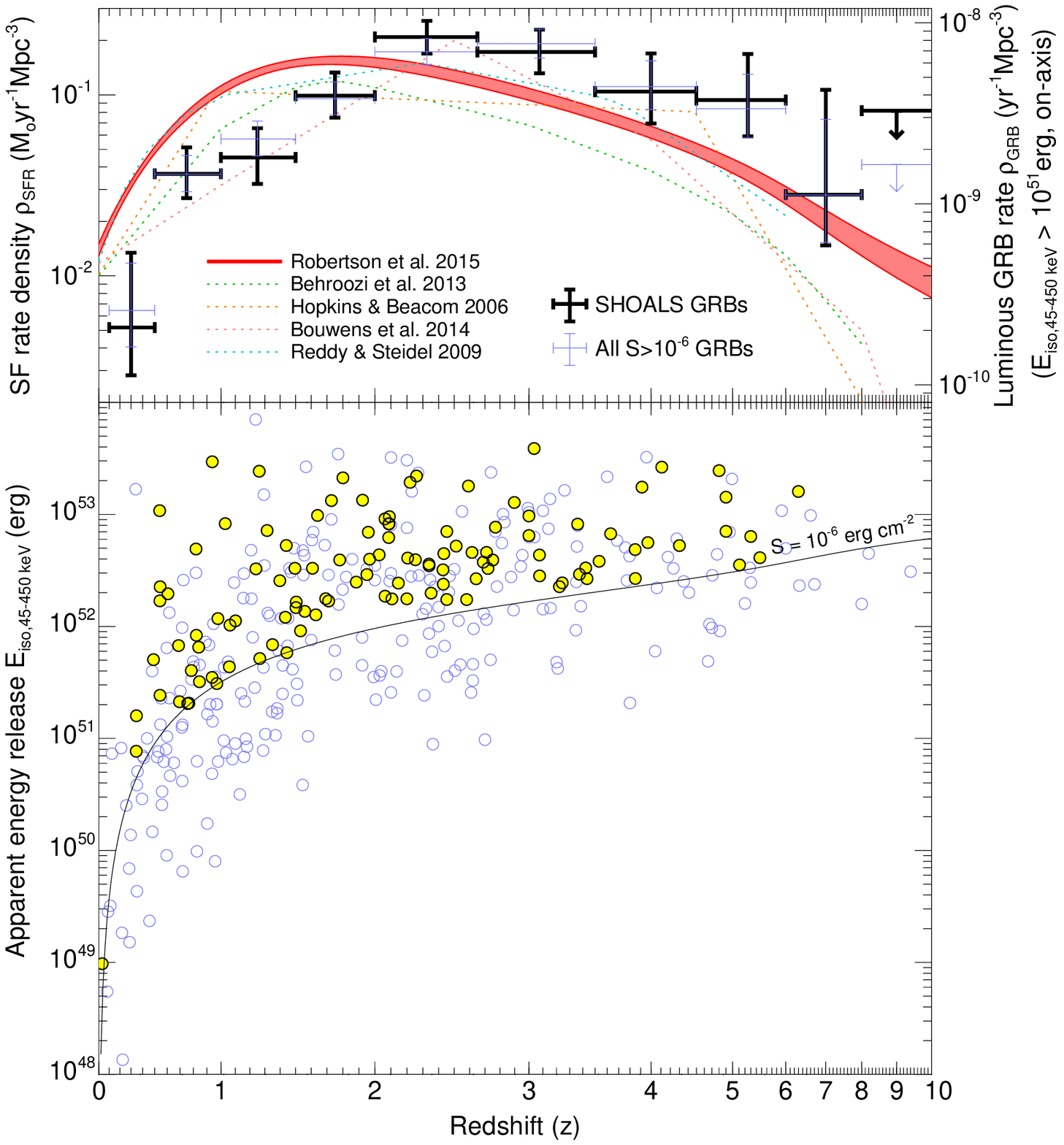}} 
\caption{The comoving GRB rate density versus redshift compared to the comoving field-survey star-formation rate density (top panel), as inferred from our redshift and $E_{\rm iso}$ distribution (lower panel).  Curves plot field-galaxy star-formation histories from various sources (\citealt{Madau+2014} via \citealt{Robertson+2015}; \citealt{Behroozi+2013,Hopkins+2006,Reddy+2009,Robertson+2012,Bouwens+2014}).  The scaling of the GRB rate (thick data points from the SHOALS sample, thin measurements from all $S>10^{-6}$ erg cm$^{-2}$ \textit{Swift} GRBs; error bars denote 10--90\% binomial confidence interval) is arbitrary, normalized against the star-formation rate at $z\sim2$.  The behavior of the GRB rate history is qualitatively similar to the star-formation rate history, but shows a modest (factor of $\sim5$) excess at high redshifts (or equivalently, a depression at low redshift) compared to the most recent SFR density measurements.}
\label{fig:ratez}
\end{figure}

\begin{deluxetable*}{llllllll}  
\tabletypesize{\small}
\tablecaption{GRB Rate Measurements}
\tablecolumns{8}
\tablehead{
\colhead{} & 
\colhead{} &
\colhead{} &
\colhead{} &
\multicolumn{2}{c}{\underline{SHOALS}} &
\multicolumn{2}{c}{\underline{All \textit{Swift}}} \\
\colhead{$z_{\rm min}$} & 
\colhead{$z_{\rm max}$} &
\colhead{$E_{\rm iso,sens}$\tablenotemark{a}} &
\colhead{$C_{\rm complete}$\tablenotemark{b}} &
\colhead{$N$\tablenotemark{c}} &
\colhead{$\rho_{\rm GRB,51}$\tablenotemark{d}} &
\colhead{$N$\tablenotemark{c}} &
\colhead{$\rho_{\rm GRB,51}$\tablenotemark{d}}
}
\startdata
 0.10 & 0.50  &   0.90 & 0.91 &   2 &$0.21_{-0.09}^{+0.33}$ &    5  &  $0.26_{-0.10}^{+0.21}$ \\
 0.50 & 1.00  &   3.19 & 2.83 &  13 &$1.47_{-0.39}^{+0.59}$ &   26  &  $1.46_{-0.29}^{+0.39}$ \\
 1.00 & 1.50  &   6.22 & 5.17 &  11 &$1.80_{-0.51}^{+0.81}$ &   28  &  $2.28_{-0.44}^{+0.58}$ \\
 1.50 & 2.00  &   9.60 & 7.65 &  16 &$3.95_{-0.96}^{+1.36}$ &   31  &  $3.81_{-0.71}^{+0.91}$ \\
 2.00 & 2.65  &  14.19 & 10.8 &  27 &$8.34_{-1.58}^{+1.91}$ &   45  &  $6.92_{-1.06}^{+1.26}$ \\
 2.65 & 3.50  &  20.23 & 14.9 &  17 &$6.90_{-1.64}^{+2.27}$ &   38  &  $7.67_{-1.28}^{+1.58}$ \\
 3.50 & 4.50  &  27.19 & 19.5 &   7 &$4.17_{-1.39}^{+2.61}$ &   15  &  $4.44_{-1.13}^{+1.73}$ \\
 4.50 & 6.00  &  37.15 & 25.8 &   5 &$3.75_{-1.39}^{+2.98}$ &    9  &  $3.36_{-1.03}^{+1.85}$ \\
 6.00 & 8.00  &  49.54 & 33.5 &   1 &$1.12_{-0.53}^{+3.14}$ &    2  &  $1.12_{-0.51}^{+1.81}$ \\
 8.00 &10.00  &  61.03 & 40.4 &   0 &$0.00_{-0.00}^{+3.27}$ &    0  &  $0.00_{-0.00}^{+1.65}$ \\
\enddata
\label{tab:ratez}
\tablenotetext{a}{$E_{\rm iso}$ sensitivity threshold for inclusion of a GRB in the count, in units of $10^{51}$ erg as measured in the 45--450 keV rest-frame band.}
\tablenotetext{b}{Completeness correction applied to scaling observed GRB counts to a common threshold of $E_{\rm iso} > 10^{51}$ erg.}
\tablenotetext{c}{Number of GRBs observed with  $E_{\rm iso} > E_{\rm iso,sens}$}
\tablenotetext{d}{On-axis luminous ($E_{\rm iso,45-450} >10^{51}$ erg) GRB rate density over this redshift interval, in units of $10^{-9}$ yr$^{-1}$ Mpc$^{-1}$.}
\end{deluxetable*}

The observed GRB redshift distribution can be used to infer the \emph{intrinsic} comoving GRB rate, provided that the sensitivity of the instrument and selection of the known-redshift sample are well-understood and the GRB luminosity function can also be inferred.  As we argue above that the sample is intrinsically fluence-limited to a good approximation, these procedures can be applied to our sample also (using energetics-dependent quantities in place of the more standard luminosity-dependent quantities).

We first calculate the isotropic-equivalent energy releases for all GRBs in our sample, $E_{\rm iso,45-450 keV} = S_{\rm 15-150 keV}$ 4$\pi$ $d_L^{2} (1+z)^{-1} k(z)$ \citepeg{Bloom+2001}.  Unlike most previous authors we do not make any attempt to estimate a bolometric or wide-bandwidth value, which is fraught with uncertainty considering that \textit{Swift}'s bandpass is narrow and does not usually contain the peak energy of the burst; the BAT-to-bolometric correction can easily be orders of magnitude and it is practically unconstrained by the \textit{Swift} data.  Instead, we calculate the energy release in the 45-450 keV rest-frame band only, corresponding to the window observed by the BAT (15-150 keV observer-frame) for an event at $z\sim2$, close to the approximate sample redshift median and also where previously-reported divergences in the GRB-to-SFR ratio begin to become apparent.  This makes the $k$-correction much smaller and more reliable:  specifically, its value is $k(z) = [(1+z)/(1+2)]^{\Gamma-2}$, where $\Gamma$ is the measured photon index over the BAT band (for bursts in which $\Gamma$ is not well-constrained by the data we take the population median value of $\Gamma=1.5$).  This $k$-correction is always small (exactly unity at $z=2$ and within factor of 2 between $1<z<4$ for any common value of $\Gamma$) and the uncertainty is even smaller (typically a few percent) so these measurements should be highly reliable.

We assume an intrinsic isotropic-equivalent energetics rate function $\phi(E_{\rm iso})$ following a single power-law and measure its power-law index by fitting the observed fluence distribution of GRBs at moderate redshift ($0.5 < z < 3.5$) in the entire \textit{Swift} sample above our fluence cut; we infer $\phi_{E{\rm iso}} \propto E_{\rm iso}^{-1.7 \pm 0.2}$.  We then count the observed number of GRBs (in both the SHOALS sample and for all $S > 10^{-6}$ \textit{Swift} GRBs) with $E_{\rm iso} > 10^{51}$ erg per redshift bin, and use the integral of the energetics function to correct the number in each bin for incompleteness in $E_{\rm iso}$, scaling all bins to a common energy cutoff of $E_{\rm iso} > 10^{51}$ erg.
This number is then scaled in the usual way by ${\rm d}V/{\rm d}z/(1+z)$ to convert the density in redshift to a comoving rate density.  

The resulting redshift-dependent rate (top panel of Figure \ref{fig:ratez} and Table \ref{tab:ratez}) shows a broad peak at $z\sim1.5-3.5$ and a modest decline towards lower and higher redshifts.  This behavior is consistent with previous studies of this type \citepeg{Kistler+2008,Kistler+2009,Butler+2010,Wanderman+2010,Robertson+2012,Jakobsson+2012,Salvaterra+2012}---but, unlike these previous studies (with the exceptions for the significantly smaller samples of \citealt{Jakobsson+2012} and \citealt{Salvaterra+2012}) our results are not limited by systematics associated with the highly incomplete redshift distribution.

It is somewhat surprising that the complete, unbiased redshift distribution produced from our work is so similar to the distribution inferred from previous studies based (largely) on samples drawn from afterglow redshifts in the literature, which favor low-extinction sightlines.  As we mentioned earlier (\S \ref{sec:zrate}) this suggests a relatively uniform fraction of obscured star-formation across most of cosmic history (our sample provides good number statistics between approximately $0.5<z<4$).  It also indicates that high-$z$ GRBs, despite being uncommon (intrinsically and observationally) and difficult to follow-up, are being identified with comparable efficiency as their low-redshift counterparts.

\section{A Redshift-Dependent GRB Efficiency?}
\label{sec:discussion}

Since GRBs originate from massive stars, the comparison of the GRB rate history and the star-formation rate history as derived by traditional galaxy survey methods (e.g., \citealt{Madau+1998,Reddy+2009,Hopkins+2006,Behroozi+2013,Bouwens+2014}; see \citealt{Madau+2014} for a review) imposes a constraint on the degree to which the GRB rate is affected by other factors, as well as an independent check on the star-formation rate history itself.  Many previous studies have quantitatively compared these two relations in detail; most of these (\citealt{Daigne+2006,Le+2007,Guetta+2007b,Salvaterra+2007,Yueksel+2008,Kistler+2008,Virgili+2011,Robertson+2012,Lien+2014}, c.f. \citealt{Wanderman+2010,Elliott+2012,Yu+2015}) have concluded that the GRB rate at high redshift is significantly higher than what would be inferred from the galaxy survey-inferred star-formation history.  This suggested either that the field surveys were insufficiently accounting for the number density of low-luminosity galaxies they do not detect, or that the cosmic GRB production efficiency $\epsilon(z) = \rho_{\rm GRB}(z)$/$\rho_{\rm SFR}(z)$ is not constant (due to e.g. metallicity enrichment suppressing the GRB rate at lower redshift): see e.g., \cite{Jakobsson+2012,Trenti+2013}.  

We carried out this exercise with our own observations as well, comparing our (selection-unbiased) rate distribution to the most recent estimate of the star-formation rate density out to very high redshifts using the recent galaxy luminosity functions of \cite{Madau+2014}, integrated down to $L = 10^{-3} L_*$ galaxies by \cite{Robertson+2015}.  We confirm the high-redshift excess (or, equivalently, a low-redshift deficiency) in the GRB rate relative to the UV-inferred star-formation rate: normalizing\footnote{Note that the relative normalization of the two curves is effectively arbitrary since we do not know the fraction of stars which explode as GRBs or the beaming correction to better than an order of magnitude.} the two curves at $z=2$, the $z\sim5$ GRB rate is in excess of the star-formation rate by a factor of 2--3 and the $z<0.5$ GRB rate is below it by a factor of 2--5.

This provides further support to the notion that the cosmic GRB efficiency may vary over time.  We emphasize, however, that while the deviation is significant, it is also relatively modest in magnitude: a factor of $\sim$5--10 across the entire span of cosmic history from $z=5$ to $z\sim0$.  In fact, perhaps the more salient conclusion to be drawn from Figure \ref{fig:ratez} is that \emph{strong} evolution in the GRB-to-SFR ratio is \emph{not} observed.  This argues that the GRB rate's dependency on metallicity must have only low-to-moderate impact on the cosmological rate, in disagreement with models implying strong variations (e.g., the single-star models of \citealt{Langer+2006}, which require approximately $Z < 0.1Z_\odot$, would imply a variation by more than a factor of 40 over this period).  Earlier analyses using non-uniform afterglow-based samples \citep{Robertson+2012,Trenti+2013,Hao+2013} have reached similar conclusions but were limited by systematics associated with incomplete redshift measurement; our work confirms that these conclusions hold within an unbiased sample.

Our result is also in agreement with independent investigations of the variations of the GRB rate based on the properties of the hosts themselves.  Recent emission-line studies \citepeg{Graham+2013,Kruehler+2015} show that GRBs form readily at moderate metallicities ($\sim$0.5$Z_\odot$ or more) that are characteristic of typical star-forming galaxies throughout most of the Universe's history.  And the host stellar mass distribution we infer from IRAC observations of the host galaxies in our sample (Paper II) similarly suggests that a metallicity threshold for GRB production is likely present but is relatively high, $\sim Z_\odot$.

\section{Summary}
\label{sec:summary}

We have defined a new legacy sample of gamma-ray bursts and host galaxies spanning nearly all of cosmic history:  119 \textit{Swift} GRBs at $0.03 < z < 6.29$ drawn from the \textit{Swift} GRB catalog using a series of observability cuts, plus a requirement that the BAT fluence exceed $S > 10^{-6}$ erg cm$^{-2}$.   Combining publicly-available afterglow and host-galaxy redshift measurements with our own host-galaxy campaign, we present redshifts for 110 (92\%) of these bursts, or 112 (94\%) if we include the tentative redshift of 110709B and the lower+upper limits on 070808.  This is by far the largest, and the most redshift-complete, sample of its type, and provides the most up-to-date and unbiased view of cosmic history as seen by GRBs.  Only one event lacks a redshift upper limit and only one event with measured redshift is at $z>5.5$.  

Mapping our redshift distribution to a comoving rate density to measure the evolution of the GRB rate with cosmic time, we measure a rise in the GRB rate from $z>6$ to $z\sim2$, followed by a drop of an order of magnitude from $z\sim2$ to the present time---the same pattern seen by traditional metrics of the cosmic star-formation rate density.  Quantitatively comparing the GRB rate history to the cosmic star-formation history, we find a modest excess in the GRB rate (versus SFRD) at high redshift compared to lower redshifts.  Consistent with previous work, this suggests that the cosmic GRB efficiency was higher in the first few billion years of cosmic history relative to today and provides support to the idea of a metal-dependent progenitor, but the modest degree of this variation rules out models requiring an exclusively \emph{very}-metal-poor (e.g., $< 0.1 Z_\odot$) environment.  The small number of high-redshift GRBs in the sample places strong limits on the fraction of high-$z$ bursts detected by \textit{Swift} and on the intrinsic GRB rate at high redshifts.

Addressing the GRB rate-evolution question in detail---and actually applying our GRB population to address broader questions in astronomy---requires more than just redshifts.   In particular, examination of the afterglows and (especially) host galaxies of these events is needed to study the galaxy population giving rise to the GRBs in our sample at each redshift, providing an independent test of factors controlling the GRB rate and a means to explore directly the importance and nature of the dusty, low-luminosity, and high-redshift populations uniquely probed by GRBs.  We are collecting and analyzing these observations under the programs introduced here, and the direct study of the hosts within our sample will serve as the subject of all remaining papers.  As the first large, thorough, highly-complete, and multi-band survey of an unbiased GRB host sample, SHOALS will enable unprecedented insight into the GRB rate and progenitor as well as a unique perspective into galaxy evolution and cosmic history.

\vskip 0.02cm

\acknowledgments

This work is based in part on observations made with the  {\it Spitzer Space Telescope}, which is operated by the Jet Propulsion Laboratory, California Institute of Technology, under a contract with NASA.   It is also based in part on observations with the NASA/ESA {\it Hubble Space Telescope}, obtained from the Space Telescope Science Institute.  STScI is operated by the Association of Universities for Research in Astronomy, Inc. under NASA contract NAS 5-26555.  These observations are associated with program GO-90062.
Support for this work was provided by NASA through an award issued by JPL/Caltech, and through Hubble Fellowship grant HST-HF-51296.01-A awarded by the Space Telescope Science Institute, which is operated by the Association of Universities for Research in Astronomy, Inc., for NASA, under contract NAS 5-26555.  D.A.P. further acknowledges support from a Marie Sklodowska-Curie Individual Fellowship within the Horizon 2020 European Union (EU) Framework Programme for Research and Innovation (H2020-MSCA-IF-2014-660113).
The Dark Cosmology Centre is funded by the DNRF. 
The research leading to these results has received funding from the European Research Council under the European Union's Seventh Framework Program (FP7/2007-2013)/ERC Grant agreement no. EGGS-278202.
S.~Schulze acknowledges support from CONICYT-Chile FONDECYT 3140534, Basal-CATA PFB-06/2007, and Project IC120009 ``Millennium Institute of Astrophysics (MAS)'' of Iniciativa Cient\'{\i}fica Milenio del Ministerio de Econom\'{\i}a, Fomento y Turismo.  SK acknowledges support from FONDECYT 3130488.  AC is supported by the NASA Postdoctoral Program at the Goddard Space Flight Center, administered by Oak Ridge Associated Universities through a contract with NASA.  AdUP and CCT are supported by Ram\'on y Cajal fellowships.

Some of the data presented here were obtained at the W.M. Keck Observatory, which is operated as a scientific partnership among the California Institute of Technology, the University of California and the National Aeronautics and Space Administration. The Observatory was made possible by the generous financial support of the W.M. Keck Foundation.  The authors wish to recognize and acknowledge the very significant cultural role and reverence that the summit of Mauna Kea has always had within the indigenous Hawaiian community. 
Based on observations made with ESO Telescopes at the La Silla Paranal Observatory, and on observations made with the Gran Telescopio Canarias (GTC).  Part of the funding for GROND (both hardware as well as personnel) was generously granted from the Leibniz-Prize to Prof. G. Hasinger (DFG grant HA 1850/28-1).

It is a pleasure to thank the \textit{Swift} team for creating such a superbly prolific and successful instrument, without which this study would have been impossible.  We also wish to extend thanks to the entire ground-based GRB follow-up community for providing many of the afterglow identifications and redshifts critical to our study.  We particularly acknowledge D.~Malesani, Y.~Urata, and K.~Huang for providing improved afterglow positions and images, and thank the anonymous referee and D.~A.~Kann for useful comments.  We also thank our other collaborators with assistance with acquiring ground-based observations, including R.~Sanchez~Ramirez, F.~E.~Bauer, and P.~Schady.  We thank B.~Robertson and M.~Trenti for useful discussions and also thank B.~Robertson for providing the most up-to-date SFRD curves.


{\it Facilities:} \facility{Spitzer:IRAC, Keck-I:LRIS, Keck-I:MOSFIRE, HST:WFC3, VLT:X-shooter, GTC:OSIRIS, Max Planck:2.2m (GROND), PO:1.5m}

\clearpage
\LongTables

\begin{deluxetable*}{lllllllllllll}  
\tabletypesize{\small}
\tablecaption{The SHOALS GRB sample}
\tablecolumns{13}
\tablehead{
\colhead{GRB} &
\colhead{$S_{\rm BAT}$\tablenotemark{a}} &
\colhead{RA\tablenotemark{b}} & 
\colhead{dec\tablenotemark{b}} &
\colhead{unc.\tablenotemark{c}} &
\colhead{type.\tablenotemark{d}} &
\colhead{OT?\tablenotemark{e}} & 
\colhead{dusty?\tablenotemark{f}} & 
\colhead{ref.\tablenotemark{i}} &
\colhead{$z$} &
\colhead{early?\tablenotemark{g}} &
\colhead{origin\tablenotemark{h}} &
\colhead{ref.\tablenotemark{i}} }
\startdata
050128  &   51 & 14:38:17.68 & $-$34:45:55.4 & 1.70 & X  &    &   & 1   & $<$5.5                 & N & host & *   \\
050315  &   32 & 20:25:54.17 & $-$42:36:02.2 & 0.28 & O  & Y  &   & 2   & 1.9500                 & Y & AG   & 3   \\
050318  &   11 & 03:18:51.01 & $-$46:23:44.0 & 0.60 & O  & Y  &   & 4   & 1.4436                 & Y & AG   & 3   \\
050319  &   13 & 10:16:47.94 & $+$43:32:53.5 & 0.40 & O  & Y  &   & 5   & 3.2425                 & Y & AG   & 6   \\
050401  &   83 & 16:31:28.80 & $+$02:11:14.0 & 0.43 & O  & Y  & Y & 2   & 2.8983                 & Y & AG   & 6   \\
050525A &  151 & 18:32:32.59 & $+$26:20:22.3 & 0.31 & O  & Y  &   & 2   & 0.606                  & Y & host & 7   \\
050726  &   20 & 13:20:11.95 & $-$32:03:51.2 & 0.32 & U  & Y  &   & 2   & $<$3.5                 & N & AG   & *   \\
050730  &   24 & 14:08:17.10 & $-$03:46:17.6 & 0.37 & O  & Y  &   & 8   & 3.9693                 & Y & AG   & 6   \\
050802  &   22 & 14:37:05.84 & $+$27:47:12.4 & 0.60 & O  & Y  &   & *   & 1.7102                 & Y & AG   & 6   \\
050803  &   22 & 23:22:37.86 & $+$05:47:08.0 & 1.40 & X  & N  & Y & 1   & 4.3$^{+0.6}_{-2.4}$    & N & host & *   \\
050814  &   20 & 17:36:45.39 & $+$46:20:21.2 & 0.30 & O  & Y  &   & 9   & 5.3                    & Y & AG   & 10  \\
050820A &   34 & 22:29:38.10 & $+$19:33:36.8 & 0.31 & O  & Y  &   & 8   & 2.6147                 & Y & AG   & 6   \\
050822  &   25 & 03:24:27.22 & $-$46:02:00.1 & 0.70 & X  &    &   & 11  & 1.434                  & N & host & 12  \\
050904  &   52 & 00:54:50.88 & $+$14:05:09.5 & 0.40 & O  & Y  &   & *   & 6.295                  & Y & AG   & 13  \\
050922B &   24 & 00:23:13.38 & $-$05:36:17.5 & 1.70 & X  & N  &   & 1   & 4.9$^{+0.3}_{-0.6}$    & N & host & *   \\
050922C &   16 & 21:09:33.08 & $-$08:45:30.3 & 0.30 & O  & Y  &   & 14  & 2.1995                 & Y & AG   & 6   \\
051001  &   18 & 23:23:48.72 & $-$31:31:23.6 & 1.50 & X  & N  & Y & 11  & 2.4296                 & N & host & 15  \\
051006  &   12 & 07:23:14.10 & $+$09:30:19.4 & 1.50 & X  & Y  & Y & 1   & 1.059                  & N & host & 12  \\
060115  &   17 & 03:36:08.32 & $+$17:20:42.8 & 0.31 & O  & Y  &   & 8   & 3.5328                 & Y & AG   & 6   \\
060202  &   22 & 02:23:22.94 & $+$38:23:03.9 & 0.50 & I  & N  & Y & *   & 0.785                  & N & host & 16  \\
060204B &   29 & 14:07:14.89 & $+$27:40:36.2 & 1.10 & O  & Y  &   & 17  & 2.3393                 & N & host & *   \\
060210  &   76 & 03:50:57.38 & $+$27:01:34.2 & 0.60 & O  & Y  & Y & 18  & 3.9122                 & Y & AG   & 6   \\
060218  &   65 & 03:21:39.68 & $+$16:52:01.9 & 0.28 & O  & Y  &   & 8   & 0.0331                 & Y & host & 19  \\
060306  &   22 & 02:44:22.92 & $-$02:08:54.1 & 1.30 & X  & N  & Y & 11  & 1.559                  & N & host & 20  \\
060502A &   23 & 16:03:42.62 & $+$66:36:03.0 & 0.40 & O  & Y  &   & *   & 1.5026                 & Y & AG   & 6   \\
060510B &   40 & 15:56:29.48 & $+$78:34:12.1 & 0.20 & O  & Y  &   & *   & 4.9                    & Y & AG   & 21  \\
060522  &   11 & 21:31:44.84 & $+$02:53:09.7 & 0.42 & O  & Y  &   & 8   & 5.11                   & Y & AG   & 22  \\
060526  &   12 & 15:31:18.34 & $+$00:17:04.9 & 0.16 & O  & Y  &   & 8   & 3.2213                 & Y & AG   & 6   \\
060607A &   26 & 21:58:50.40 & $-$22:29:47.1 & 0.38 & I  & Y  &   & 8   & 3.0749                 & Y & AG   & 6   \\
060707  &   16 & 23:48:19.06 & $-$17:54:17.3 & 0.37 & O  & Y  &   & 8   & 3.4240                 & Y & AG   & 6   \\
060714  &   29 & 15:11:26.41 & $-$06:33:58.3 & 0.43 & O  & Y  &   & 8   & 2.7108                 & Y & AG   & 6   \\
060719  &   15 & 01:13:43.71 & $-$48:22:51.0 & 0.29 & O  & Y  & Y & 8   & 1.5320                 & N & host & 15  \\
060729  &   26 & 06:21:31.80 & $-$62:22:12.3 & 0.27 & O  & Y  &   & 8   & 0.5428                 & Y & AG   & 6   \\
060814  &  148 & 14:45:21.31 & $+$20:35:10.5 & 0.18 & O  & IR & Y & 8   & 1.9229                 & Y & host & 15  \\
060908  &   28 & 02:07:18.41 & $+$00:20:31.3 & 0.42 & O  & Y  &   & 8   & 1.8836                 & Y & AG   & 6   \\
060912A &   14 & 00:21:08.14 & $+$20:58:17.4 & 0.35 & O  & Y  &   & 8   & 0.937                  & Y & host & 23  \\
060927  &   11 & 21:58:12.01 & $+$05:21:48.6 & 0.19 & O  & Y  &   & 8   & 5.467                  & Y & AG   & 24  \\
061007  &  450 & 03:05:19.58 & $-$50:30:02.3 & 0.33 & O  & Y  &   & 8   & 1.2622                 & Y & AG   & 6   \\
061021  &   30 & 09:40:36.12 & $-$21:57:04.8 & 0.30 & O  & Y  &   & 8   & 0.3463                 & N & AG   & 6   \\
061110A &   11 & 22:25:09.84 & $-$02:15:31.4 & 0.37 & O  & Y  &   & 8   & 0.7578                 & Y & AG   & 6   \\
061110B &   14 & 21:35:40.39 & $+$06:52:34.0 & 0.24 & O  & Y  &   & 8   & 3.4344                 & Y & AG   & 6   \\
061121  &  139 & 09:48:54.55 & $-$13:11:42.9 & 0.36 & O  & Y  &   & 8   & 1.3145                 & Y & AG   & 6   \\
061202  &   35 & 07:02:06.09 & $-$74:41:54.7 & 1.40 & X  &    &   & 1   & 2.253                  & N & host & *   \\
061222A &   81 & 23:53:03.41 & $+$46:31:58.6 & 0.30 & I  & IR & Y & 25  & 2.088                  & N & host & 26  \\
070110  &   16 & 00:03:39.27 & $-$52:58:27.2 & 0.30 & O  & Y  &   & 8   & 2.3521                 & Y & AG   & 6   \\
070129  &   30 & 02:28:00.94 & $+$11:41:04.1 & 0.31 & O  & Y  &   & 8   & 2.3384                 & N & host & 15  \\
070223  &   19 & 10:13:48.39 & $+$43:08:00.7 & 0.40 & I  & Y  &   & 27  & 1.6295                 & N & host & *   \\
070306  &   55 & 09:52:23.30 & $+$10:28:55.2 & 0.31 & O  & IR & Y & 8   & 1.4959                 & Y & host & 28  \\
070318  &   26 & 03:13:56.81 & $-$42:56:46.1 & 0.35 & O  & Y  &   & 8   & 0.840                  & Y & AG   & 29  \\
070328  &   91 & 04:20:27.73 & $-$34:04:00.5 & 1.40 & X  &    &   & 1   & 2.0627                 & N & host & 20  \\
070419B &   75 & 21:02:49.77 & $-$31:15:49.0 & 0.39 & O  & Y  &   & 8   & 1.9588                 & N & host & 15  \\
070508  &  201 & 20:51:11.70 & $-$78:23:05.1 & 0.40 & O  & Y  & Y & *   & 0.82                   & Y & host & 30  \\
070521  &   81 & 16:10:38.61 & $+$30:15:21.9 & 1.40 & X  & N  & Y & 1   & 2.0865                 & N & host & 20  \\
070621  &   44 & 21:35:10.09 & $-$24:49:03.1 & 1.40 & X  & N  & Y & 1   & $<$5.5                 & N & host & *   \\
070721B &   36 & 02:12:32.96 & $-$02:11:40.8 & 0.38 & O  & Y  &   & 8   & 3.6298                 & Y & AG   & 6   \\
070808  &   13 & 00:27:03.36 & $+$01:10:33.9 & 1.50 & X  & N  & Y & 1   & 1.35$\pm$0.85          & N & $N_{\rm H,X}$ & *   \\
071020  &   23 & 07:58:39.78 & $+$32:51:40.4 & 0.35 & I  & Y  &   & 31  & 2.1462                 & Y & AG   & 6   \\
071021  &   14 & 22:42:34.30 & $+$23:43:06.2 & 0.60 & O  & IR & Y & 32  & 2.4520                 & Y & host & 15  \\
071025  &   73 & 23:40:17.07 & $+$31:46:42.8 & 0.35 & I  & Y  & Y & 33  & 4.8$^{+0.4}_{-0.4}$    & N & AG   & 34  \\
071112C &   30 & 02:36:50.95 & $+$28:22:16.8 & 0.41 & O  & Y  &   & 5   & 0.8227                 & Y & AG   & 6   \\
080205  &   20 & 06:33:00.63 & $+$62:47:31.7 & 0.50 & U  & Y  &   & 35  & 2.72$^{+0.24}_{-0.74}$ & N & host & *   \\
080207  &   61 & 13:50:02.97 & $+$07:30:07.3 & 0.50 & X  & N  & Y & 36  & 2.0858                 & Y & host & 15  \\
080210  &   18 & 16:45:04.01 & $+$13:49:35.6 & 0.60 & U  & Y  &   & 35  & 2.6419                 & Y & AG   & 6   \\
080310  &   23 & 14:40:13.80 & $-$00:10:30.7 & 0.40 & O  & Y  &   & 37  & 2.4274                 & Y & AG   & 6   \\
080319A &   44 & 13:45:20.01 & $+$44:04:48.4 & 0.70 & O  & Y  &   & 38  & 2.0265                 & N & host & *   \\
080319B &  850 & 14:31:40.99 & $+$36:18:08.7 & 0.30 & O  & Y  &   & *   & 0.9382                 & Y & AG   & 6   \\
080325  &   49 & 18:31:34.23 & $+$36:31:24.8 & 0.30 & I  & IR & Y & 39  & 1.78                   & N & host & 40  \\
080411  &  265 & 02:31:55.21 & $-$71:18:07.3 & 0.50 & U  & Y  &   & 35  & 1.0301                 & Y & AG   & 6   \\
080413A &   35 & 19:09:11.75 & $-$27:40:40.4 & 0.40 & O  & Y  &   & *   & 2.4330                 & Y & AG   & 6   \\
080413B &   33 & 21:44:34.66 & $-$19:58:52.4 & 0.50 & U  & Y  &   & 35  & 1.1014                 & Y & AG   & 6   \\
080430  &   12 & 11:01:14.76 & $+$51:41:08.0 & 0.40 & O  & Y  &   & 41  & 0.767                  & Y & AG   & 42  \\
080603B &   25 & 11:46:07.67 & $+$68:03:39.8 & 0.30 & U  & Y  &   & 35  & 2.6892                 & Y & AG   & 6   \\
080605  &  134 & 17:28:30.04 & $+$04:00:56.0 & 0.30 & O  & Y  & Y & *   & 1.6403                 & Y & AG   & 6   \\
080607  &  247 & 12:59:47.21 & $+$15:55:10.5 & 0.40 & O  & Y  & Y & *   & 3.0368                 & Y & AG   & 6   \\
080710  &   14 & 00:33:05.63 & $+$19:30:05.4 & 0.30 & O  & Y  &   & *   & 0.8454                 & Y & AG   & 6   \\
080721  &  141 & 14:57:55.84 & $-$11:43:24.6 & 0.40 & O  & Y  &   & 5   & 2.5914                 & Y & AG   & 6   \\
080804  &   38 & 21:54:40.18 & $-$53:11:04.8 & 0.35 & O  & Y  &   & *   & 2.2045                 & Y & AG   & 6   \\
080805  &   27 & 20:56:53.45 & $-$62:26:40.0 & 0.60 & O  & Y  &   & 43  & 1.5042                 & Y & AG   & 6   \\
080810  &   46 & 23:47:10.49 & $+$00:19:11.3 & 0.30 & O  & Y  &   & *   & 3.3604                 & Y & AG   & 6   \\
080916A &   42 & 22:25:06.23 & $-$57:01:22.9 & 0.40 & O  & Y  &   & *   & 0.6887                 & Y & AG   & 6   \\
080928  &   24 & 06:20:16.83 & $-$55:11:58.7 & 0.40 & O  & Y  &   & *   & 1.6919                 & Y & AG   & 6   \\
081008  &   43 & 18:39:49.86 & $-$57:25:53.0 & 0.60 & U  & Y  &   & 35  & 1.967                  & Y & AG   & 44  \\
081029  &   21 & 23:07:05.35 & $-$68:09:19.7 & 0.50 & U  & Y  &   & 35  & 3.8479                 & Y & AG   & 45  \\
081109A &   40 & 22:03:09.61 & $-$54:42:40.1 & 0.35 & O  & Y  & Y & 46  & 0.9787                 & N & host & 15  \\
081118  &   12 & 05:30:22.18 & $-$43:18:05.1 & 0.60 & O  & Y  &   & 47  & 2.58                   & Y & AG   & 48  \\
081121  &   42 & 05:57:06.17 & $-$60:36:09.8 & 0.60 & U  & Y  &   & 35  & 2.512                  & Y & AG   & 49  \\
081128  &   23 & 01:23:13.10 & $+$38:07:38.7 & 0.20 & O  & Y  &   & *   & $<$3.4                 & N & host & *   \\
081210  &   19 & 04:41:56.20 & $-$11:15:26.8 & 0.64 & U  & Y  &   & 35  & 2.0631                 & N & host & 20  \\
081221  &  189 & 01:03:10.16 & $-$24:32:51.6 & 0.25 & I  & IR & Y & *   & 2.26                   & N & host & 50  \\
081222  &   52 & 01:30:57.60 & $-$34:05:41.6 & 0.50 & U  & Y  &   & 35  & 2.77                   & Y & AG   & 51  \\
090313  &   15 & 13:13:36.20 & $+$08:05:49.6 & 0.70 & O  & Y  &   & 52  & 3.375                  & Y & AG   & 53  \\
090404  &   31 & 15:56:57.52 & $+$35:30:57.5 & 0.40 & R  & N  & Y & 54  & 3.0$^{+0.8}_{-1.8}$    & N & host & 16  \\
090417B &   23 & 13:58:46.58 & $+$47:01:04.8 & 1.00 & X  & N  & Y & 11  & 0.345                  & N & host & 55  \\
090418A &   47 & 17:57:15.16 & $+$33:24:20.9 & 0.40 & O  & Y  &   & 56  & 1.608                  & Y & AG   & 57  \\
090424  &  218 & 12:38:05.08 & $+$16:50:15.0 & 0.60 & O  & Y  &   & *   & 0.544                  & Y & host & 58  \\
090516A &   90 & 09:13:02.60 & $-$11:51:15.0 & 0.40 & O  & Y  &   & *   & 4.109                  & Y & AG   & 59  \\
090519  &   12 & 09:29:07.00 & $+$00:10:48.9 & 0.60 & O  & Y  &   & 60  & 3.85                   & Y & AG   & 61  \\
090530  &   11 & 11:57:40.49 & $+$26:35:37.7 & 0.40 & O  & Y  &   & *   & 1.266                  & N & host & 62  \\
090618  & 1090 & 19:35:58.73 & $+$78:21:24.3 & 0.50 & O  & Y  &   & *   & 0.54                   & Y & AG   & 63  \\
090709A &  253 & 19:19:42.64 & $+$60:43:39.3 & 0.50 & I  & Y  & Y & 64  & 1.8$^{+0.5}_{-0.7}$    & N & host & 16  \\
090715B &   57 & 16:45:21.63 & $+$44:50:21.0 & 0.40 & O  & Y  &   & *   & 3.00                   & Y & AG   & 65  \\
090812  &   57 & 23:32:48.56 & $-$10:36:17.2 & 0.40 & O  & Y  &   & 66  & 2.452                  & Y & AG   & 67  \\
090814A &   13 & 15:58:26.39 & $+$25:37:52.4 & 0.40 & O  & Y  &   & *   & 0.696                  & Y & AG   & 68  \\
090926B &   71 & 03:05:13.93 & $-$39:00:22.2 & 1.40 & X  & N  & Y & 1   & 1.24                   & Y & AG   & 69  \\
091018  &   14 & 02:08:44.63 & $-$57:32:53.8 & 0.50 & O  & Y  &   & *   & 0.971                  & Y & AG   & 70  \\
091029  &   24 & 04:00:42.62 & $-$55:57:20.0 & 0.50 & O  & Y  &   & *   & 2.752                  & Y & AG   & 71  \\
091109A &   16 & 20:37:01.82 & $-$44:09:29.6 & 0.40 & O  & Y  &   & *   & 3.076                  & Y & AG   & 72  \\
091127  &   84 & 02:26:19.89 & $-$18:57:08.5 & 0.55 & U  & Y  &   & 35  & 0.490                  & Y & host & 73  \\
091208B &   32 & 01:57:34.10 & $+$16:53:22.6 & 0.25 & O  & Y  &   & *   & 1.0633                 & Y & AG   & 74  \\
100305A &   15 & 11:13:28.07 & $+$42:24:14.3 & 1.00 & X  & N  & Y & 11  &                        & N &      &     \\
100615A &   50 & 11:48:49.34 & $-$19:28:52.0 & 0.70 & X  & N  & Y & 75  & 1.398                  & N & host & 76  \\
100621A &  210 & 21:01:13.08 & $-$51:06:22.5 & 0.40 & I  & Y  & Y & 77  & 0.542                  & Y & host & 78  \\
100728B &   17 & 02:56:13.46 & $+$00:16:52.1 & 0.52 & U  & Y  &   & 35  & 2.106                  & Y & AG   & 79  \\
100802A &   36 & 00:09:52.38 & $+$47:45:18.8 & 0.50 & O  & Y  &   & 80  & $<$3.1                 & N & host & *   \\
100814A &   90 & 01:29:53.59 & $-$17:59:43.5 & 0.40 & O  & Y  &   & *   & 1.44                   & Y & AG   & 81  \\
110205A &  170 & 10:58:31.10 & $+$67:31:30.5 & 0.30 & O  & Y  &   & 82  & 2.22                   & Y & AG   & 83  \\
110709B &   94 & 10:58:37.11 & $-$23:27:16.7 & 0.70 & R  & N  & Y & 84  & 2.09?                  & N & host & *   \\
120119A &  170 & 08:00:06.94 & $-$09:04:53.8 & 0.30 & O  & Y  & Y & 85  & 1.728                  & Y & AG   & 85  \\
120308A &   12 & 14:36:20.05 & $+$79:41:12.2 & 0.55 & U  & Y  &   & 35  & $<$3.7                 & N & host & *   \\
\enddata
\label{tab:grbinfo}
\tablenotetext{a}{Swift-BAT prompt-emission fluence (15-150 keV) in units of 10$^{-7}$ erg cm$^{-2}$.  From \citep{Sakamoto+2011} if available, otherwise from the \textit{Swift} GRB table.}
\tablenotetext{b}{Best afterglow position (J2000), relative to the 2MASS astrometric system.}
\tablenotetext{c}{Position uncertainty (arcsec), including an estimate of the systematic uncertainty.}
\tablenotetext{d}{First letter of the wavelength at which the position was reported: X-ray, UV, Optical, IR, or Radio/millimeter.}
\tablenotetext{e}{Whether or not a variable optical afterglow was reported.  ``IR'' indicates a NIR afterglow was reported but not an optical ($\lambda < 1\mu$m) afterglow.  Blank if no observations were conducted or reported limits are very shallow.}
\tablenotetext{f}{Whether the afterglow shows evidence of being dust-obscured and/or ``dark'', based on $\beta_{\rm OX}$ or the optical-NIR color.  From a variety of sources including \cite{Perley+2013a}, \cite{Zafar+2012}, \cite{Greiner+2011}.}
\tablenotetext{g}{Whether or not a redshift estimate was promptly publicly available.  (If ``N'', the redshift was measured by late-time observations of the host galaxy or only a limit could be provided.)}
\tablenotetext{h}{Source of the redshift measurement/limit:  observations of the optical afterglow (AG), of the host galaxy (host), or of the X-ray afterglow ($N_{\rm H, X}$).}
\tablenotetext{i}{References for the afterglow position (middle column) and redshift (right column), given below.}
\end{deluxetable*}

{  Sources as follows---  *: {This work},  1:~\citealt{Evans+2009},   2: {Malesani, priv.~comm.},   3:~\citealt{Berger+2005},   4:~\citealt{GCN3114},   5: {Urata, priv.~comm.}  6:~\citealt{Fynbo+2009},   7:~\citealt{GCN3483},   8: {Malesani et al., in prep.}  9:~\citealt{GCN3807},   10:~\citealt{Jakobsson+2006},   11:~\citealt{Butler2007},   12:~\citealt{Hjorth+2012},   13:~\citealt{Kawai+2006},   14:~\citealt{GCN4015},   15:~\citealt{Kruehler+2012},   16:~\citealt{Perley+2013a},   17:~\citealt{GCN4661},   18:~\citealt{GCN4726},   19:~\citealt{GCN4792},   20:~\citealt{Kruehler+2015},   21:~\citealt{GCN5104},   22:~\citealt{GCN5155},   23:~\citealt{Levan+2007},   24:~\citealt{Ruiz-Velasco+2007},   25:~\citealt{GCN5978},   26:~\citealt{Perley+2009a},   27:~\citealt{GCN6221},   28:~\citealt{Jaunsen+2008},   29:~\citealt{GCN6217},   30:~\citealt{GCN6398},   31:~\citealt{GCN6953},   32:~\citealt{GCN6971},   33:~\citealt{GCN6989},   34:~\citealt{Perley+2010},   35: {Swift GRB Table}  36:~\citealt{Svensson+2012},   37:~\citealt{GCN7381},   38:~\citealt{GCN7429},   39:~\citealt{Hashimoto+2010},   40:~\citealt{Hashimoto+2015},   41:~\citealt{GCN7670},   42:~\citealt{GCN7654},   43:~\citealt{GCN8060},   44:~\citealt{GCN8346},   45:~\citealt{GCN8438},   46: {Kr\"uhler, priv.~comm.},   47:~\citealt{GCN8528},   48:~\citealt{GCN8531},   49:~\citealt{GCN8542},   50:~\citealt{Salvaterra+2012},   51:~\citealt{GCN8713},   52:~\citealt{GCN8979},   53:~\citealt{GCN8994},   54:~\citealt{GCN9100},   55:~\citealt{GCN9156},   56:~\citealt{GCN9179},   57:~\citealt{GCN9151},   58:~\citealt{GCN9243},   59:~\citealt{GCN9383},   60:~\citealt{GCN9403},   61:~\citealt{GCN9409},   62:~\citealt{GCN15571},   63:~\citealt{GCN9518},   64:~\citealt{GCN9635},   65:~\citealt{GCN9673},   66:~\citealt{GCN9769},   67:~\citealt{GCN9771},   68:~\citealt{GCN9797},   69:~\citealt{GCN9947},   70:~\citealt{GCN10038},   71:~\citealt{GCN10100},   72:~\citealt{GCN10350},   73:~\citealt{GCN10202},   74:~\citealt{GCN10272},   75:~\citealt{GCN10915},   76:~\citealt{GCN14264},   77:~\citealt{GCN10874},   78:~\citealt{GCN10876},   79:~\citealt{GCN11317},   80:~\citealt{GCN10042},   81:~\citealt{GCN11089},   82:~\citealt{GCN11633},   83:~\citealt{GCN11638},   84:~\citealt{GCN12190},   85:~\citealt{Morgan+2014}}
\clearpage

\begin{deluxetable*}{lllll}  
\tabletypesize{\small}
\tablecaption{Host Galaxy Photometry Used in Photometric Redshifts and Upper Limits}
\tablecolumns{5}
\tablehead{
\colhead{GRB} &
\colhead{Filter} &
\colhead{mag\tablenotemark{a}} & 
\colhead{unc.} &
\colhead{Instrument}}
\startdata
050128  & R    &    24.84 &  0.10  & VLT/FORS1    \\  
050803  & g    & $>$27.74 &        & GTC/OSIRIS   \\
        & R    &    26.32 &  0.22  & VLT/FORS2    \\
        & i    &    26.45 &  0.50  & GTC/OSIRIS   \\
        & F160W &    25.79 &  0.18  & HST/WFC3-IR  \\
        & 3.6  &    25.49 &  0.42  & Spitzer/IRAC \\
        & 4.5  & $>$25.00 &        & Spitzer/IRAC \\
050922B & g    &    27.63 &  0.50  & Keck/LRIS \\
        & R    &    26.44 &  0.22  & Keck/LRIS \\
        & i    &    25.25 &  0.14  & GTC/OSIRIS \\
        & z    &    25.09 &  0.34  & GTC/OSIRIS \\
        & Ks   & $>$22.17 &        & VLT/ISAAC \\
        & 3.6  &    24.59 &  0.22  & Spitzer/IRAC \\
        & 4.5  &    24.80 &  0.42  & Spitzer/IRAC \\
060204B & u    &    25.33 &  0.15  & Keck/LRIS \\
        & B    &    24.93 &  0.06  & Keck/LRIS \\
        & g    &    24.49 &  0.03  & Keck/LRIS \\
        & R    &    24.12 &  0.07  & Keck/LRIS \\
        & i    &    23.86 &  0.07  & Keck/LRIS \\
        & z    &    24.20 &  0.18  & Keck/LRIS \\
        & Y    &    23.22 &  0.16  & Keck/MOSFIRE \\
        & J    &    22.46 &  0.40  & Keck/MOSFIRE \\
        & Ks   &    20.35 &  0.40  & Keck/MOSFIRE \\
        & 3.6  &    22.74 &  0.05  & Spitzer/IRAC \\  
        & 4.5  &    22.00 &  0.30  & Spitzer/IRAC \\
070808  & g    & $>$27.50 &        & Keck/LRIS    \\
        & R    &    26.71 &  0.33  & VLT/FORS2    \\
        & Ks   &    21.77 &  0.37  & VLT/ISAAC    \\
        & 3.6  &    23.57 &  0.10  & Spitzer/IRAC \\  
        & 4.5  &    23.80 &  0.20  & Spitzer/IRAC \\  
080319A & u    &    25.30 &  0.13  & Keck/LRIS \\
        & B    &    25.02 &  0.10  & Keck/LRIS \\
        & g    &    24.50 &  0.04  & Keck/LRIS \\
        & R    &    24.06 &  0.12  & Keck/LRIS \\
        & i    &    24.24 &  0.08  & Keck/LRIS \\
        & z    &    24.26 &  0.12  & Keck/LRIS \\
        & Y    &    23.61 &  0.17  & Keck/MOSFIRE \\
        & J    &    22.60 &  0.12  & Keck/MOSFIRE \\
        & H    &    21.64 &  0.12  & Keck/MOSFIRE \\
        & Ks   &    21.05 &  0.13  & Keck/MOSFIRE \\
        & 3.6  &    22.63 &  0.03  & Spitzer/IRAC \\
        & 4.5  &    22.60 &  0.04  & Spitzer/IRAC \\
081128  & B    &    26.14 &  0.24  & Keck/LRIS \\
        & g    &    25.80 &  0.15  & Keck/LRIS \\
        & V    &    25.37 &  0.13  & Keck/LRIS \\
        & i    &    25.24 &  0.19  & Keck/LRIS \\
        & z    &    24.97 &  0.33  & Keck/LRIS \\
        & 3.6  &    25.20 &  0.31  & Spitzer/IRAC \\
100802A & B    &    26.44 &  0.16  & Keck/LRIS \\
        & R    &    25.23 &  0.25  & Keck/LRIS \\
        & i    &    26.23 &  0.32  & Keck/LRIS \\
        & 3.6  &    25.47 &  0.43  & Spitzer/IRAC \\
120308A & B    & $>$26.67 &        &  Keck/LRIS    \\
        & g    &    26.33 &  0.17  &  Keck/LRIS    \\
        & R    &    25.77 &  0.22  &  Keck/LRIS    \\
        & I    &    24.46 &  0.23  &  Keck/LRIS    \\
        & 3.6  &    24.98 &  0.27  &  Spitzer/IRAC \\
\enddata
\label{tab:grbphot}
\tablenotetext{a}{In the SDSS (for ugriz), Vega (BRIY), or 2MASS (JHKs) magnitude systems; AB magnitudes are used for the IRAC (3.6 and 4.5) filters.  Measurements are uncorrected for foreground extinction.}
\end{deluxetable*}


\begin{thebibliography}{}
\expandafter\ifx\csname natexlab\endcsname\relax\def\natexlab#1{#1}\fi

\bibitem[{{Ahn} {et~al.}(2014){Ahn}, {Alexandroff}, {Allende Prieto}, {Anders},
  {Anderson}, {Anderton}, {Andrews}, {Aubourg}, {Bailey}, {Bastien}, \&
  et~al.}]{DR10}
{Ahn}, C.~P., {Alexandroff}, R., {Allende Prieto}, C., {et~al.} 2014, \apjs,
  211, 17

\bibitem[{{Akerlof} \& {Swan}(2007)}]{Akerlof+2007}
{Akerlof}, C.~W., \& {Swan}, H.~F. 2007, \apj, 671, 1868

\bibitem[{{Band}(2006)}]{Band2006}
{Band}, D.~L. 2006, \apj, 644, 378

\bibitem[{{Barthelmy} {et~al.}(2005){Barthelmy}, {Barbier}, {Cummings},
  {Fenimore}, {Gehrels}, {Hullinger}, {Krimm}, {Markwardt}, {Palmer},
  {Parsons}, {Sato}, {Suzuki}, {Takahashi}, {Tashiro}, \&
  {Tueller}}]{Barthelmy+2005}
{Barthelmy}, S.~D., {Barbier}, L.~M., {Cummings}, J.~R., {et~al.} 2005, \ssr,
  120, 143

\bibitem[{{Behroozi} {et~al.}(2013){Behroozi}, {Wechsler}, \&
  {Conroy}}]{Behroozi+2013}
{Behroozi}, P.~S., {Wechsler}, R.~H., \& {Conroy}, C. 2013, \apj, 770, 57

\bibitem[{{Berger}(2014)}]{Berger+2014}
{Berger}, E. 2014, \araa, 52, 43

\bibitem[{{Berger} {et~al.}(2003){Berger}, {Cowie}, {Kulkarni}, {Frail},
  {Aussel}, \& {Barger}}]{Berger+2003}
{Berger}, E., {Cowie}, L.~L., {Kulkarni}, S.~R., {et~al.} 2003, \apj, 588, 99

\bibitem[{{Berger} {et~al.}(2008){Berger}, {Foley}, {Simcoe}, \&
  {Irwin}}]{GCN8434}
{Berger}, E., {Foley}, R., {Simcoe}, R., \& {Irwin}, J. 2008, GRB Coordinates
  Network, 8434

\bibitem[{{Berger} \& {Fox}(2009)}]{GCN9156}
{Berger}, E., \& {Fox}, D.~B. 2009, GRB Coordinates Network, 9156

\bibitem[{{Berger} \& {Rauch}(2008)}]{GCN8542}
{Berger}, E., \& {Rauch}, M. 2008, GRB Coordinates Network, 8542

\bibitem[{{Berger} {et~al.}(2002){Berger}, {Kulkarni}, {Bloom}, {Price}, {Fox},
  {Frail}, {Axelrod}, {Chevalier}, {Colbert}, {Costa}, {Djorgovski},
  {Frontera}, {Galama}, {Halpern}, {Harrison}, {Holtzman}, {Hurley}, {Kimble},
  {McCarthy}, {Piro}, {Reichart}, {Ricker}, {Sari}, {Schmidt}, {Wheeler},
  {Vanderppek}, \& {Yost}}]{Berger+2002}
{Berger}, E., {Kulkarni}, S.~R., {Bloom}, J.~S., {et~al.} 2002, \apj, 581, 981

\bibitem[{{Berger} {et~al.}(2005){Berger}, {Kulkarni}, {Fox}, {Soderberg},
  {Harrison}, {Nakar}, {Kelson}, {Gladders}, {Mulchaey}, {Oemler}, {Dressler},
  {Cenko}, {Price}, {Schmidt}, {Frail}, {Morrell}, {Gonzalez}, {Krzeminski},
  {Sari}, {Gal-Yam}, {Moon}, {Penprase}, {Jayawardhana}, {Scholz}, {Rich},
  {Peterson}, {Anderson}, {McNaught}, {Minezaki}, {Yoshii}, {Cowie}, \&
  {Pimbblet}}]{Berger+2005}
{Berger}, E., {Kulkarni}, S.~R., {Fox}, D.~B., {et~al.} 2005, \apj, 634, 501

\bibitem[{{Blain} \& {Natarajan}(2000)}]{Blain+2000}
{Blain}, A.~W., \& {Natarajan}, P. 2000, \mnras, 312, L35

\bibitem[{{Bloom}(2007)}]{GCN6989}
{Bloom}, J.~S. 2007, GRB Coordinates Network, 6989

\bibitem[{{Bloom} {et~al.}(2001){Bloom}, {Frail}, \& {Sari}}]{Bloom+2001}
{Bloom}, J.~S., {Frail}, D.~A., \& {Sari}, R. 2001, \aj, 121, 2879

\bibitem[{{Bloom} {et~al.}(2007){Bloom}, {Perley}, \& {Starr}}]{GCN6953}
{Bloom}, J.~S., {Perley}, D.~A., \& {Starr}, D.~L. 2007, GRB Coordinates
  Network, 6953

\bibitem[{{Bloom} {et~al.}(2009){Bloom}, {Perley}, {Li}, {Butler}, {Miller},
  {Kocevski}, {Kann}, {Foley}, {Chen}, {Filippenko}, {Starr}, {Macomber},
  {Prochaska}, {Chornock}, {Poznanski}, {Klose}, {Skrutskie}, {Lopez}, {Hall},
  {Glazebrook}, \& {Blake}}]{Bloom+2009}
{Bloom}, J.~S., {Perley}, D.~A., {Li}, W., {et~al.} 2009, \apj, 691, 723

\bibitem[{{Boissier} {et~al.}(2013){Boissier}, {Salvaterra}, {Le Floc'h},
  {Basa}, {Buat}, {Prantzos}, {Vergani}, \& {Savaglio}}]{Boissier+2013}
{Boissier}, S., {Salvaterra}, R., {Le Floc'h}, E., {et~al.} 2013, \aap, 557,
  A34

\bibitem[{{Bouwens} {et~al.}(2014){Bouwens}, {Bradley}, {Zitrin}, {Coe},
  {Franx}, {Zheng}, {Smit}, {Host}, {Postman}, {Moustakas}, {Labb{\'e}},
  {Carrasco}, {Molino}, {Donahue}, {Kelson}, {Meneghetti}, {Ben{\'{\i}}tez},
  {Lemze}, {Umetsu}, {Broadhurst}, {Moustakas}, {Rosati}, {Jouvel},
  {Bartelmann}, {Ford}, {Graves}, {Grillo}, {Infante}, {Jimenez-Teja}, {Lahav},
  {Maoz}, {Medezinski}, {Melchior}, {Merten}, {Nonino}, {Ogaz}, \&
  {Seitz}}]{Bouwens+2014}
{Bouwens}, R.~J., {Bradley}, L., {Zitrin}, A., {et~al.} 2014, \apj, 795, 126

\bibitem[{{Brammer} {et~al.}(2008){Brammer}, {van Dokkum}, \&
  {Coppi}}]{Brammer+2008}
{Brammer}, G.~B., {van Dokkum}, P.~G., \& {Coppi}, P. 2008, \apj, 686, 1503

\bibitem[{{Bromberg} {et~al.}(2011){Bromberg}, {Nakar}, \&
  {Piran}}]{Bromberg+2011}
{Bromberg}, O., {Nakar}, E., \& {Piran}, T. 2011, \apjl, 739, L55

\bibitem[{{Bromm} \& {Loeb}(2006)}]{Bromm+2006}
{Bromm}, V., \& {Loeb}, A. 2006, \apj, 642, 382

\bibitem[{{Burrows} {et~al.}(2005){Burrows}, {Hill}, {Nousek}, {Kennea},
  {Wells}, {Osborne}, {Abbey}, {Beardmore}, {Mukerjee}, {Short}, {Chincarini},
  {Campana}, {Citterio}, {Moretti}, {Pagani}, {Tagliaferri}, {Giommi},
  {Capalbi}, {Tamburelli}, {Angelini}, {Cusumano}, {Br{\"a}uninger}, {Burkert},
  \& {Hartner}}]{Burrows+2005}
{Burrows}, D.~N., {Hill}, J.~E., {Nousek}, J.~A., {et~al.} 2005, \ssr, 120, 165

\bibitem[{{Burrows} {et~al.}(2007){Burrows}, {Kennea}, {Abbey}, {Beardmore},
  {Campana}, {Capalbi}, {Chincarini}, {Cusumano}, {Evans}, {Hill}, {Giommi},
  {Goad}, {Godet}, {Moretti}, {Morris}, {Osborne}, {Pagani}, {Page}, {Perri},
  {Racusin}, {Romano}, {Starling}, {Tagliaferri}, {Tamburelli}, {Tyler}, \&
  {Willingale}}]{Burrows+2007}
{Burrows}, D.~N., {Kennea}, J.~A., {Abbey}, A.~F., {et~al.} 2007, in Society of
  Photo-Optical Instrumentation Engineers (SPIE) Conference Series, Vol. 6686,
  Society of Photo-Optical Instrumentation Engineers (SPIE) Conference Series,
  7

\bibitem[{{Butler}(2007)}]{Butler2007}
{Butler}, N.~R. 2007, \aj, 133, 1027

\bibitem[{{Butler} {et~al.}(2010{\natexlab{a}}){Butler}, {Bloom}, \&
  {Poznanski}}]{Butler+2010}
{Butler}, N.~R., {Bloom}, J.~S., \& {Poznanski}, D. 2010{\natexlab{a}}, \apj,
  711, 495

\bibitem[{{Butler} \& {Kocevski}(2007)}]{Butler+2007}
{Butler}, N.~R., \& {Kocevski}, D. 2007, \apj, 663, 407

\bibitem[{{Butler} {et~al.}(2010{\natexlab{b}}){Butler}, {Perley}, {Cenko},
  {Bloom}, {Levan}, \& {Tanvir}}]{GCN10915}
{Butler}, N.~R., {Perley}, D.~A., {Cenko}, S.~B., {et~al.} 2010{\natexlab{b}},
  GRB Coordinates Network, 10915

\bibitem[{{Cameron}(2011)}]{Cameron2011}
{Cameron}, E. 2011, PASA, 28, 128

\bibitem[{{Campisi} {et~al.}(2011{\natexlab{a}}){Campisi}, {Maio},
  {Salvaterra}, \& {Ciardi}}]{Campisi+2011b}
{Campisi}, M.~A., {Maio}, U., {Salvaterra}, R., \& {Ciardi}, B.
  2011{\natexlab{a}}, \mnras, 416, 2760

\bibitem[{{Campisi} {et~al.}(2011{\natexlab{b}}){Campisi}, {Tapparello},
  {Salvaterra}, {Mannucci}, \& {Colpi}}]{Campisi+2011a}
{Campisi}, M.~A., {Tapparello}, C., {Salvaterra}, R., {Mannucci}, F., \&
  {Colpi}, M. 2011{\natexlab{b}}, \mnras, 417, 1013

\bibitem[{{Castro-Tirado} {et~al.}(2007){Castro-Tirado}, {de Ugarte Postigo},
  {Jelinek}, {Gorosabel}, {Marin Fernandez de Capel}, {Abia}, {Perez-Ramirez},
  {Malesani}, {Guziy}, \& {Oreiro}}]{GCN6971}
{Castro-Tirado}, A.~J., {de Ugarte Postigo}, A., {Jelinek}, M., {et~al.} 2007,
  GRB Coordinates Network, 6971

\bibitem[{{Castro-Tirado} {et~al.}(2009){Castro-Tirado}, {Bremer}, {Winters},
  {Gorosabel}, {Guziy}, {Jelinek}, {de Ugarte Postigo}, {Santiago}, \&
  {Perez-Ramirez}}]{GCN9100}
{Castro-Tirado}, A.~J., {Bremer}, M., {Winters}, J.-M., {et~al.} 2009, GRB
  Coordinates Network, 9100

\bibitem[{{Cenko}(2005)}]{GCN3807}
{Cenko}, S.~B. 2005, GRB Coordinates Network, 3807

\bibitem[{{Cenko}(2008)}]{GCN7429}
---. 2008, GRB Coordinates Network, 7429

\bibitem[{{Cenko}(2009)}]{GCN9769}
---. 2009, GRB Coordinates Network, 9769

\bibitem[{{Cenko}(2010)}]{GCN11040}
---. 2010, GRB Coordinates Network, 11040

\bibitem[{{Cenko} {et~al.}(2006{\natexlab{a}}){Cenko}, {Berger}, {Djorgovski},
  {Mahabal}, \& {Fox}}]{GCN5155}
{Cenko}, S.~B., {Berger}, E., {Djorgovski}, S.~G., {Mahabal}, A.~A., \& {Fox},
  D.~B. 2006{\natexlab{a}}, GRB Coordinates Network, 5155

\bibitem[{{Cenko} \& {Fox}(2006)}]{GCN5978}
{Cenko}, S.~B., \& {Fox}, D.~B. 2006, GRB Coordinates Network, 5978

\bibitem[{{Cenko} {et~al.}(2011){Cenko}, {Hora}, \& {Bloom}}]{GCN11638}
{Cenko}, S.~B., {Hora}, J.~L., \& {Bloom}, J.~S. 2011, GRB Coordinates Network,
  11638

\bibitem[{{Cenko} {et~al.}(2009{\natexlab{a}}){Cenko}, {Perley}, {Junkkarinen},
  {Burbidge}, {Diego}, \& {Miller}}]{GCN9518}
{Cenko}, S.~B., {Perley}, D.~A., {Junkkarinen}, V., {et~al.}
  2009{\natexlab{a}}, GRB Coordinates Network, 9518

\bibitem[{{Cenko} {et~al.}(2006{\natexlab{b}}){Cenko}, {Fox}, {Moon},
  {Harrison}, {Kulkarni}, {Henning}, {Guzman}, {Bonati}, {Smith}, {Thicksten},
  {Doyle}, {Petrie}, {Gal-Yam}, {Soderberg}, {Anagnostou}, \&
  {Laity}}]{Cenko+2006}
{Cenko}, S.~B., {Fox}, D.~B., {Moon}, D.-S., {et~al.} 2006{\natexlab{b}},
  \pasp, 118, 1396

\bibitem[{{Cenko} {et~al.}(2009{\natexlab{b}}){Cenko}, {Kelemen}, {Harrison},
  {Fox}, {Kulkarni}, {Kasliwal}, {Ofek}, {Rau}, {Gal-Yam}, {Frail}, \&
  {Moon}}]{Cenko+2009}
{Cenko}, S.~B., {Kelemen}, J., {Harrison}, F.~A., {et~al.} 2009{\natexlab{b}},
  \apj, 693, 1484

\bibitem[{{Cenko} {et~al.}(2010){Cenko}, {Butler}, {Ofek}, {Perley}, {Morgan},
  {Frail}, {Gorosabel}, {Bloom}, {Castro-Tirado}, {Cepa}, {Chandra}, {de Ugarte
  Postigo}, {Filippenko}, {Klein}, {Kulkarni}, {Miller}, {Nugent}, \&
  {Starr}}]{Cenko+2010}
{Cenko}, S.~B., {Butler}, N.~R., {Ofek}, E.~O., {et~al.} 2010, \aj, 140, 224

\bibitem[{{Chen} {et~al.}(2009){Chen}, {Helsby}, {Shectman}, {Thompson}, \&
  {Crane}}]{GCN10038}
{Chen}, H.-W., {Helsby}, J., {Shectman}, S., {Thompson}, I., \& {Crane}, J.
  2009, GRB Coordinates Network, 10038

\bibitem[{{Chen} {et~al.}(2007){Chen}, {Prochaska}, {Herbert-Fort},
  {Christlein}, \& {Cortes}}]{GCN6217}
{Chen}, H.-W., {Prochaska}, J.~X., {Herbert-Fort}, S., {Christlein}, D., \&
  {Cortes}, S. 2007, GRB Coordinates Network, 6217

\bibitem[{{Chornock} {et~al.}(2009{\natexlab{a}}){Chornock}, {Cenko},
  {Griffith}, {Kislak}, {Kleiser}, \& {Filippenko}}]{GCN9151}
{Chornock}, R., {Cenko}, S.~B., {Griffith}, C.~V., {et~al.} 2009{\natexlab{a}},
  GRB Coordinates Network, 9151

\bibitem[{{Chornock} {et~al.}(2008){Chornock}, {Foley}, {Li}, \&
  {Filippenko}}]{GCN7381}
{Chornock}, R., {Foley}, R.~J., {Li}, W., \& {Filippenko}, A.~V. 2008, GRB
  Coordinates Network, 7381

\bibitem[{{Chornock} {et~al.}(2009{\natexlab{b}}){Chornock}, {Li}, \&
  {Filippenko}}]{GCN8979}
{Chornock}, R., {Li}, W., \& {Filippenko}, A.~V. 2009{\natexlab{b}}, GRB
  Coordinates Network, 8979

\bibitem[{{Chornock} {et~al.}(2009{\natexlab{c}}){Chornock}, {Perley}, {Cenko},
  \& {Bloom}}]{GCN9243}
{Chornock}, R., {Perley}, D.~A., {Cenko}, S.~B., \& {Bloom}, J.~S.
  2009{\natexlab{c}}, GRB Coordinates Network, 9243

\bibitem[{{Chornock} {et~al.}(2009{\natexlab{d}}){Chornock}, {Perley}, {Cenko},
  {Bloom}, {Cobb}, \& {Prochaska}}]{GCN8994}
{Chornock}, R., {Perley}, D.~A., {Cenko}, S.~B., {et~al.} 2009{\natexlab{d}},
  GRB Coordinates Network, 8994

\bibitem[{{Chornock} {et~al.}(2009{\natexlab{e}}){Chornock}, {Perley}, \&
  {Cobb}}]{GCN10100}
{Chornock}, R., {Perley}, D.~A., \& {Cobb}, B.~E. 2009{\natexlab{e}}, GRB
  Coordinates Network, 10100

\bibitem[{{Covino} {et~al.}(2013){Covino}, {Melandri}, {Salvaterra}, {Campana},
  {Vergani}, {Bernardini}, {D'Avanzo}, {D'Elia}, {Fugazza}, {Ghirlanda},
  {Ghisellini}, {Gomboc}, {Jin}, {Kr{\"u}hler}, {Malesani}, {Nava},
  {Sbarufatti}, \& {Tagliaferri}}]{Covino+2013}
{Covino}, S., {Melandri}, A., {Salvaterra}, R., {et~al.} 2013, \mnras, 432,
  1231

\bibitem[{{Coward} {et~al.}(2013){Coward}, {Howell}, {Branchesi}, {Stratta},
  {Guetta}, {Gendre}, \& {Macpherson}}]{Coward+2013}
{Coward}, D.~M., {Howell}, E.~J., {Branchesi}, M., {et~al.} 2013, \mnras, 432,
  2141

\bibitem[{{Cucchiara}(2010)}]{GCN10473}
{Cucchiara}, A. 2010, GRB Coordinates Network, 10473

\bibitem[{{Cucchiara} {et~al.}(2009){Cucchiara}, {Fox}, {Levan}, \&
  {Tanvir}}]{GCN10202}
{Cucchiara}, A., {Fox}, D., {Levan}, A., \& {Tanvir}, N. 2009, GRB Coordinates
  Network, 10202

\bibitem[{{Cucchiara} \& {Fox}(2008)}]{GCN7654}
{Cucchiara}, A., \& {Fox}, D.~B. 2008, GRB Coordinates Network, 7654

\bibitem[{{Cucchiara} {et~al.}(2008{\natexlab{a}}){Cucchiara}, {Fox}, {Cenko},
  \& {Berger}}]{GCN8346}
{Cucchiara}, A., {Fox}, D.~B., {Cenko}, S.~B., \& {Berger}, E.
  2008{\natexlab{a}}, GRB Coordinates Network, 8346

\bibitem[{{Cucchiara} {et~al.}(2008{\natexlab{b}}){Cucchiara}, {Fox}, {Cenko},
  \& {Berger}}]{GCN8713}
---. 2008{\natexlab{b}}, GRB Coordinates Network, 8713

\bibitem[{{Cucchiara} {et~al.}(2015){Cucchiara}, {Fumagalli}, {Rafelski},
  {Kocevski}, {Prochaska}, {Cooke}, \& {Becker}}]{Cucchiara+2015}
{Cucchiara}, A., {Fumagalli}, M., {Rafelski}, M., {et~al.} 2015, \apj, 804, 51

\bibitem[{{Cucchiara} {et~al.}(2011){Cucchiara}, {Levan}, {Fox}, {Tanvir},
  {Ukwatta}, {Berger}, {Kr{\"u}hler}, {K{\"u}pc{\"u} Yolda{\c s}}, {Wu},
  {Toma}, {Greiner}, {Olivares}, {Rowlinson}, {Amati}, {Sakamoto}, {Roth},
  {Stephens}, {Fritz}, {Fynbo}, {Hjorth}, {Malesani}, {Jakobsson}, {Wiersema},
  {O'Brien}, {Soderberg}, {Foley}, {Fruchter}, {Rhoads}, {Rutledge}, {Schmidt},
  {Dopita}, {Podsiadlowski}, {Willingale}, {Wolf}, {Kulkarni}, \&
  {D'Avanzo}}]{Cucchiara+2011}
{Cucchiara}, A., {Levan}, A.~J., {Fox}, D.~B., {et~al.} 2011, \apj, 736, 7

\bibitem[{{Daigne} {et~al.}(2006){Daigne}, {Rossi}, \&
  {Mochkovitch}}]{Daigne+2006}
{Daigne}, F., {Rossi}, E.~M., \& {Mochkovitch}, R. 2006, \mnras, 372, 1034

\bibitem[{{D'Avanzo} {et~al.}(2008){D'Avanzo}, {D'Elia}, {Covino},
  {Piranomonte}, {Salvaterra}, {Thone}, \& {Chincarini}}]{GCN8528}
{D'Avanzo}, P., {D'Elia}, V., {Covino}, S., {et~al.} 2008, GRB Coordinates
  Network, 8528

\bibitem[{{Dav{\'e}}(2008)}]{Dave2008}
{Dav{\'e}}, R. 2008, \mnras, 385, 147

\bibitem[{{De Cia} {et~al.}(2013){De Cia}, {Ledoux}, {Savaglio}, {Schady}, \&
  {Vreeswijk}}]{DeCia+2013}
{De Cia}, A., {Ledoux}, C., {Savaglio}, S., {Schady}, P., \& {Vreeswijk}, P.~M.
  2013, \aap, 560, A88

\bibitem[{{de Ugarte Postigo} {et~al.}(2009{\natexlab{a}}){de Ugarte Postigo},
  {Gorosabel}, {Fynbo}, {Wiersema}, \& {Tanvir}}]{GCN9771}
{de Ugarte Postigo}, A., {Gorosabel}, J., {Fynbo}, J.~P.~U., {Wiersema}, K., \&
  {Tanvir}, N. 2009{\natexlab{a}}, GRB Coordinates Network, 9771

\bibitem[{{de Ugarte Postigo} {et~al.}(2009{\natexlab{b}}){de Ugarte Postigo},
  {Gorosabel}, {Malesani}, {Fynbo}, \& {Levan}}]{GCN9383}
{de Ugarte Postigo}, A., {Gorosabel}, J., {Malesani}, D., {Fynbo}, J.~P.~U., \&
  {Levan}, A.~J. 2009{\natexlab{b}}, GRB Coordinates Network, 9383

\bibitem[{{de Ugarte Postigo} {et~al.}(2009{\natexlab{c}}){de Ugarte Postigo},
  {Jakobsson}, {Malesani}, {Fynbo}, {Simpson}, \& {Barros}}]{GCN8766}
{de Ugarte Postigo}, A., {Jakobsson}, P., {Malesani}, D., {et~al.}
  2009{\natexlab{c}}, GRB Coordinates Network, 8766

\bibitem[{{de Ugarte Postigo} {et~al.}(2009{\natexlab{d}}){de Ugarte Postigo},
  {Goldoni}, {Th{\"o}ne}, {Tanvir}, {Jakobsson}, {Vergani}, {Flores},
  {Malesani}, {Levan}, \& {Fynbo}}]{GCN10042}
{de Ugarte Postigo}, A., {Goldoni}, P., {Th{\"o}ne}, C.~C., {et~al.}
  2009{\natexlab{d}}, GRB Coordinates Network, 10042

\bibitem[{{D'Elia}(2011)}]{Delia2011}
{D'Elia}, V. 2011, Astronomische Nachrichten, 332, 272

\bibitem[{{D'Elia} {et~al.}(2008{\natexlab{a}}){D'Elia}, {Covino}, \&
  {D'Avanzo}}]{GCN8438}
{D'Elia}, V., {Covino}, S., \& {D'Avanzo}, P. 2008{\natexlab{a}}, GRB
  Coordinates Network, 8438

\bibitem[{{D'Elia} {et~al.}(2008{\natexlab{b}}){D'Elia}, {Th{\"o}ne}, {de
  Ugarte Postigo}, {D'Avanzo}, {Covino}, {Piranomonte}, {Salvaterra}, \&
  {Chincarini}}]{GCN8531}
{D'Elia}, V., {Th{\"o}ne}, C.~C., {de Ugarte Postigo}, A., {et~al.}
  2008{\natexlab{b}}, GRB Coordinates Network, 8531

\bibitem[{{Djorgovski} {et~al.}(2001){Djorgovski}, {Frail}, {Kulkarni},
  {Bloom}, {Odewahn}, \& {Diercks}}]{Djorgovski+2001}
{Djorgovski}, S.~G., {Frail}, D.~A., {Kulkarni}, S.~R., {et~al.} 2001, \apj,
  562, 654

\bibitem[{{Dressler} {et~al.}(2011){Dressler}, {Bigelow}, {Hare}, {Sutin},
  {Thompson}, {Burley}, {Epps}, {Oemler}, {Bagish}, {Birk}, {Clardy},
  {Gunnels}, {Kelson}, {Shectman}, \& {Osip}}]{Dressler+2011}
{Dressler}, A., {Bigelow}, B., {Hare}, T., {et~al.} 2011, \pasp, 123, 288

\bibitem[{{El{\'{\i}}asd{\'o}ttir} {et~al.}(2009){El{\'{\i}}asd{\'o}ttir},
  {Fynbo}, {Hjorth}, {Ledoux}, {Watson}, {Andersen}, {Malesani}, {Vreeswijk},
  {Prochaska}, {Sollerman}, \& {Jaunsen}}]{Eliasdottir+2009}
{El{\'{\i}}asd{\'o}ttir}, {\'A}., {Fynbo}, J.~P.~U., {Hjorth}, J., {et~al.}
  2009, \apj, 697, 1725

\bibitem[{{Elliott} {et~al.}(2012){Elliott}, {Greiner}, {Khochfar}, {Schady},
  {Johnson}, \& {Rau}}]{Elliott+2012}
{Elliott}, J., {Greiner}, J., {Khochfar}, S., {et~al.} 2012, \aap, 539, A113

\bibitem[{{Evans} {et~al.}(2009){Evans}, {Beardmore}, {Page}, {Osborne},
  {O'Brien}, {Willingale}, {Starling}, {Burrows}, {Godet}, {Vetere}, {Racusin},
  {Goad}, {Wiersema}, {Angelini}, {Capalbi}, {Chincarini}, {Gehrels}, {Kennea},
  {Margutti}, {Morris}, {Mountford}, {Pagani}, {Perri}, {Romano}, \&
  {Tanvir}}]{Evans+2009}
{Evans}, P.~A., {Beardmore}, A.~P., {Page}, K.~L., {et~al.} 2009, \mnras, 397,
  1177

\bibitem[{{Ferrero} {et~al.}(2009){Ferrero}, {Klose}, {Kann}, {Savaglio},
  {Schulze}, {Palazzi}, {Maiorano}, {B{\"o}hm}, {Grupe}, {Oates},
  {S{\'a}nchez}, {Amati}, {Greiner}, {Hjorth}, {Malesani}, {Barthelmy},
  {Gorosabel}, {Masetti}, \& {Roth}}]{Ferrero+2009}
{Ferrero}, P., {Klose}, S., {Kann}, D.~A., {et~al.} 2009, \aap, 497, 729

\bibitem[{{Flores} {et~al.}(2010){Flores}, {Fynbo}, {de Ugarte Postigo},
  {Milvang-Jensen}, {Malesani}, {Goldoni}, {Th{\"o}ne}, {Piranomonte}, \&
  {Vergani}}]{GCN11317}
{Flores}, H., {Fynbo}, J.~P.~U., {de Ugarte Postigo}, A., {et~al.} 2010, GRB
  Coordinates Network, 11317

\bibitem[{{Foley} {et~al.}(2005){Foley}, {Chen}, {Bloom}, \&
  {Prochaska}}]{GCN3483}
{Foley}, R.~J., {Chen}, H.-W., {Bloom}, J., \& {Prochaska}, J.~X. 2005, GRB
  Coordinates Network, 3483

\bibitem[{{Fong} {et~al.}(2013){Fong}, {Berger}, {Chornock}, {Margutti},
  {Levan}, {Tanvir}, {Tunnicliffe}, {Czekala}, {Fox}, {Perley}, {Cenko},
  {Zauderer}, {Laskar}, {Persson}, {Monson}, {Kelson}, {Birk}, {Murphy},
  {Servillat}, \& {Anglada}}]{Fong+2013a}
{Fong}, W., {Berger}, E., {Chornock}, R., {et~al.} 2013, \apj, 769, 56

\bibitem[{{Friis} {et~al.}(2015){Friis}, {De Cia}, {Kr{\"u}hler}, {Fynbo},
  {Ledoux}, {Vreeswijk}, {Watson}, {Malesani}, {Gorosabel}, {Starling},
  {Jakobsson}, {Varela}, {Wiersema}, {Drachmann}, {Trotter}, {Th{\"o}ne}, {de
  Ugarte Postigo}, {D'Elia}, {Elliott}, {Maturi}, {Goldoni}, {Greiner},
  {Haislip}, {Kaper}, {Knust}, {LaCluyze}, {Milvang-Jensen}, {Reichart},
  {Schulze}, {Sudilovsky}, {Tanvir}, \& {Vergani}}]{Friis+2015}
{Friis}, M., {De Cia}, A., {Kr{\"u}hler}, T., {et~al.} 2015, \mnras, 451, 167

\bibitem[{{Fruchter} {et~al.}(2006){Fruchter}, {Levan}, {Strolger},
  {Vreeswijk}, {Thorsett}, {Bersier}, {Burud}, {Castro Cer{\'o}n},
  {Castro-Tirado}, {Conselice}, {Dahlen}, {Ferguson}, {Fynbo}, {Garnavich},
  {Gibbons}, {Gorosabel}, {Gull}, {Hjorth}, {Holland}, {Kouveliotou}, {Levay},
  {Livio}, {Metzger}, {Nugent}, {Petro}, {Pian}, {Rhoads}, {Riess}, {Sahu},
  {Smette}, {Tanvir}, {Wijers}, \& {Woosley}}]{Fruchter+2006}
{Fruchter}, A.~S., {Levan}, A.~J., {Strolger}, L., {et~al.} 2006, \nat, 441,
  463

\bibitem[{{Fryer} \& {Heger}(2005)}]{Fryer+2005}
{Fryer}, C.~L., \& {Heger}, A. 2005, \apj, 623, 302

\bibitem[{{Fynbo} {et~al.}(2009{\natexlab{a}}){Fynbo}, {Malesani}, {Jakobsson},
  \& {D'Elia}}]{GCN9947}
{Fynbo}, J.~P.~U., {Malesani}, D., {Jakobsson}, P., \& {D'Elia}, V.
  2009{\natexlab{a}}, GRB Coordinates Network, 9947

\bibitem[{{Fynbo} {et~al.}(2006){Fynbo}, {Watson}, {Th{\"o}ne}, {Sollerman},
  {Bloom}, {Davis}, {Hjorth}, {Jakobsson}, {J{\o}rgensen}, {Graham},
  {Fruchter}, {Bersier}, {Kewley}, {Cassan}, {Castro Cer{\'o}n}, {Foley},
  {Gorosabel}, {Hinse}, {Horne}, {Jensen}, {Klose}, {Kocevski}, {Marquette},
  {Perley}, {Ramirez-Ruiz}, {Stritzinger}, {Vreeswijk}, {Wijers}, {Woller},
  {Xu}, \& {Zub}}]{Fynbo+2006b}
{Fynbo}, J.~P.~U., {Watson}, D., {Th{\"o}ne}, C.~C., {et~al.} 2006, \nat, 444,
  1047

\bibitem[{{Fynbo} {et~al.}(2009{\natexlab{b}}){Fynbo}, {Jakobsson},
  {Prochaska}, {Malesani}, {Ledoux}, {de Ugarte Postigo}, {Nardini},
  {Vreeswijk}, {Wiersema}, {Hjorth}, {Sollerman}, {Chen}, {Th{\"o}ne},
  {Bj{\"o}rnsson}, {Bloom}, {Castro-Tirado}, {Christensen}, {De Cia},
  {Fruchter}, {Gorosabel}, {Graham}, {Jaunsen}, {Jensen}, {Kann},
  {Kouveliotou}, {Levan}, {Maund}, {Masetti}, {Milvang-Jensen}, {Palazzi},
  {Perley}, {Pian}, {Rol}, {Schady}, {Starling}, {Tanvir}, {Watson}, {Xu},
  {Augusteijn}, {Grundahl}, {Telting}, \& {Quirion}}]{Fynbo+2009}
{Fynbo}, J.~P.~U., {Jakobsson}, P., {Prochaska}, J.~X., {et~al.}
  2009{\natexlab{b}}, \apjs, 185, 526

\bibitem[{{Fynbo} {et~al.}(2001){Fynbo}, {Jensen}, {Gorosabel}, {Hjorth},
  {Pedersen}, {M{\o}ller}, {Abbott}, {Castro-Tirado}, {Delgado}, {Greiner},
  {Henden}, {Magazz{\`u}}, {Masetti}, {Merlino}, {Masegosa}, {{\O}stensen},
  {Palazzi}, {Pian}, {Schwarz}, {Cline}, {Guidorzi}, {Goldsten}, {Hurley},
  {Mazets}, {McClanahan}, {Montanari}, {Starr}, \& {Trombka}}]{Fynbo+2001}
{Fynbo}, J.~U., {Jensen}, B.~L., {Gorosabel}, J., {et~al.} 2001, \aap, 369, 373

\bibitem[{{Gal-Yam} {et~al.}(2006){Gal-Yam}, {Fox}, {Price}, {Ofek}, {Davis},
  {Leonard}, {Soderberg}, {Schmidt}, {Lewis}, {Peterson}, {Kulkarni}, {Berger},
  {Cenko}, {Sari}, {Sharon}, {Frail}, {Moon}, {Brown}, {Cucchiara}, {Harrison},
  {Piran}, {Persson}, {McCarthy}, {Penprase}, {Chevalier}, \&
  {MacFadyen}}]{GalYam+2006}
{Gal-Yam}, A., {Fox}, D.~B., {Price}, P.~A., {et~al.} 2006, \nat, 444, 1053

\bibitem[{{Gehrels} {et~al.}(2004){Gehrels}, {Chincarini}, {Giommi}, {Mason},
  {Nousek}, {Wells}, {White}, {Barthelmy}, {Burrows}, {Cominsky}, {Hurley},
  {Marshall}, {M{\'e}sz{\'a}ros}, {Roming}, {Angelini}, {Barbier}, {Belloni},
  {Campana}, {Caraveo}, {Chester}, {Citterio}, {Cline}, {Cropper}, {Cummings},
  {Dean}, {Feigelson}, {Fenimore}, {Frail}, {Fruchter}, {Garmire}, {Gendreau},
  {Ghisellini}, {Greiner}, {Hill}, {Hunsberger}, {Krimm}, {Kulkarni}, {Kumar},
  {Lebrun}, {Lloyd-Ronning}, {Markwardt}, {Mattson}, {Mushotzky}, {Norris},
  {Osborne}, {Paczynski}, {Palmer}, {Park}, {Parsons}, {Paul}, {Rees},
  {Reynolds}, {Rhoads}, {Sasseen}, {Schaefer}, {Short}, {Smale}, {Smith},
  {Stella}, {Tagliaferri}, {Takahashi}, {Tashiro}, {Townsley}, {Tueller},
  {Turner}, {Vietri}, {Voges}, {Ward}, {Willingale}, {Zerbi}, \&
  {Zhang}}]{Gehrels+2004}
{Gehrels}, N., {Chincarini}, G., {Giommi}, P., {et~al.} 2004, \apj, 611, 1005

\bibitem[{{Gehrels} {et~al.}(2006){Gehrels}, {Norris}, {Barthelmy}, {Granot},
  {Kaneko}, {Kouveliotou}, {Markwardt}, {M{\'e}sz{\'a}ros}, {Nakar}, {Nousek},
  {O'Brien}, {Page}, {Palmer}, {Parsons}, {Roming}, {Sakamoto}, {Sarazin},
  {Schady}, {Stamatikos}, \& {Woosley}}]{Gehrels+2006}
{Gehrels}, N., {Norris}, J.~P., {Barthelmy}, S.~D., {et~al.} 2006, \nat, 444,
  1044

\bibitem[{{Gehrels} {et~al.}(2008){Gehrels}, {Barthelmy}, {Burrows},
  {Cannizzo}, {Chincarini}, {Fenimore}, {Kouveliotou}, {O'Brien}, {Palmer},
  {Racusin}, {Roming}, {Sakamoto}, {Tueller}, {Wijers}, \&
  {Zhang}}]{Gehrels+2008}
{Gehrels}, N., {Barthelmy}, S.~D., {Burrows}, D.~N., {et~al.} 2008, \apj, 689,
  1161

\bibitem[{{Goad} {et~al.}(2007){Goad}, {Tyler}, {Beardmore}, {Evans}, {Rosen},
  {Osborne}, {Starling}, {Marshall}, {Yershov}, {Burrows}, {Gehrels}, {Roming},
  {Moretti}, {Capalbi}, {Hill}, {Kennea}, {Koch}, \& {vanden Berk}}]{Goad+2007}
{Goad}, M.~R., {Tyler}, L.~G., {Beardmore}, A.~P., {et~al.} 2007, \aap, 476,
  1401

\bibitem[{{Goldoni} {et~al.}(2013){Goldoni}, {de Ugarte Postigo}, \&
  {Fynbo}}]{GCN15571}
{Goldoni}, P., {de Ugarte Postigo}, A., \& {Fynbo}, J.~P.~U. 2013, GRB
  Coordinates Network, 15571

\bibitem[{{Goldoni} {et~al.}(2006){Goldoni}, {Royer}, {Fran{\c c}ois},
  {Horrobin}, {Blanc}, {Vernet}, {Modigliani}, \& {Larsen}}]{Goldoni+2006}
{Goldoni}, P., {Royer}, F., {Fran{\c c}ois}, P., {et~al.} 2006, in Society of
  Photo-Optical Instrumentation Engineers (SPIE) Conference Series, Vol. 6269,
  Society of Photo-Optical Instrumentation Engineers (SPIE) Conference Series,
  2

\bibitem[{{Graham} \& {Fruchter}(2013)}]{Graham+2013}
{Graham}, J.~F., \& {Fruchter}, A.~S. 2013, \apj, 774, 119

\bibitem[{{Greiner} {et~al.}(2008){Greiner}, {Bornemann}, {Clemens}, {Deuter},
  {Hasinger}, {Honsberg}, {Huber}, {Huber}, {Krauss}, {Kr{\"u}hler},
  {K{\"u}pc{\"u} Yolda{\c s}}, {Mayer-Hasselwander}, {Mican}, {Primak},
  {Schrey}, {Steiner}, {Szokoly}, {Th{\"o}ne}, {Yolda{\c s}}, {Klose}, {Laux},
  \& {Winkler}}]{Greiner+2008}
{Greiner}, J., {Bornemann}, W., {Clemens}, C., {et~al.} 2008, \pasp, 120, 405

\bibitem[{{Greiner} {et~al.}(2011){Greiner}, {Kr{\"u}hler}, {Klose}, {Afonso},
  {Clemens}, {Filgas}, {Hartmann}, {K{\"u}pc{\"u} Yolda{\c s}}, {Nardini},
  {Olivares E.}, {Rau}, {Rossi}, {Schady}, \& {Updike}}]{Greiner+2011}
{Greiner}, J., {Kr{\"u}hler}, T., {Klose}, S., {et~al.} 2011, \aap, 526, A30

\bibitem[{{Greiner} {et~al.}(2013){Greiner}, {Kr{\"u}hler}, {Nardini},
  {Filgas}, {Moin}, {de Breuck}, {Montenegro-Montes}, {Lundgren}, {Klose},
  {fonso}, {Bertoldi}, {Elliott}, {Kann}, {Knust}, {Menten}, {Nicuesa
  Guelbenzu}, {Olivares E.}, {Rau}, {Rossi}, {Schady}, {Schmidl}, {Siringo},
  {Spezzi}, {Sudilovsky}, {Tingay}, {Updike}, {Wang}, {Weiss}, {Wieringa}, \&
  {Wyrowski}}]{Greiner+2013}
{Greiner}, J., {Kr{\"u}hler}, T., {Nardini}, M., {et~al.} 2013, \aap, 560, A70

\bibitem[{{Greiner} {et~al.}(2015){Greiner}, {Fox}, {Schady}, {Kr{\"u}hler},
  {Trenti}, {Cikota}, {Bolmer}, {Elliott}, {Delvaux}, {Perna}, {Afonso},
  {Kann}, {Klose}, {Savaglio}, {Schmidl}, {Schweyer}, {Tanga}, \&
  {Varela}}]{Greiner+2015}
{Greiner}, J., {Fox}, D.~B., {Schady}, P., {et~al.} 2015, \apj, 809, 76

\bibitem[{{Groot} {et~al.}(1998){Groot}, {Galama}, {van Paradijs},
  {Kouveliotou}, {Wijers}, {Bloom}, {Tanvir}, {Vanderspek}, {Greiner},
  {Castro-Tirado}, {Gorosabel}, {von Hippel}, {Lehnert}, {Kuijken}, {Hoekstra},
  {Metcalfe}, {Howk}, {Conselice}, {Telting}, {Rutten}, {Rhoads}, {Cole},
  {Pisano}, {Naber}, \& {Schwarz}}]{Groot+1998}
{Groot}, P.~J., {Galama}, T.~J., {van Paradijs}, J., {et~al.} 1998, \apjl, 493,
  L27

\bibitem[{{Grupe} {et~al.}(2007){Grupe}, {Nousek}, {vanden Berk}, {Roming},
  {Burrows}, {Godet}, {Osborne}, \& {Gehrels}}]{Grupe+2007}
{Grupe}, D., {Nousek}, J.~A., {vanden Berk}, D.~E., {et~al.} 2007, \aj, 133,
  2216

\bibitem[{{Guetta} \& {Della Valle}(2007)}]{Guetta+2007a}
{Guetta}, D., \& {Della Valle}, M. 2007, \apjl, 657, L73

\bibitem[{{Guetta} \& {Piran}(2007)}]{Guetta+2007b}
{Guetta}, D., \& {Piran}, T. 2007, Journal of Cosmology and Astroparticle
  Physics, 7, 3

\bibitem[{{Guidorzi} {et~al.}(2006){Guidorzi}, {Mundell}, {Smith},
  {Monfardini}, {Gomboc}, {Steele}, {Mottram}, {Bode}, {Rol}, {O'Brien}, \&
  {Bannister}}]{GCN4661}
{Guidorzi}, C., {Mundell}, C.~G., {Smith}, R.~J., {et~al.} 2006, GRB
  Coordinates Network, 4661

\bibitem[{{Hao} \& {Yuan}(2013)}]{Hao+2013}
{Hao}, J.-M., \& {Yuan}, Y.-F. 2013, \apj, 772, 42

\bibitem[{{Hashimoto} {et~al.}(2015){Hashimoto}, {Perley}, {Ohta}, {Aoki},
  {Tanaka}, {Niino}, {Yabe}, \& {Kawai}}]{Hashimoto+2015}
{Hashimoto}, T., {Perley}, D.~A., {Ohta}, K., {et~al.} 2015, \apj, 806, 250

\bibitem[{{Hashimoto} {et~al.}(2010){Hashimoto}, {Ohta}, {Aoki}, {Tanaka},
  {Yabe}, {Kawai}, {Aoki}, {Furusawa}, {Hattori}, {Iye}, {Kawabata},
  {Kobayashi}, {Komiyama}, {Kosugi}, {Minowa}, {Mizumoto}, {Niino}, {Nomoto},
  {Noumaru}, {Ogasawara}, {Pyo}, {Sakamoto}, {Sekiguchi}, {Shirasaki},
  {Suzuki}, {Tajitsu}, {Takata}, {Tamagawa}, {Terada}, {Totani}, {Watanabe},
  {Yamada}, \& {Yoshida}}]{Hashimoto+2010}
{Hashimoto}, T., {Ohta}, K., {Aoki}, K., {et~al.} 2010, \apj, 719, 378

\bibitem[{{Hirschi} {et~al.}(2005){Hirschi}, {Meynet}, \&
  {Maeder}}]{Hirschi+2005}
{Hirschi}, R., {Meynet}, G., \& {Maeder}, A. 2005, \aap, 443, 581

\bibitem[{{Hjorth} {et~al.}(2012){Hjorth}, {Malesani}, {Jakobsson}, {Jaunsen},
  {Fynbo}, {Gorosabel}, {Kr{\"u}hler}, {Levan}, {Micha{\l}owski},
  {Milvang-Jensen}, {M{\o}ller}, {Schulze}, {Tanvir}, \&
  {Watson}}]{Hjorth+2012}
{Hjorth}, J., {Malesani}, D., {Jakobsson}, P., {et~al.} 2012, \apj, 756, 187

\bibitem[{{Hogg} {et~al.}(1997){Hogg}, {Pahre}, {McCarthy}, {Cohen},
  {Blandford}, {Smail}, \& {Soifer}}]{Hogg+1997}
{Hogg}, D.~W., {Pahre}, M.~A., {McCarthy}, J.~K., {et~al.} 1997, \mnras, 288,
  404

\bibitem[{{Hook} {et~al.}(2004){Hook}, {J{\o}rgensen}, {Allington-Smith},
  {Davies}, {Metcalfe}, {Murowinski}, \& {Crampton}}]{Hook+2004}
{Hook}, I.~M., {J{\o}rgensen}, I., {Allington-Smith}, J.~R., {et~al.} 2004,
  \pasp, 116, 425

\bibitem[{{Hopkins} \& {Beacom}(2006)}]{Hopkins+2006}
{Hopkins}, A.~M., \& {Beacom}, J.~F. 2006, \apj, 651, 142

\bibitem[{{Hunt} {et~al.}(2014){Hunt}, {Palazzi}, {Micha{\l}owski}, {Rossi},
  {Savaglio}, {Basa}, {Berta}, {Bianchi}, {Covino}, {D'Elia}, {Ferrero},
  {G{\"o}tz}, {Greiner}, {Klose}, {Le Borgne}, {Le Floc'h}, {Pian},
  {Piranomonte}, {Schady}, \& {Vergani}}]{Hunt+2014}
{Hunt}, L.~K., {Palazzi}, E., {Micha{\l}owski}, M.~J., {et~al.} 2014, \aap,
  565, A112

\bibitem[{{Jakobsson} {et~al.}(2009){Jakobsson}, {de Ugarte Postigo},
  {Gorosabel}, {Tanvir}, {Christensen}, \& {Fynbo}}]{GCN9797}
{Jakobsson}, P., {de Ugarte Postigo}, A., {Gorosabel}, J., {et~al.} 2009, GRB
  Coordinates Network, 9797

\bibitem[{{Jakobsson} {et~al.}(2004){Jakobsson}, {Hjorth}, {Fynbo}, {Watson},
  {Pedersen}, {Bj{\"o}rnsson}, \& {Gorosabel}}]{Jakobsson+2004}
{Jakobsson}, P., {Hjorth}, J., {Fynbo}, J.~P.~U., {et~al.} 2004, \apjl, 617,
  L21

\bibitem[{{Jakobsson} {et~al.}(2005){Jakobsson}, {Paraficz}, {Telting},
  {Fynbo}, {Jensen}, {Hjorth}, \& {Castro Cer{\'o}n}}]{GCN4015}
{Jakobsson}, P., {Paraficz}, D., {Telting}, J., {et~al.} 2005, GRB Coordinates
  Network, 4015

\bibitem[{{Jakobsson} {et~al.}(2006){Jakobsson}, {Levan}, {Fynbo}, {Priddey},
  {Hjorth}, {Tanvir}, {Watson}, {Jensen}, {Sollerman}, {Natarajan},
  {Gorosabel}, {Castro Cer{\'o}n}, {Pedersen}, {Pursimo}, {{\'A}rnad{\'o}ttir},
  {Castro-Tirado}, {Davis}, {Deeg}, {Fiuza}, {Mikolaitis}, \&
  {Sousa}}]{Jakobsson+2006}
{Jakobsson}, P., {Levan}, A., {Fynbo}, J.~P.~U., {et~al.} 2006, \aap, 447, 897

\bibitem[{{Jakobsson} {et~al.}(2007){Jakobsson}, {Fynbo}, {Andersen},
  {Jaunsen}, {Th{\"o}ne}, {Watson}, {Hjorth}, {Milvang-Jensen}, {Tanvir}, \&
  {Moller}}]{GCN6398}
{Jakobsson}, P., {Fynbo}, J.~P.~U., {Andersen}, M.~I., {et~al.} 2007, GRB
  Coordinates Network, 6398

\bibitem[{{Jakobsson} {et~al.}(2012){Jakobsson}, {Hjorth}, {Malesani},
  {Chapman}, {Fynbo}, {Tanvir}, {Milvang-Jensen}, {Vreeswijk}, {Letawe}, \&
  {Starling}}]{Jakobsson+2012}
{Jakobsson}, P., {Hjorth}, J., {Malesani}, D., {et~al.} 2012, \apj, 752, 62

\bibitem[{{Jaunsen} {et~al.}(2008){Jaunsen}, {Rol}, {Watson}, {Malesani},
  {Fynbo}, {Milvang-Jensen}, {Hjorth}, {Vreeswijk}, {Ovaldsen}, {Wiersema},
  {Tanvir}, {Gorosabel}, {Levan}, {Schirmer}, \&
  {Castro-Tirado}}]{Jaunsen+2008}
{Jaunsen}, A.~O., {Rol}, E., {Watson}, D.~J., {et~al.} 2008, \apj, 681, 453

\bibitem[{{Kann} {et~al.}(2010){Kann}, {Klose}, {Zhang}, {Malesani}, {Nakar},
  {Pozanenko}, {Wilson}, {Butler}, {Jakobsson}, {Schulze}, {Andreev},
  {Antonelli}, {Bikmaev}, {Biryukov}, {B{\"o}ttcher}, {Burenin}, {Castro
  Cer{\'o}n}, {Castro-Tirado}, {Chincarini}, {Cobb}, {Covino}, {D'Avanzo},
  {D'Elia}, {Della Valle}, {de Ugarte Postigo}, {Efimov}, {Ferrero}, {Fugazza},
  {Fynbo}, {G{\aa}lfalk}, {Grundahl}, {Gorosabel}, {Gupta}, {Guziy}, {Hafizov},
  {Hjorth}, {Holhjem}, {Ibrahimov}, {Im}, {Israel}, {Je{\'l}inek}, {Jensen},
  {Karimov}, {Khamitov}, {Kizilo{\v g}lu}, {Klunko}, {Kub{\'a}nek}, {Kutyrev},
  {Laursen}, {Levan}, {Mannucci}, {Martin}, {Mescheryakov}, {Mirabal},
  {Norris}, {Ovaldsen}, {Paraficz}, {Pavlenko}, {Piranomonte}, {Rossi},
  {Rumyantsev}, {Salinas}, {Sergeev}, {Sharapov}, {Sollerman}, {Stecklum},
  {Stella}, {Tagliaferri}, {Tanvir}, {Telting}, {Testa}, {Updike}, {Volnova},
  {Watson}, {Wiersema}, \& {Xu}}]{Kann+2010}
{Kann}, D.~A., {Klose}, S., {Zhang}, B., {et~al.} 2010, \apj, 720, 1513

\bibitem[{{Kawai} {et~al.}(2006){Kawai}, {Kosugi}, {Aoki}, {Yamada}, {Totani},
  {Ohta}, {Iye}, {Hattori}, {Aoki}, {Furusawa}, {Hurley}, {Kawabata},
  {Kobayashi}, {Komiyama}, {Mizumoto}, {Nomoto}, {Noumaru}, {Ogasawara},
  {Sato}, {Sekiguchi}, {Shirasaki}, {Suzuki}, {Takata}, {Tamagawa}, {Terada},
  {Watanabe}, {Yatsu}, \& {Yoshida}}]{Kawai+2006}
{Kawai}, N., {Kosugi}, G., {Aoki}, K., {et~al.} 2006, \nat, 440, 184

\bibitem[{{Kelly} {et~al.}(2014){Kelly}, {Filippenko}, {Modjaz}, \&
  {Kocevski}}]{Kelly+2014}
{Kelly}, P.~L., {Filippenko}, A.~V., {Modjaz}, M., \& {Kocevski}, D. 2014,
  \apj, 789, 23

\bibitem[{{Kistler} {et~al.}(2009){Kistler}, {Y{\"u}ksel}, {Beacom}, {Hopkins},
  \& {Wyithe}}]{Kistler+2009}
{Kistler}, M.~D., {Y{\"u}ksel}, H., {Beacom}, J.~F., {Hopkins}, A.~M., \&
  {Wyithe}, J.~S.~B. 2009, \apjl, 705, L104

\bibitem[{{Kistler} {et~al.}(2008){Kistler}, {Y{\"u}ksel}, {Beacom}, \&
  {Stanek}}]{Kistler+2008}
{Kistler}, M.~D., {Y{\"u}ksel}, H., {Beacom}, J.~F., \& {Stanek}, K.~Z. 2008,
  \apjl, 673, L119

\bibitem[{{Kocevski} \& {West}(2011)}]{Kocevski+2011}
{Kocevski}, D., \& {West}, A.~A. 2011, \apjl, 735, L8

\bibitem[{{Kohn} {et~al.}(2015){Kohn}, {Micha{\l}owski}, {Bourne}, {Baes},
  {Fritz}, {Cooray}, {De Looze}, {De Zotti}, {Dannerbauer}, {Dunne}, {Dye},
  {Eales}, {Furlanetto}, {Gonzalez-Nuevo}, {Ibar}, {Ivison}, {Maddox}, {Scott},
  {Smith}, {Smith}, {Symeonidis}, \& {Valiante}}]{Kohn+2015}
{Kohn}, S.~A., {Micha{\l}owski}, M.~J., {Bourne}, N., {et~al.} 2015, \mnras,
  448, 1494

\bibitem[{{Kr{\"u}hler} {et~al.}(2013){Kr{\"u}hler}, {Malesani}, {Xu}, {Fynbo},
  {Levan}, {Tanvir}, {D'Elia}, \& {Perley}}]{GCN14264}
{Kr{\"u}hler}, T., {Malesani}, D., {Xu}, D., {et~al.} 2013, GRB Coordinates
  Network, 14264

\bibitem[{{Kr{\"u}hler} {et~al.}(2008){Kr{\"u}hler}, {Schrey}, {Greiner},
  {Clemens}, {Yoldas}, {Kupcu Yoldas}, \& {Szokoly}}]{GCN8060}
{Kr{\"u}hler}, T., {Schrey}, F., {Greiner}, J., {et~al.} 2008, GRB Coordinates
  Network, 8060

\bibitem[{{Kr{\"u}hler} {et~al.}(2011){Kr{\"u}hler}, {Greiner}, {Schady},
  {Savaglio}, {Afonso}, {Clemens}, {Elliott}, {Filgas}, {Gruber}, {Kann},
  {Klose}, {K{\"u}pc{\"u}-Yolda{\c s}}, {McBreen}, {Olivares}, {Pierini},
  {Rau}, {Rossi}, {Nardini}, {Nicuesa Guelbenzu}, {Sudilovsky}, \&
  {Updike}}]{Kruehler+2011}
{Kr{\"u}hler}, T., {Greiner}, J., {Schady}, P., {et~al.} 2011, \aap, 534, A108

\bibitem[{{Kr{\"u}hler} {et~al.}(2012){Kr{\"u}hler}, {Malesani},
  {Milvang-Jensen}, {Fynbo}, {Hjorth}, {Jakobsson}, {Levan}, {Sparre},
  {Tanvir}, \& {Watson}}]{Kruehler+2012}
{Kr{\"u}hler}, T., {Malesani}, D., {Milvang-Jensen}, B., {et~al.} 2012, \apj,
  758, 46

\bibitem[{{Kr{\"u}hler} {et~al.}(2015){Kr{\"u}hler}, {Malesani}, {Fynbo},
  {Hartoog}, {Hjorth}, {Jakobsson}, {Perley}, {Rossi}, {Schady}, {Schulze},
  {Tanvir}, {Vergani}, {Wiersema}, {Afonso}, {Bolmer}, {Cano}, {Covino},
  {D'Elia}, {de Ugarte Postigo}, {Filgas}, {Friis}, {Graham}, {Greiner},
  {Goldoni}, {Gomboc}, {Hammer}, {Japelj}, {Kann}, {Kaper}, {Klose}, {Levan},
  {Leloudas}, {Milvang-Jensen}, {Nicuesa Guelbenzu}, {Palazzi}, {Pian},
  {Piranomonte}, {S{\'a}nchez-Ram{\'{\i}}rez}, {Savaglio}, {Selsing},
  {Tagliaferri}, {Vreeswijk}, {Watson}, \& {Xu}}]{Kruehler+2015}
{Kr{\"u}hler}, T., {Malesani}, D., {Fynbo}, J.~P.~U., {et~al.} 2015, \aap, 581,
  A125

\bibitem[{{Lamb} \& {Reichart}(2000)}]{Lamb+2000}
{Lamb}, D.~Q., \& {Reichart}, D.~E. 2000, \apj, 536, 1

\bibitem[{{Langer} \& {Norman}(2006)}]{Langer+2006}
{Langer}, N., \& {Norman}, C.~A. 2006, \apjl, 638, L63

\bibitem[{{Le} \& {Dermer}(2007)}]{Le+2007}
{Le}, T., \& {Dermer}, C.~D. 2007, \apj, 661, 394

\bibitem[{{Le Floc'h} {et~al.}(2006){Le Floc'h}, {Charmandaris}, {Forrest},
  {Mirabel}, {Armus}, \& {Devost}}]{LeFloch+2006}
{Le Floc'h}, E., {Charmandaris}, V., {Forrest}, W.~J., {et~al.} 2006, \apj,
  642, 636

\bibitem[{{Levan} {et~al.}(2007){Levan}, {Jakobsson}, {Hurkett}, {Tanvir},
  {Gorosabel}, {Vreeswijk}, {Rol}, {Chapman}, {Gehrels}, {O'Brien}, {Osborne},
  {Priddey}, {Kouveliotou}, {Starling}, {vanden Berk}, \&
  {Wiersema}}]{Levan+2007}
{Levan}, A.~J., {Jakobsson}, P., {Hurkett}, C., {et~al.} 2007, \mnras, 378,
  1439

\bibitem[{{Levesque} {et~al.}(2010){Levesque}, {Kewley}, {Berger}, \&
  {Zahid}}]{Levesque+2010mz}
{Levesque}, E.~M., {Kewley}, L.~J., {Berger}, E., \& {Zahid}, H.~J. 2010, \aj,
  140, 1557

\bibitem[{{Lien} {et~al.}(2014){Lien}, {Sakamoto}, {Gehrels}, {Palmer},
  {Barthelmy}, {Graziani}, \& {Cannizzo}}]{Lien+2014}
{Lien}, A., {Sakamoto}, T., {Gehrels}, N., {et~al.} 2014, \apj, 783, 24

\bibitem[{{MacFadyen} \& {Woosley}(1999)}]{MacFadyen+1999}
{MacFadyen}, A.~I., \& {Woosley}, S.~E. 1999, \apj, 524, 262

\bibitem[{{Madau} \& {Dickinson}(2014)}]{Madau+2014}
{Madau}, P., \& {Dickinson}, M. 2014, \araa, 52, 415

\bibitem[{{Madau} {et~al.}(1998){Madau}, {Pozzetti}, \&
  {Dickinson}}]{Madau+1998}
{Madau}, P., {Pozzetti}, L., \& {Dickinson}, M. 1998, \apj, 498, 106

\bibitem[{{McLean} {et~al.}(2012){McLean}, {Steidel}, {Epps}, {Konidaris},
  {Matthews}, {Adkins}, {Aliado}, {Brims}, {Canfield}, {Cromer}, {Fucik},
  {Kulas}, {Mace}, {Magnone}, {Rodriguez}, {Rudie}, {Trainor}, {Wang}, {Weber},
  \& {Weiss}}]{McLean+2012}
{McLean}, I.~S., {Steidel}, C.~C., {Epps}, H.~W., {et~al.} 2012, in Society of
  Photo-Optical Instrumentation Engineers (SPIE) Conference Series, Vol. 8446,
  Society of Photo-Optical Instrumentation Engineers (SPIE) Conference Series,
  0

\bibitem[{{Melandri} {et~al.}(2008){Melandri}, {Mundell}, {Kobayashi},
  {Guidorzi}, {Gomboc}, {Steele}, {Smith}, {Bersier}, {Mottram}, {Carter},
  {Bode}, {O'Brien}, {Tanvir}, {Rol}, \& {Chapman}}]{Melandri+2008}
{Melandri}, A., {Mundell}, C.~G., {Kobayashi}, S., {et~al.} 2008, \apj, 686,
  1209

\bibitem[{{Melandri} {et~al.}(2009){Melandri}, {Mundell}, {Cano}, {Smith},
  {Steele}, {Kobayashi}, {Mottram}, {Bersier}, {Gomboc}, \&
  {Guidorzi}}]{GCN9726}
{Melandri}, A., {Mundell}, C.~G., {Cano}, Z., {et~al.} 2009, GRB Coordinates
  Network, 9726

\bibitem[{{Melandri} {et~al.}(2012){Melandri}, {Sbarufatti}, {D'Avanzo},
  {Salvaterra}, {Campana}, {Covino}, {Vergani}, {Nava}, {Ghisellini},
  {Ghirlanda}, {Fugazza}, {Mangano}, {Capalbi}, \&
  {Tagliaferri}}]{Melandri+2012}
{Melandri}, A., {Sbarufatti}, B., {D'Avanzo}, P., {et~al.} 2012, \mnras, 421,
  1265

\bibitem[{{M{\'e}sz{\'a}ros} \& {Rees}(1997)}]{Meszaros+1997}
{M{\'e}sz{\'a}ros}, P., \& {Rees}, M.~J. 1997, \apj, 476, 232

\bibitem[{{Micha{\l}owski} {et~al.}(2008){Micha{\l}owski}, {Hjorth}, {Castro
  Cer{\'o}n}, \& {Watson}}]{Michalowski+2008}
{Micha{\l}owski}, M.~J., {Hjorth}, J., {Castro Cer{\'o}n}, J.~M., \& {Watson},
  D. 2008, \apj, 672, 817

\bibitem[{{Micha{\l}owski} {et~al.}(2012){Micha{\l}owski}, {Kamble}, {Hjorth},
  {Malesani}, {Reinfrank}, {Bonavera}, {Castro Cer{\'o}n}, {Ibar}, {Dunlop},
  {Fynbo}, {Garrett}, {Jakobsson}, {Kaplan}, {Kr{\"u}hler}, {Levan},
  {Massardi}, {Pal}, {Sollerman}, {Tanvir}, {van der Horst}, {Watson}, \&
  {Wiersema}}]{Michalowski+2012}
{Micha{\l}owski}, M.~J., {Kamble}, A., {Hjorth}, J., {et~al.} 2012, \apj, 755,
  85

\bibitem[{{Milvang-Jensen} {et~al.}(2012){Milvang-Jensen}, {Fynbo}, {Malesani},
  {Hjorth}, {Jakobsson}, \& {M{\o}ller}}]{MilvangJensen+2012}
{Milvang-Jensen}, B., {Fynbo}, J.~P.~U., {Malesani}, D., {et~al.} 2012, \apj,
  756, 25

\bibitem[{{Milvang-Jensen} {et~al.}(2010){Milvang-Jensen}, {Goldoni}, {Tanvir},
  {Wiersema}, {Malesani}, {de Ugarte Postigo}, {D'Elia}, {Vergani}, {Fynbo},
  {Kaper}, {Sollerman}, {Updike}, {Greiner}, \& {Krhler}}]{GCN10876}
{Milvang-Jensen}, B., {Goldoni}, P., {Tanvir}, N.~R., {et~al.} 2010, GRB
  Coordinates Network, 10876

\bibitem[{{Mirabal} \& {Halpern}(2006)}]{GCN4792}
{Mirabal}, N., \& {Halpern}, J.~P. 2006, GRB Coordinates Network, 4792

\bibitem[{{Modjaz} {et~al.}(2008){Modjaz}, {Kewley}, {Kirshner}, {Stanek},
  {Challis}, {Garnavich}, {Greene}, {Kelly}, \& {Prieto}}]{Modjaz+2008}
{Modjaz}, M., {Kewley}, L., {Kirshner}, R.~P., {et~al.} 2008, \aj, 135, 1136

\bibitem[{{Morgan}(2014)}]{Morgan+2014b}
{Morgan}, A.~N. 2014, PhD thesis, University of California, Berkeley

\bibitem[{{Morgan} {et~al.}(2009){Morgan}, {Bloom}, \& {Klein}}]{GCN9635}
{Morgan}, A.~N., {Bloom}, J.~S., \& {Klein}, C.~R. 2009, GRB Coordinates
  Network, 9635

\bibitem[{{Morgan} {et~al.}(2014){Morgan}, {Perley}, {Cenko}, {Bloom},
  {Cucchiara}, {Richards}, {Filippenko}, {Haislip}, {LaCluyze}, {Corsi},
  {Melandri}, {Cobb}, {Gomboc}, {Horesh}, {James}, {Li}, {Mundell}, {Reichart},
  \& {Steele}}]{Morgan+2014}
{Morgan}, A.~N., {Perley}, D.~A., {Cenko}, S.~B., {et~al.} 2014, \mnras, 440,
  1810

\bibitem[{{Mulchaey} \& {Berger}(2005)}]{GCN3114}
{Mulchaey}, J., \& {Berger}, E. 2005, GRB Coordinates Network, 3114

\bibitem[{{Mundell} {et~al.}(2011){Mundell}, {Smith}, \& {Cano}}]{GCN11633}
{Mundell}, C.~G., {Smith}, R.~J., \& {Cano}, Z. 2011, GRB Coordinates Network,
  11633

\bibitem[{{Mundell} {et~al.}(2006){Mundell}, {Melandri}, {Gomboc}, {Guidorzi},
  {Steele}, {Mottram}, {Monfardini}, {Smith}, {Bode}, {Rol}, {O'Brien}, \&
  {Bannister}}]{GCN4726}
{Mundell}, C.~G., {Melandri}, A., {Gomboc}, A., {et~al.} 2006, GRB Coordinates
  Network, 4726

\bibitem[{{Nysewander} {et~al.}(2009){Nysewander}, {Fruchter}, \&
  {Pe'er}}]{Nysewander+2009}
{Nysewander}, M., {Fruchter}, A.~S., \& {Pe'er}, A. 2009, \apj, 701, 824

\bibitem[{{Oates} \& {Markwardt}(2008)}]{GCN7253}
{Oates}, S.~R., \& {Markwardt}, C.~B. 2008, GRB Coordinates Network, 7253

\bibitem[{{Oke} {et~al.}(1995){Oke}, {Cohen}, {Carr}, {Cromer}, {Dingizian},
  {Harris}, {Labrecque}, {Lucinio}, {Schaal}, {Epps}, \& {Miller}}]{Oke+1995}
{Oke}, J.~B., {Cohen}, J.~G., {Carr}, M., {et~al.} 1995, \pasp, 107, 375

\bibitem[{{O'Meara} {et~al.}(2010){O'Meara}, {Chen}, \& {Prochaska}}]{GCN11089}
{O'Meara}, J., {Chen}, H.-W., \& {Prochaska}, J.~X. 2010, GRB Coordinates
  Network, 11089

\bibitem[{{Pavlenko} {et~al.}(2009){Pavlenko}, {Rumyantsev}, \&
  {Pozanenko}}]{GCN9179}
{Pavlenko}, E., {Rumyantsev}, V., \& {Pozanenko}, A. 2009, GRB Coordinates
  Network, 9179

\bibitem[{{Peek} \& {Graves}(2010)}]{Peek+2010}
{Peek}, J.~E.~G., \& {Graves}, G.~J. 2010, \apj, 719, 415

\bibitem[{{Perley} {et~al.}(2009{\natexlab{a}}){Perley}, {Prochaska}, {Kalas},
  {Howard}, {Fitzgerald}, {Marcy}, \& {Graham}}]{GCN10272}
{Perley}, D.~A., {Prochaska}, J.~X., {Kalas}, P., {et~al.} 2009{\natexlab{a}},
  GRB Coordinates Network, 10272

\bibitem[{{Perley} {et~al.}(2009{\natexlab{b}}){Perley}, {Metzger}, {Granot},
  {Butler}, {Sakamoto}, {Ramirez-Ruiz}, {Levan}, {Bloom}, {Miller}, {Bunker},
  {Chen}, {Filippenko}, {Gehrels}, {Glazebrook}, {Hall}, {Hurley}, {Kocevski},
  {Li}, {Lopez}, {Norris}, {Piro}, {Poznanski}, {Prochaska}, {Quataert}, \&
  {Tanvir}}]{Perley+2009b}
{Perley}, D.~A., {Metzger}, B.~D., {Granot}, J., {et~al.} 2009{\natexlab{b}},
  \apj, 696, 1871

\bibitem[{{Perley} {et~al.}(2009{\natexlab{c}}){Perley}, {Cenko}, {Bloom},
  {Chen}, {Butler}, {Kocevski}, {Prochaska}, {Brodwin}, {Glazebrook},
  {Kasliwal}, {Kulkarni}, {Lopez}, {Ofek}, {Pettini}, {Soderberg}, \&
  {Starr}}]{Perley+2009a}
{Perley}, D.~A., {Cenko}, S.~B., {Bloom}, J.~S., {et~al.} 2009{\natexlab{c}},
  \aj, 138, 1690

\bibitem[{{Perley} {et~al.}(2010){Perley}, {Bloom}, {Klein}, {Covino},
  {Minezaki}, {Wo{\'z}niak}, {Vestrand}, {Williams}, {Milne}, {Butler},
  {Updike}, {Kr{\"u}hler}, {Afonso}, {Antonelli}, {Cowie}, {Ferrero},
  {Greiner}, {Hartmann}, {Kakazu}, {K{\"u}pc{\"u} Yolda{\c s}}, {Morgan},
  {Price}, {Prochaska}, \& {Yoshii}}]{Perley+2010}
{Perley}, D.~A., {Bloom}, J.~S., {Klein}, C.~R., {et~al.} 2010, \mnras, 406,
  2473

\bibitem[{{Perley} {et~al.}(2011){Perley}, {Morgan}, {Updike}, {Yuan},
  {Akerlof}, {Miller}, {Bloom}, {Cenko}, {Li}, {Filippenko}, {Prochaska},
  {Kann}, {Tanvir}, {Levan}, {Butler}, {Christian}, {Hartmann}, {Milne},
  {Rykoff}, {Rujopakarn}, {Wheeler}, \& {Williams}}]{Perley+2011}
{Perley}, D.~A., {Morgan}, A.~N., {Updike}, A., {et~al.} 2011, \aj, 141, 36

\bibitem[{{Perley} {et~al.}(2013){Perley}, {Levan}, {Tanvir}, {Cenko}, {Bloom},
  {Hjorth}, {Kr{\"u}hler}, {Filippenko}, {Fruchter}, {Fynbo}, {Jakobsson},
  {Kalirai}, {Milvang-Jensen}, {Morgan}, {Prochaska}, \&
  {Silverman}}]{Perley+2013a}
{Perley}, D.~A., {Levan}, A.~J., {Tanvir}, N.~R., {et~al.} 2013, \apj, 778, 128

\bibitem[{{Perley} {et~al.}(2015{\natexlab{a}}){Perley}, {Perley}, {Hjorth},
  {Micha{\l}owski}, {Cenko}, {Jakobsson}, {Kr{\"u}hler}, {Levan}, {Malesani},
  \& {Tanvir}}]{Perley+2015a}
{Perley}, D.~A., {Perley}, R.~A., {Hjorth}, J., {et~al.} 2015{\natexlab{a}},
  \apj, 801, 102

\bibitem[{{Perley} {et~al.}(2016){Perley}, {Tanvir}, {Hjorth},
  {Laskar}, {Berger}, {Chary}, {de Ugarte Postigo}, {Fynbo}, {Kr{\"u}hler},
  {Levan}, {Micha{\l}owski}, \& {Schulze}}]{Paper2}
{Perley}, D.~A., {Tanvir}, N.~R., {Hjorth}, J., {et~al.} 2016,
  \apj, 817, 8

\bibitem[{{Pian} {et~al.}(2006){Pian}, {Mazzali}, {Masetti}, {Ferrero},
  {Klose}, {Palazzi}, {Ramirez-Ruiz}, {Woosley}, {Kouveliotou}, {Deng},
  {Filippenko}, {Foley}, {Fynbo}, {Kann}, {Li}, {Hjorth}, {Nomoto}, {Patat},
  {Sauer}, {Sollerman}, {Vreeswijk}, {Guenther}, {Levan}, {O'Brien}, {Tanvir},
  {Wijers}, {Dumas}, {Hainaut}, {Wong}, {Baade}, {Wang}, {Amati}, {Cappellaro},
  {Castro-Tirado}, {Ellison}, {Frontera}, {Fruchter}, {Greiner}, {Kawabata},
  {Ledoux}, {Maeda}, {M{\o}ller}, {Nicastro}, {Rol}, \& {Starling}}]{Pian+2006}
{Pian}, E., {Mazzali}, P.~A., {Masetti}, N., {et~al.} 2006, \nat, 442, 1011

\bibitem[{{Poole} {et~al.}(2005){Poole}, {Moretti}, {Holland}, {Chester},
  {Angelini}, \& {Gehrels}}]{GCN3698}
{Poole}, T., {Moretti}, A., {Holland}, S.~T., {et~al.} 2005, GRB Coordinates
  Network, 3698

\bibitem[{{Porciani} \& {Madau}(2001)}]{Porciani+2001}
{Porciani}, C., \& {Madau}, P. 2001, \apj, 548, 522

\bibitem[{{Price}(2006)}]{GCN5104}
{Price}, P.~A. 2006, GRB Coordinates Network, 5104

\bibitem[{{Prochaska} {et~al.}(2007){Prochaska}, {Chen}, {Dessauges-Zavadsky},
  \& {Bloom}}]{Prochaska+2007}
{Prochaska}, J.~X., {Chen}, H.-W., {Dessauges-Zavadsky}, M., \& {Bloom}, J.~S.
  2007, \apj, 666, 267

\bibitem[{{Prochaska} {et~al.}(2009){Prochaska}, {Sheffer}, {Perley}, {Bloom},
  {Lopez}, {Dessauges-Zavadsky}, {Chen}, {Filippenko}, {Ganeshalingam}, {Li},
  {Miller}, \& {Starr}}]{Prochaska+2009}
{Prochaska}, J.~X., {Sheffer}, Y., {Perley}, D.~A., {et~al.} 2009, \apjl, 691,
  L27

\bibitem[{{Ramirez-Ruiz} {et~al.}(2002){Ramirez-Ruiz}, {Trentham}, \&
  {Blain}}]{RamirezRuiz+2002}
{Ramirez-Ruiz}, E., {Trentham}, N., \& {Blain}, A.~W. 2002, \mnras, 329, 465

\bibitem[{{Rau} {et~al.}(2010){Rau}, {Fynbo}, \& {Greiner}}]{GCN10350}
{Rau}, A., {Fynbo}, J., \& {Greiner}, J. 2010, GRB Coordinates Network, 10350

\bibitem[{{Reddy} \& {Steidel}(2009)}]{Reddy+2009}
{Reddy}, N.~A., \& {Steidel}, C.~C. 2009, \apj, 692, 778

\bibitem[{{Rinner} \& {Kugel}(2008)}]{GCN7670}
{Rinner}, C., \& {Kugel}, F. 2008, GRB Coordinates Network, 7670

\bibitem[{{Robertson} \& {Ellis}(2012)}]{Robertson+2012}
{Robertson}, B.~E., \& {Ellis}, R.~S. 2012, \apj, 744, 95

\bibitem[{{Robertson} {et~al.}(2015){Robertson}, {Ellis}, {Furlanetto}, \&
  {Dunlop}}]{Robertson+2015}
{Robertson}, B.~E., {Ellis}, R.~S., {Furlanetto}, S.~R., \& {Dunlop}, J.~S.
  2015, \apjl, 802, L19

\bibitem[{{Rol} {et~al.}(2007){Rol}, {Tanvir}, {Mirabal}, {Wiersema},
  {Halpern}, {Levan}, {Chapman}, {Melandri}, \& {Pinfield}}]{GCN6221}
{Rol}, E., {Tanvir}, N., {Mirabal}, N., {et~al.} 2007, GRB Coordinates Network,
  6221

\bibitem[{{Roming} {et~al.}(2005){Roming}, {Kennedy}, {Mason}, {Nousek}, {Ahr},
  {Bingham}, {Broos}, {Carter}, {Hancock}, {Huckle}, {Hunsberger}, {Kawakami},
  {Killough}, {Koch}, {McLelland}, {Smith}, {Smith}, {Soto}, {Boyd},
  {Breeveld}, {Holland}, {Ivanushkina}, {Pryzby}, {Still}, \&
  {Stock}}]{Roming+2005}
{Roming}, P.~W.~A., {Kennedy}, T.~E., {Mason}, K.~O., {et~al.} 2005, \ssr, 120,
  95

\bibitem[{{Rossi} {et~al.}(2012){Rossi}, {Klose}, {Ferrero}, {Greiner},
  {Arnold}, {Gonsalves}, {Hartmann}, {Updike}, {Kann}, {Kr{\"u}hler},
  {Palazzi}, {Savaglio}, {Schulze}, {Afonso}, {Amati}, {Castro-Tirado},
  {Clemens}, {Filgas}, {Gorosabel}, {Hunt}, {K{\"u}pc{\"u} Yolda{\c s}},
  {Masetti}, {Nardini}, {Nicuesa Guelbenzu}, {Olivares}, {Pian}, {Rau},
  {Schady}, {Schmidl}, {Yolda{\c s}}, \& {de Ugarte Postigo}}]{Rossi+2012}
{Rossi}, A., {Klose}, S., {Ferrero}, P., {et~al.} 2012, \aap, 545, A77

\bibitem[{{Ruiz-Velasco} {et~al.}(2007){Ruiz-Velasco}, {Swan}, {Troja},
  {Malesani}, {Fynbo}, {Starling}, {Xu}, {Aharonian}, {Akerlof}, {Andersen},
  {Ashley}, {Barthelmy}, {Bersier}, {Castro Cer{\'o}n}, {Castro-Tirado},
  {Gehrels}, {G{\"o}{\v g}{\"u}{\c s}}, {Gorosabel}, {Guidorzi}, {G{\"u}ver},
  {Hjorth}, {Horns}, {Huang}, {Jakobsson}, {Jensen}, {K{\i}z{\i}lo{\v g}lu},
  {Kouveliotou}, {Krimm}, {Ledoux}, {Levan}, {Marsh}, {McKay}, {Melandri},
  {Milvang-Jensen}, {Mundell}, {O'Brien}, {{\"O}zel}, {Phillips}, {Quimby},
  {Rowell}, {Rujopakarn}, {Rykoff}, {Schaefer}, {Sollerman}, {Tanvir},
  {Th{\"o}ne}, {Urata}, {Vestrand}, {Vreeswijk}, {Watson}, {Wheeler}, {Wijers},
  {Wren}, {Yost}, {Yuan}, {Zhai}, \& {Zheng}}]{Ruiz-Velasco+2007}
{Ruiz-Velasco}, A.~E., {Swan}, H., {Troja}, E., {et~al.} 2007, \apj, 669, 1

\bibitem[{{Sakamoto} {et~al.}(2011){Sakamoto}, {Barthelmy}, {Baumgartner},
  {Cummings}, {Fenimore}, {Gehrels}, {Krimm}, {Markwardt}, {Palmer}, {Parsons},
  {Sato}, {Stamatikos}, {Tueller}, {Ukwatta}, \& {Zhang}}]{Sakamoto+2011}
{Sakamoto}, T., {Barthelmy}, S.~D., {Baumgartner}, W.~H., {et~al.} 2011, \apjs,
  195, 2

\bibitem[{{Salvaterra} \& {Chincarini}(2007)}]{Salvaterra+2007}
{Salvaterra}, R., \& {Chincarini}, G. 2007, \apjl, 656, L49

\bibitem[{{Salvaterra} {et~al.}(2009){Salvaterra}, {Della Valle}, {Campana},
  {Chincarini}, {Covino}, {D'Avanzo}, {Fern{\'a}ndez-Soto}, {Guidorzi},
  {Mannucci}, {Margutti}, {Th{\"o}ne}, {Antonelli}, {Barthelmy}, {de Pasquale},
  {D'Elia}, {Fiore}, {Fugazza}, {Hunt}, {Maiorano}, {Marinoni}, {Marshall},
  {Molinari}, {Nousek}, {Pian}, {Racusin}, {Stella}, {Amati}, {Andreuzzi},
  {Cusumano}, {Fenimore}, {Ferrero}, {Giommi}, {Guetta}, {Holland}, {Hurley},
  {Israel}, {Mao}, {Markwardt}, {Masetti}, {Pagani}, {Palazzi}, {Palmer},
  {Piranomonte}, {Tagliaferri}, \& {Testa}}]{Salvaterra+2009}
{Salvaterra}, R., {Della Valle}, M., {Campana}, S., {et~al.} 2009, \nat, 461,
  1258

\bibitem[{{Salvaterra} {et~al.}(2012){Salvaterra}, {Campana}, {Vergani},
  {Covino}, {D'Avanzo}, {Fugazza}, {Ghirlanda}, {Ghisellini}, {Melandri},
  {Nava}, {Sbarufatti}, {Flores}, {Piranomonte}, \&
  {Tagliaferri}}]{Salvaterra+2012}
{Salvaterra}, R., {Campana}, S., {Vergani}, S.~D., {et~al.} 2012, \apj, 749, 68

\bibitem[{{Sari} \& {Piran}(1997)}]{Sari+1997}
{Sari}, R., \& {Piran}, T. 1997, \apj, 485, 270

\bibitem[{{Schady} {et~al.}(2014){Schady}, {Savaglio}, {M{\"u}ller},
  {Kr{\"u}hler}, {Dwelly}, {Palazzi}, {Hunt}, {Greiner}, {Linz},
  {Micha{\l}owski}, {Pierini}, {Piranomonte}, {Vergani}, \&
  {Gear}}]{Schady+2014}
{Schady}, P., {Savaglio}, S., {M{\"u}ller}, T., {et~al.} 2014, \aap, 570, A52

\bibitem[{{Schlafly} \& {Finkbeiner}(2011)}]{Schlafly+2011}
{Schlafly}, E.~F., \& {Finkbeiner}, D.~P. 2011, \apj, 737, 103

\bibitem[{{Schlegel} {et~al.}(1998){Schlegel}, {Finkbeiner}, \&
  {Davis}}]{Schlegel+1998}
{Schlegel}, D.~J., {Finkbeiner}, D.~P., \& {Davis}, M. 1998, \apj, 500, 525

\bibitem[{{Schulze} {et~al.}(2015){Schulze}, {Chapman}, {Hjorth}, {Levan},
  {Jakobsson}, {Bj{\"o}rnsson}, {Perley}, {Kr{\"u}hler}, {Gorosabel}, {Tanvir},
  {de Ugarte Postigo}, {Fynbo}, {Milvang-Jensen}, {M{\o}ller}, \&
  {Watson}}]{Schulze+2015}
{Schulze}, S., {Chapman}, R., {Hjorth}, J., {et~al.} 2015, \apj, 808, 73

\bibitem[{{Skrutskie} {et~al.}(2006){Skrutskie}, {Cutri}, {Stiening},
  {Weinberg}, {Schneider}, {Carpenter}, {Beichman}, {Capps}, {Chester},
  {Elias}, {Huchra}, {Liebert}, {Lonsdale}, {Monet}, {Price}, {Seitzer},
  {Jarrett}, {Kirkpatrick}, {Gizis}, {Howard}, {Evans}, {Fowler}, {Fullmer},
  {Hurt}, {Light}, {Kopan}, {Marsh}, {McCallon}, {Tam}, {Van Dyk}, \&
  {Wheelock}}]{2MASS}
{Skrutskie}, M.~F., {Cutri}, R.~M., {Stiening}, R., {et~al.} 2006, \aj, 131,
  1163

\bibitem[{{Sparre} {et~al.}(2014){Sparre}, {Hartoog}, {Kr{\"u}hler}, {Fynbo},
  {Watson}, {Wiersema}, {D'Elia}, {Zafar}, {Afonso}, {Covino}, {de Ugarte
  Postigo}, {Flores}, {Goldoni}, {Greiner}, {Hjorth}, {Jakobsson}, {Kaper},
  {Klose}, {Levan}, {Malesani}, {Milvang-Jensen}, {Nardini}, {Piranomonte},
  {Sollerman}, {S{\'a}nchez-Ram{\'{\i}}rez}, {Schulze}, {Tanvir}, {Vergani}, \&
  {Wijers}}]{Sparre+2014}
{Sparre}, M., {Hartoog}, O.~E., {Kr{\"u}hler}, T., {et~al.} 2014, \apj, 785,
  150

\bibitem[{{Stanek} {et~al.}(2006){Stanek}, {Gnedin}, {Beacom}, {Gould},
  {Johnson}, {Kollmeier}, {Modjaz}, {Pinsonneault}, {Pogge}, \&
  {Weinberg}}]{Stanek+2006}
{Stanek}, K.~Z., {Gnedin}, O.~Y., {Beacom}, J.~F., {et~al.} 2006, ActaA, 56,
  333

\bibitem[{{Svensson} {et~al.}(2010){Svensson}, {Levan}, {Tanvir}, {Fruchter},
  \& {Strolger}}]{Svensson+2010}
{Svensson}, K.~M., {Levan}, A.~J., {Tanvir}, N.~R., {Fruchter}, A.~S., \&
  {Strolger}, L.-G. 2010, \mnras, 405, 57

\bibitem[{{Svensson} {et~al.}(2012){Svensson}, {Levan}, {Tanvir}, {Perley},
  {Michalowski}, {Page}, {Bloom}, {Cenko}, {Hjorth}, {Jakobsson}, {Watson}, \&
  {Wheatley}}]{Svensson+2012}
{Svensson}, K.~M., {Levan}, A.~J., {Tanvir}, N.~R., {et~al.} 2012, \mnras, 421,
  25

\bibitem[{{Tanvir} {et~al.}(2004){Tanvir}, {Barnard}, {Blain}, {Fruchter},
  {Kouveliotou}, {Natarajan}, {Ramirez-Ruiz}, {Rol}, {Smith}, {Tilanus}, \&
  {Wijers}}]{Tanvir+2004}
{Tanvir}, N.~R., {Barnard}, V.~E., {Blain}, A.~W., {et~al.} 2004, \mnras, 352,
  1073

\bibitem[{{Tanvir} {et~al.}(2009){Tanvir}, {Fox}, {Levan}, {Berger},
  {Wiersema}, {Fynbo}, {Cucchiara}, {Kr{\"u}hler}, {Gehrels}, {Bloom},
  {Greiner}, {Evans}, {Rol}, {Olivares}, {Hjorth}, {Jakobsson}, {Farihi},
  {Willingale}, {Starling}, {Cenko}, {Perley}, {Maund}, {Duke}, {Wijers},
  {Adamson}, {Allan}, {Bremer}, {Burrows}, {Castro-Tirado}, {Cavanagh}, {de
  Ugarte Postigo}, {Dopita}, {Fatkhullin}, {Fruchter}, {Foley}, {Gorosabel},
  {Kennea}, {Kerr}, {Klose}, {Krimm}, {Komarova}, {Kulkarni}, {Moskvitin},
  {Mundell}, {Naylor}, {Page}, {Penprase}, {Perri}, {Podsiadlowski}, {Roth},
  {Rutledge}, {Sakamoto}, {Schady}, {Schmidt}, {Soderberg}, {Sollerman},
  {Stephens}, {Stratta}, {Ukwatta}, {Watson}, {Westra}, {Wold}, \&
  {Wolf}}]{Tanvir+2009}
{Tanvir}, N.~R., {Fox}, D.~B., {Levan}, A.~J., {et~al.} 2009, \nat, 461, 1254

\bibitem[{{Tanvir} {et~al.}(2012){Tanvir}, {Levan}, {Fruchter}, {Fynbo},
  {Hjorth}, {Wiersema}, {Bremer}, {Rhoads}, {Jakobsson}, {O'Brien}, {Stanway},
  {Bersier}, {Natarajan}, {Greiner}, {Watson}, {Castro-Tirado}, {Wijers},
  {Starling}, {Misra}, {Graham}, \& {Kouveliotou}}]{Tanvir+2012}
{Tanvir}, N.~R., {Levan}, A.~J., {Fruchter}, A.~S., {et~al.} 2012, \apj, 754,
  46

\bibitem[{{Th{\"o}ne} {et~al.}(2009{\natexlab{a}}){Th{\"o}ne}, {Malesani},
  {Levan}, {Jakobsson}, {Hjorth}, \& {Tanvir}}]{GCN9403}
{Th{\"o}ne}, C.~C., {Malesani}, D., {Levan}, A.~J., {et~al.}
  2009{\natexlab{a}}, GRB Coordinates Network, 9403

\bibitem[{{Th{\"o}ne} {et~al.}(2009{\natexlab{b}}){Th{\"o}ne}, {Jakobsson}, {De
  Cia}, {Levan}, {Fynbo}, {Hjorth}, {Malesani}, {Tanvir}, {Fugazza}, \&
  {D'Avanzo}}]{GCN9409}
{Th{\"o}ne}, C.~C., {Jakobsson}, P., {De Cia}, A., {et~al.} 2009{\natexlab{b}},
  GRB Coordinates Network, 9409

\bibitem[{{Th{\"o}ne} {et~al.}(2013){Th{\"o}ne}, {Fynbo}, {Goldoni}, {de
  Ugarte}, {Campana}, {Vergani}, {Covino}, {Kr{\"u}hler}, {Kaper}, {Tanvir},
  {Zafar}, {D'Elia}, {Gorosabel}, {Greiner}, {Groot}, {Hammer}, {Jakobsson},
  {Klose}, {Levan}, {Milvang-Jensen}, {Nicuesa}, {Palazzi}, {Piranomonte},
  {Tagliaferri}, {Watson}, {Wiersema}, \& {Wijers}}]{Thoene+2013}
{Th{\"o}ne}, C.~C., {Fynbo}, J.~P.~U., {Goldoni}, P., {et~al.} 2013, \mnras,
  428, 3590

\bibitem[{{Totani}(1997)}]{Totani1997}
{Totani}, T. 1997, \apjl, 486, L71

\bibitem[{{Trenti} {et~al.}(2012){Trenti}, {Perna}, {Levesque}, {Shull}, \&
  {Stocke}}]{Trenti+2012}
{Trenti}, M., {Perna}, R., {Levesque}, E.~M., {Shull}, J.~M., \& {Stocke},
  J.~T. 2012, \apjl, 749, L38

\bibitem[{{Trenti} {et~al.}(2013){Trenti}, {Perna}, \&
  {Tacchella}}]{Trenti+2013}
{Trenti}, M., {Perna}, R., \& {Tacchella}, S. 2013, \apjl, 773, L22

\bibitem[{{Updike} {et~al.}(2010){Updike}, {Nicuesa}, {Nardini}, {Kr{\"u}hler},
  \& {Greiner}}]{GCN10874}
{Updike}, A., {Nicuesa}, A., {Nardini}, M., {Kr{\"u}hler}, T., \& {Greiner}, J.
  2010, GRB Coordinates Network, 10874

\bibitem[{{van den Heuvel} \& {Portegies Zwart}(2013)}]{vandenHeuvel+2013}
{van den Heuvel}, E.~P.~J., \& {Portegies Zwart}, S.~F. 2013, \apj, 779, 114

\bibitem[{{van der Horst} {et~al.}(2009){van der Horst}, {Kouveliotou},
  {Gehrels}, {Rol}, {Wijers}, {Cannizzo}, {Racusin}, \&
  {Burrows}}]{vanderHorst+2009}
{van der Horst}, A.~J., {Kouveliotou}, C., {Gehrels}, N., {et~al.} 2009, \apj,
  699, 1087

\bibitem[{{Vergani} {et~al.}(2015){Vergani}, {Salvaterra}, {Japelj}, {Le
  Floc'h}, {D'Avanzo}, {Fernandez-Soto}, {Kr{\"u}hler}, {Melandri}, {Boissier},
  {Covino}, {Puech}, {Greiner}, {Hunt}, {Perley}, {Petitjean}, {Vinci},
  {Hammer}, {Levan}, {Mannucci}, {Campana}, {Flores}, {Gomboc}, \&
  {Tagliaferri}}]{Vergani+2015}
{Vergani}, S.~D., {Salvaterra}, R., {Japelj}, J., {et~al.} 2015, \aap, 581,
  A102

\bibitem[{{Vernet} {et~al.}(2011){Vernet}, {Dekker}, {D'Odorico}, {Kaper},
  {Kjaergaard}, {Hammer}, {Randich}, {Zerbi}, {Groot}, {Hjorth}, {Guinouard},
  {Navarro}, {Adolfse}, {Albers}, {Amans}, {Andersen}, {Andersen}, {Binetruy},
  {Bristow}, {Castillo}, {Chemla}, {Christensen}, {Conconi}, {Conzelmann},
  {Dam}, {de Caprio}, {de Ugarte Postigo}, {Delabre}, {di Marcantonio},
  {Downing}, {Elswijk}, {Finger}, {Fischer}, {Flores}, {Fran{\c c}ois},
  {Goldoni}, {Guglielmi}, {Haigron}, {Hanenburg}, {Hendriks}, {Horrobin},
  {Horville}, {Jessen}, {Kerber}, {Kern}, {Kiekebusch}, {Kleszcz}, {Klougart},
  {Kragt}, {Larsen}, {Lizon}, {Lucuix}, {Mainieri}, {Manuputy}, {Martayan},
  {Mason}, {Mazzoleni}, {Michaelsen}, {Modigliani}, {Moehler}, {M{\o}ller},
  {Norup S{\o}rensen}, {N{\o}rregaard}, {P{\'e}roux}, {Patat}, {Pena}, {Pragt},
  {Reinero}, {Rigal}, {Riva}, {Roelfsema}, {Royer}, {Sacco}, {Santin},
  {Schoenmaker}, {Spano}, {Sweers}, {Ter Horst}, {Tintori}, {Tromp}, {van
  Dael}, {van der Vliet}, {Venema}, {Vidali}, {Vinther}, {Vola}, {Winters},
  {Wistisen}, {Wulterkens}, \& {Zacchei}}]{Vernet+2011}
{Vernet}, J., {Dekker}, H., {D'Odorico}, S., {et~al.} 2011, \aap, 536, A105

\bibitem[{{Virgili} {et~al.}(2009){Virgili}, {Liang}, \&
  {Zhang}}]{Virgili+2009}
{Virgili}, F.~J., {Liang}, E.-W., \& {Zhang}, B. 2009, \mnras, 392, 91

\bibitem[{{Virgili} {et~al.}(2011){Virgili}, {Zhang}, {Nagamine}, \&
  {Choi}}]{Virgili+2011}
{Virgili}, F.~J., {Zhang}, B., {Nagamine}, K., \& {Choi}, J.-H. 2011, \mnras,
  417, 3025

\bibitem[{{Wanderman} \& {Piran}(2010)}]{Wanderman+2010}
{Wanderman}, D., \& {Piran}, T. 2010, \mnras, 406, 1944

\bibitem[{{Wang} \& {Dai}(2011)}]{Wang+2011}
{Wang}, F.~Y., \& {Dai}, Z.~G. 2011, \apjl, 727, L34

\bibitem[{{Wiersema} {et~al.}(2009){Wiersema}, {Levan}, {Kamble}, {Tanvir}, \&
  {Malesani}}]{GCN9673}
{Wiersema}, K., {Levan}, A., {Kamble}, A., {Tanvir}, N., \& {Malesani}, D.
  2009, GRB Coordinates Network, 9673

\bibitem[{{Wiersema} {et~al.}(2007){Wiersema}, {Savaglio}, {Vreeswijk},
  {Ellison}, {Ledoux}, {Yoon}, {M{\o}ller}, {Sollerman}, {Fynbo}, {Pian},
  {Starling}, \& {Wijers}}]{Wiersema+2007}
{Wiersema}, K., {Savaglio}, S., {Vreeswijk}, P.~M., {et~al.} 2007, \aap, 464,
  529

\bibitem[{{Wijers} {et~al.}(1998){Wijers}, {Bloom}, {Bagla}, \&
  {Natarajan}}]{Wijers+1998}
{Wijers}, R.~A.~M.~J., {Bloom}, J.~S., {Bagla}, J.~S., \& {Natarajan}, P. 1998,
  \mnras, 294, L13

\bibitem[{{Woosley} \& {Heger}(2006)}]{Woosley+2006}
{Woosley}, S.~E., \& {Heger}, A. 2006, \apj, 637, 914

\bibitem[{{Yoon} \& {Langer}(2005)}]{Yoon+2005}
{Yoon}, S.-C., \& {Langer}, N. 2005, \aap, 443, 643

\bibitem[{{Yu} {et~al.}(2015){Yu}, {Wang}, {Dai}, \& {Cheng}}]{Yu+2015}
{Yu}, H., {Wang}, F.~Y., {Dai}, Z.~G., \& {Cheng}, K.~S. 2015, \apjs, 218, 13

\bibitem[{{Y{\"u}ksel} {et~al.}(2008){Y{\"u}ksel}, {Kistler}, {Beacom}, \&
  {Hopkins}}]{Yueksel+2008}
{Y{\"u}ksel}, H., {Kistler}, M.~D., {Beacom}, J.~F., \& {Hopkins}, A.~M. 2008,
  \apjl, 683, L5

\bibitem[{{Zafar} {et~al.}(2012){Zafar}, {Watson}, {El{\'{\i}}asd{\'o}ttir},
  {Fynbo}, {Kr{\"u}hler}, {Schady}, {Leloudas}, {Jakobsson}, {Th{\"o}ne},
  {Perley}, {Morgan}, {Bloom}, \& {Greiner}}]{Zafar+2012}
{Zafar}, T., {Watson}, D., {El{\'{\i}}asd{\'o}ttir}, {\'A}., {et~al.} 2012,
  \apj, 753, 82

\bibitem[{{Zauderer} \& {Berger}(2011)}]{GCN12190}
{Zauderer}, A., \& {Berger}, E. 2011, GRB Coordinates Network, 12190

\bibitem[{{Zauderer} {et~al.}(2013){Zauderer}, {Berger}, {Margutti}, {Levan},
  {Olivares E.}, {Perley}, {Fong}, {Horesh}, {Updike}, {Greiner}, {Tanvir},
  {Laskar}, {Chornock}, {Soderberg}, {Menten}, {Nakar}, {Carpenter}, {Chandra},
  {Castro-Tirado}, {Bremer}, {Gorosabel}, {Guziy}, {P{\'e}rez-Ram{\'{\i}}rez},
  \& {Winters}}]{Zauderer+2013}
{Zauderer}, B.~A., {Berger}, E., {Margutti}, R., {et~al.} 2013, \apj, 767, 161

\end{thebibliography}

\bibliographystyle{apj}

\end{document}